\documentclass[10pt,aps,prc,floatfix,twocolumn,nofootinbib]{revtex4-2}

\usepackage[labelfont={small},subrefformat=parens,caption=false]{subfig}
\usepackage[dvipsnames]{xcolor}
\usepackage{amsfonts,amsmath,amssymb,bm}
\usepackage{graphicx}
\usepackage[utf8]{inputenc}
\usepackage{xspace}
\usepackage{isotope}
\usepackage{xparse}
\usepackage{babel}

\usepackage[pdfpagelabels, pdfencoding=auto, psdextra]{hyperref}
 \hypersetup{%
  pdfsubject=Paper,
  pdfkeywords={nuclear physics} {few-body} {many-body} {scattering} {reaction} {continuum}{NHQM} {emulator} {surrogate model},
  unicode = true,
  breaklinks = true,
  colorlinks = true,
  linkcolor = blue,
  citecolor = blue,
  menucolor = blue,
  citecolor = blue,
  urlcolor = blue
 }

\usepackage{orcidlink}
\graphicspath{{./}}

\newcommand\schro{Schr\"{o}dinger\xspace}
\newcommand{\adj}[1]{\tilde{#1}}
\newcommand{\soket}[1]{|{S}_{#1}\rangle}

\newcommand{\soadjket}[1]{|\tilde{{S}}_{#1}\rangle} 
 
\newcommand{\so}{{S}}
\newcommand{\soadj}{\tilde{{S}}}
\newcommand{\psiadj}[1]{\tilde{\psi}_{#1}}

\newcommand{\wf}{\Psi}
\newcommand{\wfadj}{\tilde{\Psi}}

\newcommand{\matrixform}[1]{\begin{bmatrix} #1   \end{bmatrix}}

\newcommand{\resolventamp}{\mathcal{A}}

\newcommand{\vectheta}{\bm{\theta}}
\newcommand{\LAMtwo}{\Lambda_2}
\newcommand{\lamtwo}{\lambda}
\newcommand{\lamfour}{\lambda_4}
\newcommand{\Pin}{P_\mathrm{rel}}
\newcommand{\Ein}{E_\mathrm{rel}}
\newcommand{\bsren}{\xi}
\newcommand{\perm}{\mathbb{P}}
\renewcommand{\Re}{\mathrm{Re}\,}
\renewcommand{\Im}{\mathrm{Im}\,}
\newcommand{\Neff}{N_\mathrm{eff}}
\newcommand{\RIR}{R_{IR}}

\newcommand\BCT{BCT\xspace}
\newcommand{\Treversal}{\mathbb{T}}

\DeclareMathOperator\erfc{Erfc}

\usepackage{amsmath,environ}
\NewEnviron{EqS}{%
\begin{equation}\begin{split}
  \BODY
\end{split}\end{equation}
}
\NewEnviron{EqSn}{%
\begin{equation*}\begin{split}
  \BODY
\end{split}\end{equation*}
}

\begin{document}

\title{Non-Hermitian quantum mechanics approach for extracting and emulating continuum physics based on bound-state-like calculations: Detailed description} 

\author{Xilin Zhang\,\orcidlink{0000-0001-9278-5359}} 
\email{zhangx@frib.msu.edu}
\affiliation{\href{https://ror.org/03r4g9w46}{Facility for Rare Isotope Beams}, \href{https://ror.org/05hs6h993}{Michigan State University}, East Lansing, MI~48824, USA}

\date{\today}

\begin{abstract}
This work applies a reduced basis method to study the continuum physics of a finite quantum system---either few or many-body. Specifically, I develop reduced-order models, or emulators, for the underlying inhomogeneous Schr\"{o}dinger equation and train the emulators against the equation's bound-state-like solutions at complex energies. The emulators rapidly and accurately interpolate and extrapolate the matrix elements of the Hamiltonian resolvent operator (Green's function) across a parameter space that includes both complex energy and other real-valued physical inputs in the Schr\"{o}dinger equation. The spectra, discretized and compressed as the result of emulation, and the associated resolvent matrix elements (or amplitudes), have the defining characteristics of non-Hermitian quantum mechanics calculations, featuring complex eigenenergies with negative imaginary parts and  branch cuts moved below the real axis in the complex energy plane. Therefore, one now has a method that extracts continuum physics from bound-state-like calculations and emulates those extractions in the input parameter space. Building on a prior Letter [arXiv:2408.03309], this article provides the full theoretical details, a comprehensive analysis of the method's performance, and a brief discussion of how it can be coupled with existing continuum approaches to perform emulations in their input parameter spaces.

\end{abstract}

\maketitle

\section{Introduction} \label{sec:introduction}

In a recent article~\cite{Zhang:2024ril}, I reported on a new application of the reduced basis method (RBM)~\cite{Duguet:2023wuh, Drischler:2022ipa, Melendez:2022kid} in the study of continuum physics of \emph{finite} quantum systems. Technical details of the study are provided in this article. Readers are advised to read the short report~\cite{Zhang:2024ril} first, which could facilitate reading the current article.

By continuum physics, I mean the part of the spectrum of a Hamiltonian operator where the system can break up into subsystems (i.e., above the system's lowest threshold) and the states and observables associated with that sector of the spectrum. 
To be quantitative, let $H(\vectheta)$ be the Hamiltonian operator.
An important operator is $H$'s resolvent or Green's function that varies with the total energy $E$. Its matrix element between two source states $|\so(\vectheta)\rangle$ and $|\soadj(\vectheta)\rangle$ is named amplitude $\resolventamp$, with  

\begin{align}
    \resolventamp(E,\vectheta) \equiv \left\langle \soadj(\vectheta) \left\vert \frac{1}{E - H(\vectheta)} \right\vert \so(\vectheta) \right\rangle \, .\label{eq:resolventdef}
\end{align}
The vector $\vectheta$ collects the parameters of $H$, $\so$ and $\soadj$. 

$\resolventamp$ is related to an array of continuum physics observables, such as response functions and scattering amplitudes, depending on the construction of the source states. See Sec.~\ref{subsec:observables} for the details. 
The analytical properties of $\resolventamp$ in the complex $E$ plane, such as isolated poles and branch cuts ({\BCT}s), are directly connected to the basic features of $H$'s spectrum~\cite{newton2002scattering}.\footnote{$\resolventamp$'s behavior in the complex $E$ plane is different from that of the scattering $S$-matrix. The latter (see e.g.,~\cite{HUMBLET1961529,JoachainQCT1975,newton2002scattering,RamirezJimenez:2018dge,Yamada:2022xam}) could be more complicated even in the first Riemann sheet.} This informs an interesting numeric computational framework, called non-Hermitian quantum mechanics (NHQM)~\cite{ReinhardtComlexScaling2007,reinhardt1982complex, Moiseyev_2011,Myo:2014ypa, afnan1991resonances}, as discussed in Sec.~\ref{subsec:NHQM}. Since the approximations of $\resolventamp$ produced in a broad class of calculations (including NHQM) could be viewed as rational approximations\footnote{The rational approximation refers to approximating a function in terms of a ratio between two polynomials. Here, I focus on a particular type that treats a univariate function as a finite sum of simple  poles.}~\cite{Trefethen_approximation_theory_book, Trefethen2023review} in terms of the variable $E$ (with $\vectheta$ fixed), a relevant recent development on this subject is mentioned in Sec.~\ref{subsec:optimarational}.

In principle, one can solve the inhomogeneous \schro equation, 

\begin{equation}
    (E - H ) | \wf \rangle  = | S \rangle  \, ,  \label{eq:inhomeeqn_1} 
\end{equation}
or

\begin{equation} 
    \langle  \wfadj | (E - H )   = \langle \soadj |  \, , \label{eq:inhomeeqn_2}
\end{equation}
and compute 

\begin{align}
\resolventamp = \langle \soadj | \wf \rangle = \langle \wfadj| \so \rangle \, . \label{eq:source_psi_overlaps}   %
\end{align}
To simplify the notation, I assume the $E$ and $\vectheta$ dependence is implicit unless otherwise stated.
These equations and $\resolventamp$ are the main targets of this study. 

Solving these equations directly poses severe numeric challenges, when $H$ is \emph{Hermitian} and $E$ is above continuum thresholds with $\Im E \to 0^+$ ($0^-$), because the solutions are not spatially localized (or integrable) but instead satisfies the outgoing (incoming) boundary conditions in spatial coordinate. However, when away from that energy region (e.g.,~at complex $E$s), the solutions are bound-state like~\cite{Schlessinger:1966zz, efros1985computation}, which are easier to compute. 

Here, I aim to develop RBM-based reduced-order models (ROMs), or emulators~\footnote{Broadly speaking, emulators, also known as surrogate models, are the tools used to rapidly interpolate and even extrapolate complex calculations or simulations in their parameter spaces~\cite{Melendez:2022kid,Drischler:2022ipa, Duguet:2023wuh}.}, for the solutions of Eqs.~\eqref{eq:inhomeeqn_1} and~\eqref{eq:inhomeeqn_2} and $\resolventamp$ so that one can extrapolate them in the complex $E$ plane and interpolate them in the space of $\vectheta$. Extrapolation in $E$ is a crucial capability, based on which $\resolventamp$ at real $E$s can be inferred from the solutions at complex $E$s---as mentioned, the latter are much easier to compute. The $\vectheta$-emulation  enables rapid explorations of the continuum physics calculations in the parameter space, another helpful functionality.

As a model-order-reduction (MOR) tool for a \emph{parameterized} equation system~\cite{hesthaven2015certified, Quarteroni:218966, Benner_2017aa, Benner2017modelRedApprox, benner2015survey}, the RBM first constructs a subspace spanned by the equation's full (or high-fidelity) solution at a sample of parameter sets, called snapshots, during the so-called offline training stage. Afterward, the equation system is projected into the subspace to form a ROM, which can be used to emulate the solutions and the associated observables in the parameter space. The dimension of the subspace is typically low and scales mildly with the number of parameters~\cite{Duguet:2023wuh}. Consequently, the computing cost for running emulators at the online emulation stage is dramatically reduced compared to simply repeating high-fidelity calculations, e.g., when exploring the parameter space of the calculations.

The basic principle of the RBM was recently rediscovered in nuclear theory as the eigenvector continuation method~\cite{Frame:2017fah}, where the focus was solving the eigenvalue problem. The RBM-based emulators have gained much attention and further development, including for nuclear-bound states~\cite{Sarkar:2020mad, Sarkar:2021fpz, Konig:2019adq, Demol:2019yjt, Ekstrom:2019lss, Demol:2020mzd, Yoshida:2021jbl, Anderson:2022jhq, Giuliani:2022yna, Yapa:2023xyf,Liu:2024pqp}, resonant states~\cite{Yapa:2023xyf,Yapa:2024lya} and general continuum scattering states~\cite{Furnstahl:2020abp, Drischler:2021qoy, Melendez:2021lyq, Zhang:2021jmi, Bai:2021xok, Drischler:2022yfb, Melendez:2022kid, Drischler:2022ipa, Bai:2022hjg, Garcia:2023slj, Odell:2023cun}.  

One difference in my work's RBM aspect is emulating inhomogeneous linear equations with continuous spectra. In contrast, most previous quantum physics-related studies have considered emulating a specific eigenstate of a Hamiltonian operator, either a bound, resonance, or scattering state at a real energy $E$. Note that Ref.~\cite{Melendez:2022kid} surveyed different ROMs, including for the inhomogeneous equations. However, the continuous spectrum aspect was not illuminated. More significantly, this is the first time to consider the complex $E$ plane as part of the parameter space, in addition to the other model input parameters, such as $\vectheta$ in Eqs.~\eqref{eq:inhomeeqn_1} and~\eqref{eq:inhomeeqn_2}. My RBM formalism is discussed in detail in Sec.~\ref{sec:interiorview}.

If one only considers the $E$ variable, the complex-$E$ emulation (CEE) is superficially similar to the rational Krylov methods~\cite{Antoulas2005book, Beeumen2017, Peng2019} applied in studying finite linear equation systems. However, the CEE generalizes the Krylov methods to studying linear systems with continuous spectra. One also gains insights about a potential connection, mentioned throughout this paper, between the CEE  and the (near-)optimal rational approximations~\cite{Trefethen2023review} of a univariate function with branch points.  Such a connection does not exist in the case of a linear system with only a discrete spectrum.  When including emulation in other real-valued parameters, my study further extends the rational Krylov methods to the case with higher-dimensional parameter spaces; it also generalizes the univariate rational approximation to a multivariate one. I call it CERPE, an abbreviation for ``complex-energy real-parameter emulator.'' Further discussions on the related works can be found in Sec.~\ref{subsec:otherworks}.

On the physics front, my CEE is a new NHQM method for computing continuum states and observables. This is demonstrated numerically with two and three-body systems in Secs.~\ref{sec:two-body} and~\ref{sec:three-body}. Some analytic understanding of the NHQM aspect of my CEE and the existing NHQM methods, including integration-contour deformation~\cite{Glockle:1983}, complex scaling of different variants~\cite{reinhardt1982complex, Moiseyev_2011, Myo:2014ypa, Lazauskas:2011uj, Lazauskas:2012jc, Papadimitriou:2015rca, Lazauskas:2015ula, Lazauskas:2019hil}, and  Berggren-basis based methods~\cite{Berggren:1968zz, Berggren1993, Michel:2021jkx}, 
are presented in Sec.~\ref{subsec:NHQM}. The generic strategies behind these methods are elaborated in that section using a simple model. 

However, my method differs significantly from  existing NHQM approaches. The fundamental distinction is in constructing a finite-dimensional non-Hermitian Hamiltonian $H$ matrix---a step I call ``non-Hermitization.''
This difference and its implications are discussed in Sec.~\ref{subsec:NHQM_differences}.  

The CERPE component of this study is also useful for continuum physics studies. Both Hamiltonian spectra and $\resolventamp$ can be interpolated, extrapolated, or emulated in the space of $\vectheta$ in the inhomogeneous \schro equations. The functionality of  CERPE follows the same argument of existing emulators: they provide efficient interfaces for the users to access computationally expensive calculations with dramatically reduced computing costs~\cite{Zhang:2021jmi}. For example, with this emulator technology, model calibration and uncertainty quantification, particularly those based on Bayesian statistics, would become feasible for complex models and expensive calculations. 

In short, with CEE, continuum physics can potentially be extracted from bound-state-like calculations---an advantage of the NHQM methods. The CERPE further expands the functionality of such continuum physics calculations by reaching more users. 

Another new physics insight is concerned with complex-energy (CE)~\cite{Schlessinger:1966zz, Schlessinger:1968vsk, Schlessinger:1968zz, McDonald:1969zza, Uzu:2003ms, Deltuva:2012fa, Deltuva:2013qf, Deltuva:2013mda, Deltuva:2014pda} and Lorentz integral transformation (LIT) methods~\cite{efros1985computation, Efros:1994iq, Efros:2007nq, Orlandini:2013eya, Sobczyk:2021dwm, Sobczyk:2023sxh, Bonaiti:2024fft}. The inhomogeneous \schro equations are also solved at complex energies in the LIT calculations and effectively in the CE calculations.\footnote {In the existing implementation of the CE method~\cite{Uzu:2003ms, Deltuva:2012fa, Deltuva:2013qf, Deltuva:2013mda, Deltuva:2014pda}, the Lippmann-Schwinger and the Faddeev equations are solved for the on- and off-shell scattering amplitudes. However, the wave functions can be computed with those amplitudes and vice versa~\cite{Schlessinger:1966zz}.} Using their procedures, these methods connect the complex-$E$ results to the real-$E$ ones. 
my results suggest that these existing calculations can be viewed from the lens of the general NHQM framework. Perhaps more importantly, the CERPE developed here can be applied directly to emulate these existing calculations. The general procedures for achieving this can be found in Sec.~\ref{sec:couplingwithothers}.  

One counterintuitive understanding of computing continuum physics is worth a brief mention. The results shown in this work suggest that the CEE, which approximates $\resolventamp$ using a small non-Hermitian $H$-matrix in Eq.~\eqref{eq:resolventdef}, produces better results at \emph{real} energies than the high-fidelity calculations based on a large Hermitian $H$-matrix. What is puzzling is that the two agree numerically in the instances with complex energies, i.e., the emulator is trained by these high-fidelity calculations. In contrast, in the existing emulator studies, emulators are supposed to reproduce high-fidelity calculations, including during extrapolations. One inevitably concludes that the CEE (and thus CERPE) as an extrapolant for $E$ is \emph{biased} to the physical continuum physics instead of the high-fidelity results based on a discrete spectrum. The non-Hermitization of $H$ plays a key role here. This is further discussed in Sec.~\ref{sec:summary}.

I emphasize that although numerical results are only presented for simple two- and three-body systems with short-range interactions, both CEE and CERPE should work for general finite systems, as the applicability of the RBM method and the working of non-Hermitization are general, without specific reference to the size and the interaction nature of the system. The RBM method requires smooth dependence of the solution on the input parameters, which have been found to hold up in few and many-body studies. Meanwhile, as explained later, the non-Hermitization depends on the spectrum's continuous nature and the training points' setup in the complex $E$ plane. 

However, I also need to point out that all the numeric calculations here are performed with high accuracy, with relative errors on the order of $10^{-12}$ in the training calculations. The presented  understanding of my methods is based on such calculations.  In practice, the training calculations, although attainable using bound-state methods, can have more significant errors. How the errors impact the performance of the emulator's extrapolation, and ways to stabilize the extrapolation need to be studied in the future. 

A summary of the organization of the rest of the paper is as follows. In Sec.~\ref{sec:exteriorview}, a general discussion about $H$'s spectrum, its resolvent operator, and their connections to the continuum observables are provided. Recent developments in rational approximation studies are mentioned in light of their relevance in this work. Section~\ref{sec:interiorview} discusses the RBM framework used in this study. Numerical experiments of the CEE and CERPE in both two and three-body systems are discussed and analyzed carefully in Secs.~\ref{sec:two-body} and~\ref{sec:three-body}. In Sec.~\ref{sec:couplingwithothers}, I discuss the potential couplings between the methods presented here and other calculation methods. In Sec.~\ref{sec:summary}, a summary is provided. The appendices collect some detailed information needed to reproduce the numerical calculations in this work. Source codes for generating the results of this work can be accessed via the companion website~\cite{BUQEYEsoftware}.
 
\section{Continuum physics, NHQM methods and rational approximations} \label{sec:exteriorview}

Here, I discuss the NHQM approaches for continuum physics and their connections with rational approximations of univariate functions with branch points. My emulator-based NHQM approach will be briefly motivated towards the end of Sec.~\ref{subsec:NHQM}, but the formalism is elaborated in Sec.~\ref{sec:interiorview}.

\subsection{Scattering and reaction amplitudes} \label{subsec:observables}

In this work, I focus on the matrix elements of the $H$-resolvent operator with the spatially \emph{localized} $|S\rangle$ and $|\tilde{S}\rangle$ sources. Per Eq.~\eqref{eq:source_psi_overlaps}, the matrix elements at real $E$ above thresholds can be computed relying only on the state $|\wf\rangle$ or $|\wfadj\rangle$ at the finite ranges comparable to the spatial size of the sources, without the need for knowing the correct asymptotic behaviors in the wave functions. These matrix elements directly represent or relate to various continuum physics observables. They are discussed separately in the following bullets.  

\begin{enumerate}
\item 
If $\soadjket{} = \soket{}  =  O | \wf_\text{bound} \rangle$ with $O$ as a transition operator and $|\wf_\text{bound}\rangle$ as a bound state, $-\Im \resolventamp(E)/\pi$ is the $O$-induced response function of the $ |\wf_\text{bound}\rangle$ state~\cite{FETTER71, Efros:2007nq}, when $\Im E = 0^+$. Note the parameters in the sources (i.e., part of $\vectheta$) could be those associated with current operators, such as two-body current parameters (see e.,g., Ref.~\cite{Pastore:2008ui}), or the momentum transfer from the probe~\cite{Walecka1995}.   

\item 
The connections between on-shell $\resolventamp$ and scattering and reaction $T$-matrices  have been discussed in depth in Chap.~16 of Ref.~\cite{Goldberger1964} and Chap.~5 of Ref.~\cite{newton1982scattering} (also see, e.g., Refs.~\cite{efros1985computation, Efros:2007nq}). Suppose $ | \Phi_i \rangle$ and $ | \Phi_f \rangle$ are the direct product states of cluster internal \emph{full} wave functions and relative-motion wave functions corresponding to the initial and final states, properly (anti)symmetrized if needed. Both satisfy the full homogeneous \schro equation $(E - H) | \Phi_{i,f} \rangle = 0$ with real-valued $E$ and large inter-cluster separations. With $E$ being the same as the $E$ variable in the resolvent, $ | S \rangle = (H - E) | \Phi_i\rangle$, and similarly $ | \tilde{S} \rangle = (H - E) | \Phi_f\rangle$, $\resolventamp$ becomes the \emph{non-Born} term in the corresponding $T$-matrix. The non-Born term is defined here as the full $T$-matrix with the Born term, $\langle \tilde{S}| \Phi_i\rangle = \langle \Phi_f | S \rangle$, subtracted.  Note these $|S\rangle$ and $|\tilde{S}\rangle$ states are spatially localized for systems with only two fragments in the initial and final states, because  $ | \Phi_i \rangle$ and $ | \Phi_f \rangle$ satisfy the full \schro equations when the clusters are separated beyond their interaction ranges. Later, I would treat the $E$ variable in the sources and that in the resolvent operator as separate variables if necessary. Note that these discussions hold for charged systems as well, if Coulomb wave functions are used for the relative motion between fragments. 

\item For describing the \emph{exclusive} transition amplitude induced by a perturbative probe $O$ (e.g., electroweak currents) or its time reversal process, one can use a formalism that mixes up those discussed in the previous two bullet points. I.e., $ | S \rangle = O |\wf_\text{bound} \rangle$ and $
 |\tilde{S} \rangle = (H - E) |\Phi_f \rangle$ or with $\so$ and $\soadj$ exchanged in the assignments.  $\resolventamp(E)$ gives the full transition amplitude in this situation.  

\item $\resolventamp$ is also an essential and costly component in microscopically computing (nuclear) optical potentials~\cite{Rotureau:2016jpf, Burrows:2023ygq}. The sources here can be chosen to be spatially localized, as the optical potential is intrinsically a quantity defined in the region in which the systems interact.

\end{enumerate}

\subsection{NHQM methods} \label{subsec:NHQM}

\begin{figure*}
	    \centering
	        \includegraphics{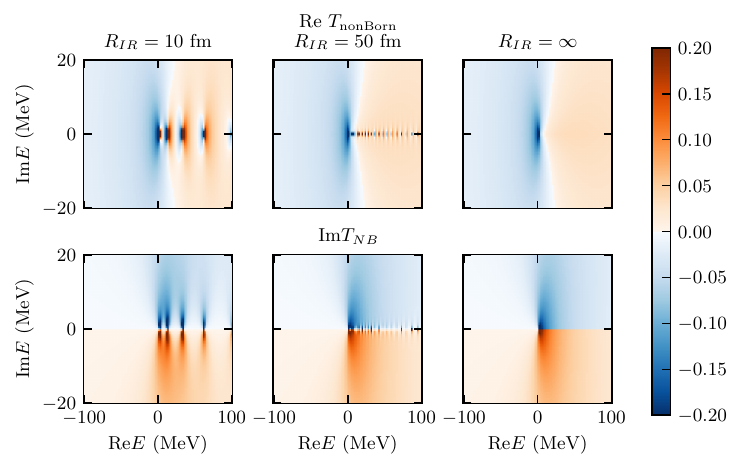}
		    \caption{A resolvent matrix element $\resolventamp(E)$ in the complex $E$ plane with fixed sources. Here,  the off-shell behavior  of a $T$-matrix, specifically its non-Born term, is explored. The two rows show the real and imaginary parts of the amplitude. From left to right, three different calculations are shown. The first two are based on bound-state-like calculations, where I enforce a Dirichlet boundary condition on the relative motion, forcing the wave functions to be zero at $r = R_{IR}$; the third is the exact calculation, corresponding to the $R_{IR}\to \infty$ limit.}
		        \label{fig:Tnb_in_CE_plane}
\end{figure*}

To illustrate the NHQM methods, including the one developed here, I use a two-body toy model with finite-range interactions. The toy system mimics a two-nucleon system in $s$-wave, for which I compute $\resolventamp$ in various ways. The sources are chosen such that $\resolventamp$ corresponds to the non-Born term in the scattering $T$-matrix, previously discussed as one option in Sec.~\ref{subsec:observables}. My focus here is on the generic features of the numerical results, while the numeric details can be found in the beginning of Sec.~\ref{sec:two-body}. 

Figure~\ref{fig:Tnb_in_CE_plane} is intended to (1) show how the \BCT of the exact $\resolventamp$ in the complex $E$ plane emerges from a series of finite-Hermitian-matrix-based calculations,
and (2) by comparing the numeric results to the exact result, identify the failure of those numeric calculations, upon which the NHQM methods aim to improve.

The figure plots the real (top row) and imaginary (bottom row) parts of various $\resolventamp(E)$s in the complex $E$ plane; from left to right, the first two columns are from the $\resolventamp$ calculations where finite $H$ matrices are inserted in Eq.~\eqref{eq:resolventdef}. To get such $H$ matrices,  I enforce a boundary condition such that all the states are zero at $r = \RIR$ (10 and 50 fm in the 1st and 2nd column), a particular long-distance (infrared, IR) regulator, while a series of short-distance (ultra-violet, UV) regulators~\footnote{In a finite basis, such as a coordinate-space Lagrange mesh~\cite{Baye:2015xoi} used in my two-body calculations or a plane-wave basis, the resolution of the small-distance scale, i.e., mesh-point spacing or  momentum cut-off in the respective bases, is finite, which is known as a short-distance (UV) regularization. In general, this resolution improves with basis size.} are tested so that the results shown in the figure converge with respect to increasing the UV resolution. The last column shows the exact results, essentially $\resolventamp$'s limit at $\RIR \to \infty$. 

The sharp features around the real axis, as seen in the first two columns, can be understood via the spectrum of the $H$ matrices. Specifically, the approximated $\resolventamp$ with finite $\RIR$ and UV cut-off is a finite sum of poles if one consider it as a function of $E$: 

\begin{align}
 \matrixform{\resolventamp} & =  \sum_i \frac{\langle \tilde{S} | \wf_i \rangle \langle \wf_i | S \rangle}{E - E_i}  
 \ . \label{eq:resolvent_approx}
\end{align}
$|\wf_i\rangle$ is the $i$th eigenstate of $\matrixform{H}$ and the associated eigenenergy $E_i$ determines a pole location.  Note here and later, I label the finite-matrix-based approximations of a quantity, e.g., $\resolventamp$, by $\matrixform{\resolventamp}$. Since $\matrixform{H}$ is kept Hermitian while I construct a basis to project $H$, the poles are only on the real axis. 

One would have three observations: (1) above and below the real axis, the exact $\resolventamp$ (in the 3rd column) is smooth almost every where, except the $E=0$ point where the function is not as well behaved. When crossing the positive real axis, the amplitude experiences a discontinuity. This is a reflection of $E=0$ being $\resolventamp$'s branch point and the positive real axis as the \BCT; (2) when taking $\RIR \to \infty$ limit, the poles to the right of the $E=0$ point become more and more densely distributed and eventually merge into a \BCT along the real axis; and (3) to the left of $E=0$, there are no densely distributed poles except possible isolated bound-state poles (absent in this toy model). $E=0$ is such a nontrivial location because it is the model's continuum threshold, below which the kinetic phase space for the scattering eigenstates is zero and above which the phase space starts increasing from zero\footnote{This can also be visualized by plotting $-\Im \resolventamp(E)$ with $\Im E \to 0^+$. The function is zero below the threshold and grows starting at the threshold. For example, see the 3rd panel in Fig.~\ref{fig:spectra_fixSfixpot_BE2_10_lam4_-0.2}.}. 

That is, at $\RIR \to \infty$ limit, the summing over the index $i$ in Eq.~\eqref{eq:resolvent_approx} turns into integrating over the continuous eigenenergy variable $\tilde{E}$:

\begin{align}
 \resolventamp & = \sum_B \frac{\langle \tilde{S} | \wf_B \rangle \langle \wf_B | S \rangle}{E - E_B} \notag  \\ 
 & + \int_{-\infty}^{\infty} d\tilde{E} \frac{\langle \tilde{S} | \wf_{\tilde{E}} \rangle \langle \wf_{\tilde{E}}| S \rangle}{E - \tilde{E}} \Theta(\tilde{E} - E_{\text{th}} )
 \ . \label{eq:resolvent_approx_inf}
\end{align}
Here, $\wf_B$ and $\wf_{\tilde{E}}$ are, respectively, the bound state in the discrete spectrum and the other eigenstates in the continuous spectrum;  $E_\text{th}$ is the location of the threshold (located at $0$ here).  It is the integration support defined by the step function $\Theta(\tilde{E} - E_\text{th})$ that  gives rise to the branch point at $E = E_\text{th}$. Again, the \BCT defined along the real axis is due to the continuous real eigenenergies.  

The above discussions can be generalized to more complex situations where multiple thresholds (or channels) exist. At infinite-$\RIR$ limit, there are the same number of eigenstate groups as the number of thresholds\footnote{The eigenstates of the full Hamiltonian provide a complete basis to expand the full space. See Chap.~7 in Ref.~\cite{newton1982scattering}. At each threshold, a new group of eigenstates emerges as initiated by the corresponding incoming channel.}. Each group starts to contribute to $\resolventamp$ above the corresponding threshold  (labeled by the associated channel $c$) in a simple additive way, i.e., the integration in Eq.~\eqref{eq:resolvent_approx_inf} can be broken into individual groups, each with an integral support $\Theta(\tilde{E} - E_{\text{th}, c} )$. Repeating the discussion in the single-channel case, one see that  $\resolventamp(E)$ has a branching point at each $E = E_{\text{th},c}$. It is worth pointing out that at finite $\RIR$, the discussion could be tricky, as each discrete eigenstate can't simply be labeled by a particular incoming channel $c$.   

In addition, it will be interesting and also possible to generalize these conclusions to the cases with long-range \emph{repulsive} Coulomb interactions (needed in nuclear physics), in which the branch points are also essential singularities due to the Coulomb-barrier-induced Gamow suppression factors. 

To understand the failure of the finite matrix calculations, one returns to the first two columns in Fig.~\ref{fig:Tnb_in_CE_plane}. The regions with the largest errors are around the real energy axis, where the poles are located\footnote{It is worth pointing out that the discretized \BCT poles can be partially understood through quantization conditions for the corresponding eigenvalues. At least for simple systems, the continuum physics can be extracted at the poles' locations using L{\"{u}}scher-type approaches~\cite{Luscher:1990ux, Busch1998, Stetcu:2010xq, Rotureau:2011vf, Zhang:2019cai, Zhang:2020rhz, Guo:2021uig,Zhang:2024mot,Zhang:2024vmz,Zhang:2024vch, Bagnarol:2024rhq, Kievsky:2011gp, Furnstahl:2013vda, Shirokov:2018nlj, Yu:2023ucq}.  These methods also extract the continuum physics by computing bound-state-like problems.}. 
That is, unfortunately, where the continuum physics observables can be physically measured. At the same time, the calculations potentially already converge in the other regions. This is the essential feature of the rational approximation~\cite{Trefethen2023review}, to which Eq.~\eqref{eq:resolvent_approx} belongs, applied to a function with branch points: poles are assigned to mimic branch cut so that the approximation works in the region \emph{away} from the {\BCT}s but not around the {\BCT}s. Therefore, the cause of the continuum-physics difficulty is the Hermiticity of the $\matrixform{H}$, although Hermitian Hamiltonians have been used extensively in computing discrete bound states. 

Existing NHQM methods, such as the complex scaling method~\cite{ReinhardtComlexScaling2007,reinhardt1982complex,afnan1991resonances,Moiseyev_2011,Myo:2014ypa}, solve this problem by giving up $\matrixform{H}$'s Hermiticity so that the {\BCT}s and the problematic region of the rational approximation are moved away from the real axis. As a result, $\matrixform{\resolventamp}$ can converge with finite $\RIR$ around the real energy region. Moreover, if the \BCT is moved further away enough that the resonance poles, which were on the second Riemann sheet\footnote{A second Riemann sheet is sometimes called as an unphysical sheet, which could lead to the wrong impression that it is not important. On the contrary, the function's behavior in that part of the complex plane is directly related to the function's behaviors relevant to observations. } when the {\BCT}s are defined as along the real axis, are now in the region enclosed by the old and new {\BCT}s. I.e., these resonance poles become the poles on the 1st Riemann sheet. Therefore, the resonance poles can now be seen simultaneously in $\resolventamp$ with bound state poles. Eq.~\eqref{eq:resolventdef} then dictates that the resonances, as the bound state, are part of the $H$'s eigenstates. In addition, all the eigenstates, including the bound and resonance states and the so-called discretized continuum (DC) states that give rise to the discretized \BCT poles in $\resolventamp$, are integrable. Therefore, the NHQM methods enable computing the continuum physics from the bound-state-like calculations.

The other solution, which I follow here, is based on the observation that  $\matrixform{\resolventamp(E)}$ can converge to a certain required precision when the dimension of the $\matrixform{H}$ and the size of $|\Im(E)|$ are large enough. As shown in Fig.~\ref{fig:Tnb_in_CE_plane}, with $R_{IR} = 10$ fm, the computed $\matrixform{\resolventamp}$ with $|\Im E| > 10$ MeV already converges nicely to the exact results. This statement applies to general cases that the wave function solutions in Eqs.~\eqref{eq:inhomeeqn_1} and~\eqref{eq:inhomeeqn_2} for complex $E$ and compact sources satisfy the bound-state-like boundary conditions~\cite{efros1985computation}, and therefore can be computed with a finite-dimensional Hermitian $\matrixform{H}$---even when the matrix is Hermitian. Equivalently, in the momentum space, the singular functions in solving those equations with $\Im E\to 0^+$ are smeared out when $\Im E$ is nonzero. 

After attaining converged calculations for the smallest $\Im E$ with given computing resources,  my second step is to analytically continue the computed $\resolventamp(E)$ to the other domains of the complex $E$ plane, including the real axis. One main goal of this work is to develop RBM-based emulators to perform such analytical continuation. As demonstrated later, this approach effectively moves the {\BCT}s below the real energy axis and could expose resonance states as eigenstates. It is 
closely connected to the so-called (near-)optimal-rational-approximation studied recently~\cite{Trefethen2021NodeClustering, Trefethen2023review}. The other goal of this work is to use RBM-based emulators to efficiently interpolate this analytical continuation in the $\vectheta$ space.

The strategy of utilizing those complex-$E$ calculations has already been employed elsewhere. Its simplest form would be a quick remedy used extensively in linear response function calculations (see e.g.,~\cite{Nakatsukasa:2016nyc}). Here, I use $-\Im \resolventamp(E_0)$ with real $E_0$ as a response-like~\footnote{The term ``response-like'' emphasized that $-\Im\resolventamp(E_0)$ is generally not a physical response function (unless the sources are properly chosen), but they do share the same structure as the imaginary part of a resolvent matrix element evaluated at $E_0 + i0^+$.} function to explain the procedure. As just discussed, a straightforward calculation with a Hermitian $H$ matrix produces $-\Im \matrixform{\resolventamp(E_0)}$ as a summation of $\delta$ functions on the real energy axis. The remedy is to smear out $-\Im \matrixform{\resolventamp(E_0)}$ by adding a finite imaginary part $i\eta$ to $E_0$---believed to be due to the so-called damping factor. This amounts to $-\Im \resolventamp(E_0)\approx -\Im\matrixform{\resolventamp(E_0 + i\eta)}$. From Fig.~\ref{fig:Tnb_in_CE_plane}, it is reasonable to assume $-\Im\matrixform{\resolventamp(E_0 + i\eta)} \approx -\Im \resolventamp(E_0 + i\eta)$. Therefore, the remedy essentially assumes that $-\Im \resolventamp(E_0 + i\eta) \approx -\Im \resolventamp(E_0 )$, i.e., the exact result $-\Im \resolventamp(E)$ wouldn't change when it is extrapolated from $E = E_0 + i\eta$ to $ E = E_0$. Empirically, such extrapolation works better than using $-\Im \matrixform{\resolventamp(E_0)}$ directly! On the one hand, this shows the transition of the continuum physics from $E_0 + i \eta$ to $E_0$ is smoother or simpler than what is produced by a large-matrix-based calculation. On the other hand, it points out the need for an extrapolation in $E$ that is at least better than assuming no change in that extrapolation.    

The other more sophisticated methods of this type include the CE method~\cite{Schlessinger:1966zz, Schlessinger:1968vsk, Schlessinger:1968zz, McDonald:1969zza, Uzu:2003ms, Deltuva:2012fa, Deltuva:2013qf, Deltuva:2013mda, Deltuva:2014pda} and the LIT methods~\cite{efros1985computation, Efros:1994iq, Efros:2007nq, Orlandini:2013eya, Sobczyk:2021dwm, Sobczyk:2023sxh, Bonaiti:2024fft}.
The two types of calculations extrapolate the complex-$E$ results to the real-$E$ axis using their specific procedures. Moreover, the $\vectheta$-emulation discussed in this work could be helpful for the CE and LIT calculations. Sec.~\ref{sec:couplingwithothers} will go into detail about these aspects.

\subsection{Exponentially pole clustering near branch points and discretized {\BCT}s in (near-)optimal rational approximation} \label{subsec:optimarational}

Lloyd N.~Trefethen and others, in recent years, have done a series of numerical studies on the rational approximations of a univariate function ($f(z)$) with branch points and other singularities, which approximates $f$ as a sum of simple poles: 

\begin{align}
  f(z) \approx r(z) = \sum_{i=1}^{N_p} \frac{w_i}{z - z_i} \ , \label{eq:rationaldef}
\end{align}
with $z_{i=1,2,,,N_p}$ as the locations of the poles. One of the poles   could be located at $\infty$---meaning a polynomial could be added here.  

A good review can be found in Ref.~\cite{Trefethen2023review}. In particular, they~\cite{Trefethen2021NodeClustering} noticed that to achieve an efficient (or near-optimal) approximation, the poles  cluster exponentially near the branch points of $f(z)$, such as the $\resolventamp(z)$.  Such pole distribution has been known in the case of approximating $|z|$ function (or equivalently $\sqrt{z}$)~\cite{Newman1964} for some time, as mentioned in Ref.~\cite{Trefethen2021NodeClustering}. Still, these recent works emphasize the ``widespread'' of this pole clustering phenomenon in the near-optimal rational approximations. To have a visual impression of the clustering pattern of the poles or eigenenergies, see, for example, Figs.~\ref{fig:Swave_BE_10_LAM2_200_error_2dim} and ~\ref{fig:Swave_BE_10_LAM2_200_poles}. 

These works also found that the distribution of the logarithms of the modulus between poles and the corresponding branch points is not precisely uniform. Instead, tapering off that distribution to zero when approaching the branch points is needed for more efficient approximation. Through numerical experimentation with various functions having branch points and heuristic arguments in the limit of dense distribution of the training (or interpolating) points, they~\cite{Trefethen2021NodeClustering,Trefethen2023review} show that the approximation errors generally scale as $e^{- c \sqrt{N_p}}$ with $c$ being positive.

Since the rational approximation performs the worst in the region where poles are located, those poles should lie away from the region of the training (or interpolating) data. Heuristically speaking, the area of the poles should be minimized to further reduce the errors in the complex plane. Therefore, when approximating a function with branch points, the poles ``line up'' to form discretized \BCT lines, another general behavior observed and supported with some general understandings in these recent studies (see Chap.~7 in Ref.~\cite{Trefethen2023review}). However, rigorously speaking, a rational function doesn't have branch points and thus no {\BCT}s. I.e., although the definition of the true {\BCT}s is a convention for the exact function $f(z)$, the discretized {\BCT}s emerge such that $f$'s rational approximation achieves the best performance. Effectively, one can consider (near-)optimal rational approximation chooses its own {\BCT}s for the $f(z)$, a narrative I will use.

Note that the definition of an adequate approximation depends on how close to the threshold one wants the calculation to be correct. A rational approximation of $\resolventamp$, to which any finite-matrix-based calculations belong, has increasing errors towards the branch points because the approximant is infinitely differentiable there, but the exact function is not. However, the problematic regions keep shrinking when the number of effective poles increases towards the branch points. This illustrates the challenge of a near-threshold calculation if one don't build the known analytical structure into the calculation at its start.

A so-called AAA algorithm has been used in these demonstrations~\cite{Nakatsukasa_2018, nakatsukasa2023years}\footnote{In this work, I utilize its Python implementation in the \texttt{baryrat} package~\cite{baryrat_2021}.}, which is considered as (near-)optimal rational approximation~\cite{Trefethen2021NodeClustering}. It can be viewed as a ``data-driven'' emulator~\cite{Duguet:2023wuh}, as it only requires training data $f(z)$ at training (interpolating) points $z_j$ without knowing the underlying physics. It has been used in recent quantum physics studies~\cite{Xu:2022ybc, Huang_2023}, for example, to analytically continue the many-body Green's function in the complex energies or frequencies to the real domain~\cite{Goswami_2024, Ying_2022}. 

In this algorithm, $f(z)$ is constructed in the barycentric representation of Eq.~\eqref{eq:rationaldef}:
\begin{align}
    r(z) = \frac{n(z)}{d(z)} = \left. \sum_{j=1}^m \frac{w_j f_j}{z-z_j} \middle/ \sum_{j=1}^m \frac{w_j }{z-z_j}  \right. \ . \label{eq:AAAdef}
\end{align}
This representation provides numerical stability. Here $z_j$ are the training points and $f_j = f(z_j)$. It is easy to see that $r(z_j) = f_j$. To find the weights, $w_j$, the method minimizes the errors in terms of $f(z) d(z) - n(z)$, with other constraints. The method requires input about expected errors/noise in the data so that it doesn't overfit and is stable against noise. As can be seen, $r(z)$ has only simple poles and approaches to constant at infinity. In contrast, Eq.~\eqref{eq:resolvent_approx} suggests that at $z\to \infty$, $\matrixform{\resolventamp}$s in the NHQM approximations, including my emulator-based approach,  goes to zero at the limit. However, this difference is not substantial, as I only focus on a finite domain, and poles far away in all these approximations effectively behave as polynomials of $z$. Also, note that the AAA algorithm provides the positions of its poles. 

As will be shown through the numeric results in Secs.~\ref{sec:two-body} and~\ref{sec:three-body}, there are intriguing \emph{similarities} between my CEE results and the (near-)optimal-rational approximations concerning the location of the {\BCT}s, the distribution of the discretized \BCT poles, the approximation of $\resolventamp$, and the scaling of the approximation errors in terms of the number of poles. Specifically, I will compare my RBM-based emulators and the AAA approximations in several cases. It is tempting to equate the optimality of the rational approximation with the effectiveness of the RBM subspace projection in constructing the ROMs for a single variable.

\section{RBM-based emulations} \label{sec:interiorview}

This section centers around the RBM emulators used to interpolate, extrapolate, or emulate $\resolventamp$ in a combined space of complex $E$ plane and the $\vectheta$ dimensions. 
Note that certain parameters inside $\vectheta$ could be shared among the $H$ operator and the two sources. The training points, labeled as $(E^\mathrm{tr}_\alpha, \vectheta^\mathrm{tr}_\alpha)$ with $\alpha =1,,,N_b$, are those with finite (most of the time positive) $\Im E^\mathrm{tr}$. With those parameter values or at those points in the parameter space, I solve Eqs.~\eqref{eq:inhomeeqn_1} and~\eqref{eq:inhomeeqn_2} to get the training solution snapshots, $ |\wf^\mathrm{tr}_\alpha\rangle \equiv |\wf(E_\alpha^\mathrm{tr},\vectheta_\alpha^\mathrm{tr})\rangle$ and $ \langle \wfadj^\mathrm{tr}_\alpha| \equiv \langle \wfadj(E_\alpha^\mathrm{tr},\vectheta_\alpha^\mathrm{tr})|$. This step is called the training or off-line step. In the next emulation or online step, the RBM proposes a general solution at emulation point $(E, \vectheta)$, 

\begin{align}
| \wf(E,\vectheta) \rangle  &  = \sum_{\alpha=1}^{N_b} c_\alpha(E,\vectheta) | \wf_\alpha^\mathrm{tr} \rangle   \ . \label{eq:RBM_ansatz}
\end{align}
As shown later, the inhomogeneous \schro equations are effectively projected into a subspace spanned by the training solutions, $|\wf^\mathrm{tr}_\alpha \rangle$ and  $\langle \wfadj^\mathrm{tr}_\alpha |$.

According to previous studies, the RBM-based emulators have robust extrapolation capabilities because the emulators learn a problem's solution manifold---mapped out by varying the input parameters---by sampling solution vectors in the subspace~\cite{Duguet:2023wuh}. Using the CEE, I extrapolate $\resolventamp$ at complex training energies to the real energy axis in the complex $E$ plane and further into the region where potential resonances could exist. Therefore, this procedure extracts $\resolventamp$ at real energies and resonance properties from the training solutions, which satisfy bound-state-like boundary conditions~\cite{efros1985computation}. In addition, the CERPE emulates those continuum physics extractions in the $\vectheta$ space. 

When emulating at real energies with $\Im(E)\to 0^+$, the constructed wave function in Eq.~\eqref{eq:RBM_ansatz} has bound-state-like asymptotic behavior, which is incorrect; the correct one should be outgoing waves oscillating to infinity spatially (see Fig.~\ref{fig:emulated_wfs} for a visual demonstration). How is it possible, then, for the emulated $\resolventamp$ at real energies to be sensible? The answer comes from the fact that the sources are spatially localized; $\resolventamp(E)$ can thus be determined by the overlap of the source with the internal part of the emulated solution $|\wf\rangle$ as shown in Eq.~\eqref{eq:source_psi_overlaps}. This basic setup allows me to emulate continuum physics, including $\resolventamp(E)$ at real $E$ and  spectra (bound, resonance, and DC states), without reproducing the intricate wave function asymptotics correctly. This point will be illustrated further using an example in Fig.~\ref{fig:emulated_wfs} and the discussion around it.  Bear in mind that a spectrum in emulation is in its low-dimensional representation. Thus, I deem emulation as an efficient way to compress an infinite-dimensional continuous spectrum---at least a portion of it. 

From the discussion at the end of Sec.~\ref{subsec:optimarational}, one can see that  $\matrixform{\resolventamp}$'s discretized {\BCT}s (equivalently, the DC states in the compressed spectrum) are responsible for $\matrixform{\resolventamp}$'s near-threshold behavior. As will be shown in the Secs.~\ref{sec:two-body} and~\ref{sec:three-body}, these poles exponentially cluster toward the branch points or thresholds, similar to the pole distribution in the (near-)optimal rational approximation. The threshold locations depend only on a particular $H$, but will the distribution of the discretized \BCT poles be sensitive to the sources, i.e., different dynamics?  The numeric results in the Secs.~\ref{sec:two-body} and~\ref{sec:three-body} suggest the answer is no, which is consistent with the distribution being the property of a spectrum (albeit compressed) but not a property unique to specific dynamics. This is also in parallel with the observation that the ``widespread'' behavior of the pole distribution in the (near-)optimal rational approximation of functions with branch points is determined mainly by the locations of the branch points but not significantly by the other details of the functions~\cite{Trefethen2021NodeClustering, Trefethen2023review}. 

Moreover, the exponential clustering of the discretized \BCT poles also means one needs a small number of them (i.e., low-dim subspace) to approximate the smooth component of $\resolventamp$---defined as the full $\resolventamp$ with the resonance contributions subtracted---over an extensive range of real energies (see e.g., Fig.~\ref{fig:Swave_BE_10_LAM2_200_poles}). This explains the feasibility of compressing the DC component of the continuum spectra. Of course, the bound and resonance states are the other spectra components, but their number can not be reduced. 

Now, the task is to determine $c_\alpha(E,\vectheta)$'s dependence on $E$ and $\vectheta$, as addressed below. As a reminder, if I fix $\vectheta$ to a specific value without varying them, the emulation is called CEE; otherwise, it is named CERPE.  

\subsection{A variational method for solving a linear system and the associated RBM ROM} \label{sec:var_and_rbm}

Suppose I need to solve an inhomogeneous linear equation:   

\begin{align}
    M \, |\wf\rangle & =  | S\rangle  \ , \label{eqn:BasicEq}
\end{align}
and compute $\langle \tilde{S}| \wf \rangle $. A variational approach exists for this task~\cite{Pomraning1965}. The involved functional, depending on both trial solution $|\wf_t\rangle$ and an auxiliary variable  $|\wfadj_t\rangle$, is expressed as 

\begin{align}
    \mathcal{F}[|\wf_t\rangle, |\wfadj_t\rangle] & = \langle \tilde{S} |\wf_t\rangle + \langle \wfadj_t |  (S - M \wf_t ) \rangle  . 
\end{align}

If a trial solution pair is close to the exact solutions, i.e.,  $ | \wf_t \rangle  = | \wf\rangle  + | \delta\rangle $ and $ | \wfadj_t \rangle  = | \wfadj \rangle  + | \tilde{\delta} \rangle$, with $|\wf \rangle$ as Eq.~\eqref{eqn:BasicEq}'s exact solution  and $|\wfadj\rangle$ as the exact solution of another equation to be defined, I will have 

\begin{align}
    \delta \mathcal{F}[|\wf_t\rangle, |\adj{\wf}_t\rangle] & = \langle \tilde{S} | \delta\rangle + \langle \adj{\delta} | (S - M \wf ) \rangle  + \langle \adj{\wf} | (- M \delta) \rangle \notag \\
    & \quad + O(\delta^2)  \notag \\ 
    & = \langle (\tilde{S}-\adj{\wf}  M) | \delta \rangle + O(\delta^2)  \ . 
\end{align}
Therefore, I require $|\wfadj \rangle$ satisfy

\begin{align}
    \langle \tilde{S} |  & =  \langle \adj{\wf} |  M  \leftrightarrow  
    | \tilde{S} \rangle   =  M^\dagger | \adj{\wf} \rangle \ , 
\end{align}
so that the functional $\mathcal{F}$ is stationary around the exact solution and provides an estimate of $\langle \tilde{S} | \wf \rangle$ on the second order of $| \delta\rangle$ and $|\tilde{\delta}\rangle$. This equation defines the adjoint state vector $| \tilde{\wf} \rangle$, which is why $|\wfadj \rangle$ and Eq.~\eqref{eq:inhomeeqn_2} were introduced in Sec.~\ref{sec:introduction}.

Now, one can use the $\mathcal{F}$ functional to determine $c_\alpha$ as functions of $E$ and $\vectheta$ in Eq.~\eqref{eq:RBM_ansatz}. This procedure leads to a choice of test functions in the framework of Galerkin projection~\cite{Melendez:2022kid}.  To develop a ROM for Eq.~\eqref{eqn:BasicEq} with parameters in $M$ and the sources, I construct\footnote{In the previous report~\cite{Zhang:2024ril}, I use the convention of the implicit sum over the repeated indices. Here, that sum is expressed explicitly.} 

\begin{align}
    | \wf_t \rangle & = \sum_\alpha c_\alpha | \wf_{\alpha}^\mathrm{tr} \rangle \ , \label{eq:trial} \\ 
    \langle \wfadj_t |  & = \sum_\alpha \adj{c}_\alpha \langle \wfadj_\alpha^\mathrm{tr} | \ , \label{eq:trialadj}
\end{align}
with $\alpha$ (and later $\beta$) indexing the training points. At each training point,

\begin{align}
    M_{\alpha} | \wf_{\alpha}^\mathrm{tr} \rangle & = | \so_{\alpha}\rangle  \ , \label{eq:training_eq}\\ 
    \langle \wfadj_{\alpha}^\mathrm{tr}| M_{\alpha}  & = \langle \soadj_{\alpha}|   \leftrightarrow    M_{\alpha}^{\dagger} | \wfadj^\mathrm{tr}_{\alpha} \rangle = | \soadj_{\alpha} \rangle \ . \label{eq:training_adjeq} 
\end{align}

With the RBM trial solutions, the functional turns into

\begin{align}
    \mathcal{F}(c, \tilde{c}) & = \sum_\alpha \bigg(c_\alpha \langle \soadj| \wf_{\alpha}^\mathrm{tr} \rangle + \adj{c}_\alpha \langle \wfadj_{\alpha}^\mathrm{tr}| \so \rangle \bigg) \notag \\ 
    & \quad - \sum_{\alpha,\beta}  \adj{c}_\alpha c_\beta \langle \adj{\wf}_{\alpha}^\mathrm{tr}| M | \wf_{\beta}^\mathrm{tr} \rangle  \ . 
\end{align}
Note when $\so$, $\soadj$, or $M$ is without $\alpha.\beta$ indices, they are evaluated at the general emulation point. To find the stationary points, I look for the solutions such that the first derivatives of $\mathcal{F}$ with respect to $c$ and $\tilde{c}$ are zero. One then obtains a low-dimensional linear equation system:

\begin{align}
    \begin{pmatrix}
     0 &  \matrixform{ M } ^T \\ 
    \matrixform{ M } & 0 
    \end{pmatrix}
    \begin{pmatrix}
     c \\  
    \adj{c}
    \end{pmatrix} =
    \begin{pmatrix}
     \matrixform{\tilde{S}^\dagger}  \\  
    \matrixform{S } 
    \end{pmatrix}
     \ . \label{eq:Eq_stationarypoint}
\end{align}
The upper right corner of the matrix on the left side uses transpose, $^T$, not complex conjugate, due to my definition of $\adj{c}$ in Eq.~\eqref{eq:trial}. 
The block matrix has its matrix elements defined as 

\begin{align}
    \matrixform{ M}_{\alpha, \beta} \equiv \langle \adj{\wf}_{\alpha}^\mathrm{tr}|  M |  \wf_{\beta}^\mathrm{tr} \rangle\,,
\end{align}
and similarly, the projected source vectors have 

\begin{align}
     \matrixform{\soadj^\dagger }_\alpha &  \equiv  \langle \soadj | \wf_{\alpha}^\mathrm{tr} \rangle\,, \\
    \matrixform{ \so}_\alpha & \equiv \langle \wfadj_{\alpha}^\mathrm{tr} | \so \rangle \ . 
\end{align}

It is  easy to see  

\begin{align}
\sum_{\alpha} c_\alpha \langle \soadj| \wf_{\alpha}^\mathrm{tr}\rangle & = \sum_{\alpha} \adj{c}_\alpha \langle \wfadj_{\alpha}^\mathrm{tr} | \so \rangle \notag \\ 
   & = \sum_{\alpha,\beta} \adj{c}_\alpha c_\beta \langle \wfadj_{\alpha}^\mathrm{tr} | M | \wf_{\beta}^\mathrm{tr}\rangle \ .
\end{align}
One thus has 

\begin{align}
    \mathcal{F}\vert_\mathrm{stationary} & = \sum_\alpha c_\alpha \langle \tilde{S} | \wf_{\alpha}^\mathrm{tr} \rangle   \label{eq:varF_stationary}
\end{align}

Therefore, after obtaining, $|\wf^\mathrm{tr}_\alpha \rangle $ and $\langle \wfadj^\mathrm{tr}_\alpha|$, I construct the low-dimensional matrix $\matrixform{M}$ and vectors $\matrixform{\soadj^\dagger}$ and $\matrixform{\so}$---called as emulator components---at emulation stage. Note 
\begin{align}
\matrixform{M}_{\alpha \beta} 
 & = \langle \wfadj_\alpha^\mathrm{tr} | M - M_\beta |\wf_\beta^\mathrm{tr} \rangle + \langle \wfadj_\alpha^\mathrm{tr} | S_\beta \rangle  \\ 
  &  = \langle \wfadj_\alpha^\mathrm{tr} | M - M_\alpha |\wf_\beta^\mathrm{tr} \rangle + \langle \tilde{S}_\alpha | \wf_\beta^\mathrm{tr} \rangle   \ . 
\end{align}
Since the training solution basis is not orthonormal, I need to deal with the norm matrix: 

\begin{align}
    \matrixform{N}_{\alpha\beta} & \equiv \langle \tilde{\wf}_\alpha^\mathrm{tr} |  \wf_\beta^\mathrm{tr} \rangle \ . 
\end{align}

Now, specifically for the inhomogeneous \schro equation, there, $M = E - H$ with $E$ as one important parameter. 
The projected $H$ matrix, defined as  

\begin{align}
    \matrixform{H}_{\alpha\beta} = \langle \tilde{\wf}_\alpha^\mathrm{tr} | H | \wf_\beta^\mathrm{tr} \rangle  \ , 
\end{align}
can then be inferred via 

\begin{align}
\matrixform{H} = E \matrixform{N} - \matrixform{M} \ . 
\end{align}

If the parameter dependencies in $\matrixform{M}$, $\matrixform{\tilde{S}^\dagger}$, and $\matrixform{S}$ are factorized from the high-dimensional tensors, the expensive calculations for computing the emulator components can be performed once at the training stage. Later, one can scale them appropriately at the emulation stage to get those emulator components at each point with little cost. For the parameters without such affine dependence, various solutions exist, such as interpolating and extrapolating the emulator components in $\vectheta$  using data-driven methods (e.g.~the emulator-in-emulator method in Ref.~\cite{Zhang:2021jmi}) or applying empirical interpolation methods~\cite{Benner_2017aa, Odell:2023cun} to approximate the non-affine dependence with affine structures. One then have fast emulations. 

Note the variational approach informs the test function space---i.e., the space spanned by the $\langle \wfadj^\mathrm{tr}_\alpha|$---used in this work. However, other choices for the test function space have been studied in the MOR and will be explored in the future. One crucial ingredient, keeping the analyticity of $\matrixform{N}_{\alpha,\beta}$ and $\matrixform{H}_{\alpha,\beta}$ in terms of the $E^\mathrm{tr}_\alpha$ variable, is critical for realizing non-Hermitization, which should be considered when exploring new test function space in studying quantum continuum physics.

\subsection{Compressed spectra} \label{subsec:spectra}

When constructing ROMs for the inhomogeneous \schro equations, Eq.~\eqref{eqn:BasicEq} turns into Eq.~\eqref{eq:Eq_stationarypoint}, with $M = E - H$. Per Eqs.~\eqref{eq:Eq_stationarypoint} and~\eqref{eq:varF_stationary}, the resulting $\matrixform{\resolventamp(E)}$ has simple poles located at the zeros of $\det(\matrixform{M})$ in the complex $E$ plane. I.e., these pole positions are determined by the eigenvalues of the following generalized eigenvalue problem:

\begin{align}
    \matrixform{H} v  = E \matrixform{N} v  \label{eq:gev1} \,. 
\end{align}

When three conditions are satisfied: $| \so_\alpha \rangle = | \soadj_\alpha \rangle$ and being invariant under time-reversal $\Treversal$ transformation, and $H$ being Hermitian in the training calculations, $\matrixform{H}$ and $\matrixform{N}$ become complex symmetric matrices. This property is the shared (but not a required) characteristic in the existing NHQM methods~\cite{Moiseyev_2011, Michel:2021jkx}. The training equations turn into, 

\begin{align}
    (E_\alpha - H_\alpha  ) | \wf_\alpha^\mathrm{tr} \rangle & = | \so_\alpha \rangle  \ , \label{eq:training_wf} \\
     (E_\alpha^\ast - H_\alpha  ) | \adj{\wf}_\alpha^\mathrm{tr} \rangle & = | \so_\alpha \rangle  \ . \label{eq:training_wfadj}
\end{align}
Since $\Treversal | S_\alpha \rangle = | S_\alpha \rangle $, I have 

\begin{align}
    | \adj{\wf}_\alpha^\mathrm{tr} \rangle & = \Treversal | \wf_\alpha^\mathrm{tr}\rangle \ , \label{eq:solution_connection_T_1_special_case} \\
    |  {\wf}_\alpha^\mathrm{tr} \rangle & = \Treversal | \adj{\wf}_\alpha^\mathrm{tr}\rangle \ .  \label{eq:solution_connection_T_2_special_case}
\end{align}
It is then easy to show that 

\begin{align}
    \matrixform{N}_{\alpha\beta} = \langle \Treversal \wf_\alpha^\mathrm{tr} |  \wf_\beta^\mathrm{tr} \rangle = \left(\langle \wf_\alpha^\mathrm{tr} |  \Treversal^\dagger \wf_\beta^\mathrm{tr} \rangle\right)^\ast =  \matrixform{N}_{\beta\alpha}
\end{align}
and similarly $\matrixform{H}_{\alpha\beta} = \matrixform{H}_{\beta\alpha}$. This proves the previous assertion. Note that $H$ is Hermitian in the training calculations, but $\matrixform{H}$ and $\matrixform{N}$ are non-Hermitian in general and complex symmetric in this particular case.

As briefly mentioned at the beginning of Sec.~\ref{sec:interiorview} and elaborated later in Secs.~\ref{sec:two-body} and~\ref{sec:three-body}, these eigenvalues are distributed in a pattern similar to the pole distribution of a (near-)optimal rational approximation discussed in Sec.~\ref{subsec:optimarational}. 
The emulation performance certainly depends on whether $\matrixform{\resolventamp}$ can reproduce the physics, including the pole positions, in the region where the bound states and resonances are located.  

However, the DC states in the emulator's compressed spectrum depend on the setup of the training points, in particular, the distribution of $E^\mathrm{tr}_\alpha$ in the complex $E$ plane. As one interesting specific case, when (1) $\matrixform{H}$ and $\matrixform{N}$ are complex symmetric and (2) both $E^\mathrm{tr}_\alpha$ and $E^{\mathrm{tr}\ast}_\alpha$ are included as the training points associated with a given $\vectheta_\alpha$, the eigenenergies are always real and thus the discretized {\BCT}s are along the real axis. See Fig.~\ref{fig:Swave_BE_10_LAM2_200_error_2dim} for a visual presentation, which is a special case with $\vectheta$ fixed. This point is further elaborated in the Appendix~\ref{app:proof_pseduo_hermicity}.  Such a setup could be beneficial for discrete bound spectrum calculations. However, the discretized {\BCT}s should be away from the real axis for the continuum calculations. Therefore, I focus on the emulations with $\Im E^\mathrm{tr}_\alpha > 0$ in the training points. 

\subsection{Difference between the existing NHQM and the CEE methods} \label{subsec:NHQM_differences}

I am now at a good point to discuss the differences between the existing NHQM methods and the CEE. The previous approaches must analytically continue the conventional single-particle basis (e.g., harmonic oscillator basis) in which $H$ is Hermitian to a basis that $H$ turns non-Hermitian. These methods then construct a many-body basis as a direct product of the single-particle basis. The numerical difficulty could arise during the analytical continuation, limiting how far away from the real axis the {\BCT}s can be moved.  

In contrast, CEE achieves the ``non-Hermitization'' of $H$ by building the many-body basis from the solution snapshots of the inhomogeneous \schro equations with complex energies, i.e., by the RBM-based subspace projection (or viewed as spectrum compression) during the on-line emulation stage. Note these snapshot basis build in inter-particle correlations.

At the off-line training stage, one does not need to---but one can---perform analytical continuation of the $H$ matrix to obtain the training solution snapshots, as the solutions to the equation with a Hermitian $H$ already have bound-state-like boundary conditions. As a result, the non-Hermitian $\matrixform{H}$ matrices presented in this work generally have much lower dimensions than the corresponding matrix in the other NHQM methods. 

It is also relevant to note that my approach targets a specific portion of the spectrum, as the eigenenergies of the compressed spectrum are distributed in a finite domain of the complex $E$ plane. The portion is not precisely defined. However, when the $E$ variable is extrapolated too far away from the range of the training energies during the emulation stage, the emulation errors become significant. In contrast, in the existing NHQM methods, the spectrum range is defined by the energy cut-off applied on a single-particle basis.

\subsection{Some numerical details} \label{sec:numerical_details}

At the training stages, I typically over-sampled training points in the parameter space for simplicity. I.e., $N_b$ is larger than the true dimension of the solution subspace. As the result, during emulations,  the system of Eq.~\eqref{eq:Eq_stationarypoint} becomes ill-conditioned. This is not a deficiency of the framework but, in fact, tied to the efficiency of the dimension reduction. However, regularization needs to be implemented to remove the redundancy. Here, I apply the singular value decomposition (SVD) to the $\matrixform{M}$ and eliminate the singular values $\sigma_{.}$ that satisfy $|\sigma_i/\sigma_{max}| < 10^{-13}$, a threshold similar to the relative errors of the training calculation, to perform pseudo-inverse to compute $c$ and $\tilde{c}$. The truncation on the singular values leads to an effective number of poles in the emulated $\matrixform{\resolventamp}$ in the complex-$E$ plane, which is labeled as $\Neff$.

For the same reason, when solving the generalized eigenvalue problem to emulate spectra using Eq.~\eqref{eq:gev1}, I invert $\matrixform{N}$ also using the pseudo-inverse with the same SVD singular value truncation, which again reveal the dimension of the subspace ($\Neff$). Bear in mind that it is only when the training points are over-sampled that $\Neff$ approaches the dimension of the subspace; otherwise, $\Neff$ is the same as $N_b$.

In the future, the so-called greedy algorithms~\cite{sarkar2021selflearning, Quarteroni:218966, Maldonado:2025ftg} should be pursued to strategically and efficiently sample the training points, which stops at the right $N_b$ that is close to the dimension of the solution subspace. 

It is worth pointing out a second way to emulate $\resolventamp$ in $E$ when one has an emulated spectrum at $\vectheta$. Note that 

\begin{align}
    \matrixform{\resolventamp(E, \vectheta)} & = \sum_{\beta=1}^{N_b} \frac{W_\beta(\vectheta)}{E  - E_\beta(\vectheta)}  \label{eq:emulated_amp_eigen_decomp} \ , 
\end{align}
with $E_\beta$ as the emulated eigenvalues. To fix all the residues, $W_{\beta}$,  I first compute $\resolventamp(E_{\alpha}^\mathrm{tr}, \vectheta)$ by using the first $\resolventamp$ emulation method (i.e., solving Eq.~\eqref{eq:Eq_stationarypoint} and then evaluating Eq.~\eqref{eq:varF_stationary}). The residues can be computed by solving a linear equation system: 

\begin{align}
    \matrixform{\resolventamp(E_\alpha^\mathrm{tr}, \vectheta)} & = \sum_{\beta=1}^{N_b} \frac{W_\beta(\vectheta)}{E_\alpha^\mathrm{tr}  - E_\beta(\vectheta)} \ .  
\end{align}
Eq.~\eqref{eq:emulated_amp_eigen_decomp} then rapidly emulates $\resolventamp$ in the entire complex $E$ plane for a given $\vectheta$. This eliminates the need to repeatedly solve Eq.~\eqref{eq:Eq_stationarypoint} when varying $E$. Again, the SVD and the truncation on the singular values are applied when the equation for solving $W_\beta$ becomes ill-conditioned.

\subsection{Existing works} \label{subsec:otherworks}

In the MOR studies of dynamic linear systems, the equations in the frequency domain take the form of Eq.~\eqref{eqn:BasicEq}. The so-called rational-Krylov method (e.g., Chap.~10 and 11 in Ref.~\cite{Antoulas2005book}) is similar to the CEE.  However, these studies are about systems with finite-dimensional matrices. Similarly, recent two studies~\cite{Beeumen2017, Peng2019} in quantum chemistry also apply this method to emulate $\resolventamp(E)$ involving continuum physics. Again, the critical difference between finite-dimensional matrix and $H$ with continuum was not studied there, including the point that the continuum spectra can be efficiently compressed. It seems the emulated eigenenergies in these works are always real, i.e., their Galerkin projection, used to derive the ROMs, differs significantly from the one used in this work. In addition, I demonstrate the effectiveness of emulation for other parameters, which has not been studied in Refs.~\cite{Beeumen2017, Peng2019}. 

\section{Two-body demonstrations} \label{sec:two-body}

In this section, I work with a toy model for two-nucleon-like systems. The particle mass is $M_N = 940$ MeV in the natural units. The interaction, in the $\ell$th partial-wave channel, takes a separable form:  

\begin{align}
    V & =  \lambda | g \rangle \langle g | \ , \text{with} \ 
     \langle q, \ell | g \rangle = g_\ell(q)   \ . \label{eq:two-body_V_def}
\end{align}
The form factor in the momentum-space representation, $g_\ell(q)$, is a Gaussian function with a width parameter $\LAMtwo$; $\lamtwo$ is the coupling-strength parameter. Here, I focus on $s$ and $p$ waves. The associated $\resolventamp(E)$s are known analytically in the complex $E$ plane. The details about the model and relevant analytical formulas can be found in Appendix~\ref{app:twobody}. 

For emulation, the training calculations are carried out using a bound-state method, employing the Lagrange function of the Legendre polynomials as bases~\cite{Baye:2015xoi}. The basis function is defined in a finite interval of the \emph{spatial} radial coordinate, $[0,\RIR]$, as mentioned in discussing Fig.~\ref{fig:Tnb_in_CE_plane}. I choose $\RIR$ large enough and the UV resolution scale small enough (via increasing the basis size) so that the emulator components,  $\matrixform{N}$ and $\matrixform{H}$, all converged to a level of $10^{-12}$ or even smaller.  

\subsection{Emulation with fixed $H$ but varying sources}

\begin{figure}
    \centering
    \includegraphics[width=0.48\textwidth]{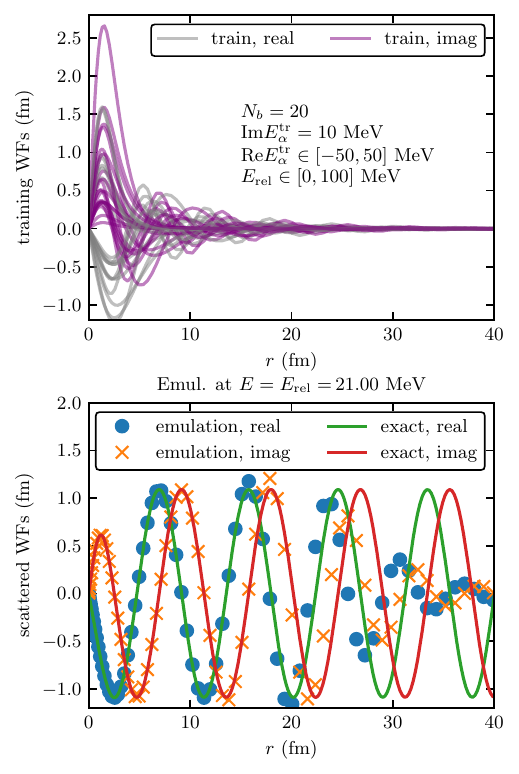}
    \caption{Emulation in $(\Re E, \Im E, \Ein)$. The training wave functions  are plotted in the top panel, and at the bottom, the emulated and the exact wave functions are compared for a particular point for on-shell scattering. 
    Note that all the plotted function are $\langle r | \wf \rangle $ multiplied by $r$.  The information about training points is listed in the top panel, while the information about the testing point is provided in the title of the bottom panel.}
    \label{fig:emulated_wfs}
\end{figure}

In this subsection, I hope to gain some understanding of my emulators by analyzing the basic behaviors of the wave function  solution of Eq.~\eqref{eq:inhomeeqn_1} in coordinate space. Here, the sources are chosen as $ | \so \rangle = | \soadj\rangle = V | \Pin,\ell \rangle$ with $\Pin \equiv \sqrt{M_N \Ein}$. Bear in mind that $\Ein$ is always positive. 

To begin with, some analytical analysis is in order here.  In my simple example, the sources can be expressed as 

\begin{align}
| S \rangle & = | g \rangle \times \left( \lamtwo \langle g|  \Pin,\ell \rangle \right) \ , \label{eq:two_body_analytical_analysis_1}
\end{align}
which are localized, as long as the overlap integral in the expression is finite---as guaranteed by keeping $\Ein$ and thus $\Pin$ real. To deal with the exact solution $|\wf(E)\rangle$, it is rewritten as 

\begin{align}
    | \wf(E) \rangle  & = \frac{1}{E - H_0} \left( |\so\rangle + T(E) \frac{1}{E - H_0} |\so\rangle \right) \notag \\ 
    & = \frac{1}{E - H_0} \bigg[ |\so\rangle + | g \rangle  \left(\tau(E)   
    \left.\langle g \right| \frac{1}{E - H_0} \left|\so\rangle\right.\right) \bigg] \notag \\ 
    & \equiv  \frac{1}{E - H_0} | \overline{\so} \rangle  \ . \label{eq:two_body_analytical_analysis_2}
\end{align} 
In the second step, both $\tau(E)$ (discussed in Appendix~\ref{app:twobody}) and the matrix element of $H_0$'s resolvent are smooth functions of $E$ when $E$ is away from the real-$E$ axis.  The compactness of  $| \overline{\so} \rangle$ then follows from that of $| g \rangle$.
Therefore, at large $r$, based on the asymptotic behavior of the free Green's function~\cite{newton2002scattering}, one can see that for $s$-wave,

\begin{align}
\langle r | \wf(E)\rangle & \xrightarrow{r\to \infty}    \frac{\exp\left( i p r  \right)}{r} F(E) \ , \label{eq:two_body_analytical_analysis_3}
\end{align} 
with the factor $F$ determined by $|\overline{S}\rangle$. Here, $ p = \sqrt{2\mu |E| } \exp(i\theta_E/2) $ and $\theta_E \equiv \arg(E) \in [0, 2\pi )$ (i.e, with the \BCT defined on the positive-$E$ axis). When $E$ is right above the \BCT, $\langle r | \wf(E)\rangle $ behaves as an outgoing wave. However,  if $\arg(E) \neq 0 $, the function is exponentially damped at large $r$ due to the $\exp(- r \Im p )$ factor. 

Now, let me discuss numerical tests on these analytical understandings, for which I consider the $s$-wave channel and choose $\LAMtwo = 200$ MeV and $\lamtwo$ such that the system has a bound ground state with a $10$ MeV binding energy. The emulation is performed in a three-dimensional parameter space for $(\Re E, \Im E, \Ein)$, while the interaction operators in the $H$ and sources are fixed. Specifically, when setting up training calculations, I fix $\Im E^\mathrm{tr}_\alpha = 10 $ MeV and sample $N_b = 20$ training points using Latin Hypercube sampling (LHS)~\cite{doi:10.1080/01621459.1993.10476423} in the two-dimensional parameter space for $(\Re E, \Ein )$ with $\Re E \in [-50, 50]$ MeV and $\Ein \in [0, 100]$ MeV. I then construct the emulator and perform emulation at a randomly chosen point corresponding to on-shell scattering: $E = \Ein = 21$ MeV. 

The emulation-related wave functions in the $s$-wave scattering are plotted in Fig.~\ref{fig:emulated_wfs} (an extra $r$ factor is included in the plotted functions). The top panel shows the training solutions, which are spatially localized (bound-state-like) states. The bottom panel compares the emulated wave function and the exact solution: the agreement extends well beyond the range of the sources---recall $\LAMtwo = 200$ MeV.  More results with other $\Ein$ values can be found in the Supplemental Material (SM)~\cite{Zhang:2024gac_SM}.

The emulated solution eventually approaches zero with $r \to \infty$ since the training solutions have this asymptotic. However, as explained in Sec.~\ref{sec:interiorview}, the incorrect asymptotics in wave function emulation pose no issue since the $\resolventamp$ is computed via the overlaps in Eq.~\eqref{eq:source_psi_overlaps}. The spatial coverage of the sources is on the order of the interaction range, in which the wave function emulation is correct, and thus, so is the $\resolventamp$ emulation. 

Although these results, including Eqs.~\eqref{eq:two_body_analytical_analysis_1}-\eqref{eq:two_body_analytical_analysis_3}, are based on a separable interaction, the asymptotic behaviors of the training and emulated wave functions, as observed here, should be applicable to  non-separable interactions as well. To illustrate this point, the so-called Minnesota potential~\cite{THOMPSON197753} is studied in the SM~\cite{Zhang:2024gac_SM}. 

Finally, the treatment of the $E$ and $\Ein$ variables is worth pointing out again, as it is relevant for later discussions. An emulator for varying only $E$ while fixing all the other parameters can provide, for example, $\matrixform{\resolventamp}$ along the real axis. Hence, it is suitable for emulating response function calculations. However, for the on-shell scattering amplitudes, if one needs to emulate in $\Ein$, the $E$ and $\Ein$ variables must be varied \emph{independently}. Otherwise, while keeping the  $E = \Ein$ condition in the training calculations, the sources are not spatially compact if $\Ein$ turns complex-valued. 

\subsection{Emulation with fixed $H$ and sources} \label{subsec:two-body-givenHandSources}

The primary purpose is to understand the spectrum of a fixed $H$, as compressed in the emulator subspace, and its dependence on the $N_b$ and the locations of $E^\mathrm{tr}_\alpha$ in the complex $E$ plane. The latter dependence suggests a general choice of $E^\mathrm{tr}_\alpha$ suitable for continuum physics. Another main observation is the similarities between the CEE and the (near-)optimal rational approximation. 

\begin{figure*}
    \centering
    \includegraphics[width=0.48\textwidth]{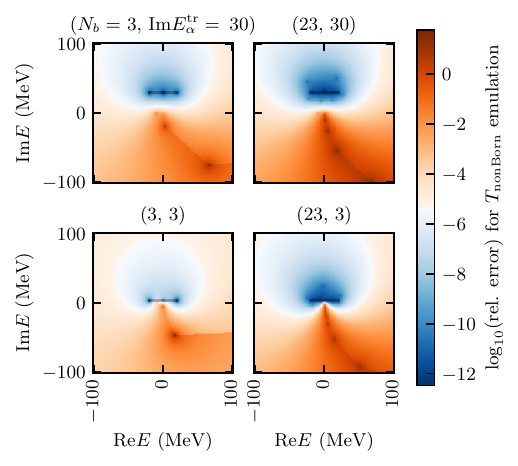}
    \includegraphics[width=0.48\textwidth]
    {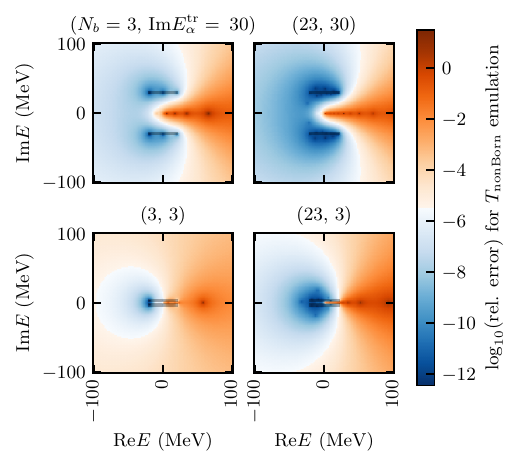}
    \caption{The relative error for emulated $\resolventamp$ in the complex $E$ plane. In each sub-figure, from left to right, $N_b$ increases from 3 to 23, while from top to bottom $\Im E^\mathrm{tr}_\alpha$ decreases from 30 to 3 MeV. As discussed in the main text, $E^\mathrm{tr}_\alpha$ are evenly separated along the black lines in each panel. 
    The two sub-figures differ in the arrangement of the training points.} 
    \label{fig:Swave_BE_10_LAM2_200_error_2dim}
\end{figure*}

I again choose  $ | \so \rangle = | \soadj \rangle = V | \Pin,\, \ell \rangle $ with $| \Pin,\, \ell \rangle$ as the incoming plane wave  ($\Pin = \sqrt{M_N \Ein}$), so that $\resolventamp$ corresponds to the non-Born term of $T$-matrix (labeled as $T_\mathrm{nonBorn}$). The sources and $H$ are fixed by fixing $\Ein$ $\lamtwo$ and $\LAMtwo$. Here, $\LAMtwo = 200$ MeV, and $\lamtwo$ is tuned to get a desired bound or resonance state in the $H$'s spectrum.
About the training points, $\Im E^\mathrm{tr}_\alpha$ is fixed to one or two values, while $\Re E^\mathrm{tr}_\alpha$ are sampled on an even mesh in a  $[-20, 20]$ MeV interval (see the black lines in e.g., Fig.~\ref{fig:Swave_BE_10_LAM2_200_error_2dim}). 

First, I aim to understand the CEE via its errors, which is also the primary approach for studying the (near-)optimal rational approximation~\cite{Trefethen2023review}\footnote{For example, the pole distribution pattern in the (near-)optimal rational approximation can be understood as the result of minimizing the approximation errors~\cite{Trefethen2023review}.}.  Fig.~\ref{fig:Swave_BE_10_LAM2_200_error_2dim} shows such relative errors for the emulated $s$-wave $\resolventamp(E)$s in the complex $E$ plane. Here, the two-body binding energy  $B_2$ is $10$ MeV and $\Ein = 10$ MeV in the sources. In the left sub-figure, the training points are lined on one black line above the real axis, but in the right, they are distributed on two black lines corresponding to a complex conjugate pair. The latter setup was mentioned previously at the end of Sec.~\ref{subsec:spectra}. Four emulations are contrasted within each sub-figure, based on different combinations of $\Im E^\mathrm{tr}_\alpha$  and $N_b $ values. 

According to the error patterns in the left sub-figure, the underlying $\resolventamp$ that the emulator approximates has its \BCT defined as away from the real axis, confirming the NHQM nature of the CEE, as asserted in Secs.~\ref{sec:introduction}, \ref{subsec:NHQM}, and~\ref{subsec:spectra}.

The general theme of the left sub-figure is that the errors are in the same order as those of the training calculations in the interpolation region (colored dark blue) between the training points or close to them. The errors increase when evaluating $\matrixform{\resolventamp}$ (i.e., the emulated $\resolventamp$) further away. The orange dots representing the largest emulation errors are the locations of the DC poles of $\matrixform{\resolventamp}$  because $\matrixform{\resolventamp}$ diverges at those locations but the exact results are finite. Those orange dots visually trace out a \BCT of the exact $\resolventamp$ that is significantly below the real $E$ axis, even though the high-fidelity calculation with the Hermitian $H$ operator would have a \BCT along the real axis (see, e.g., Fig.~\ref{fig:Tnb_in_CE_plane}). I.e., those DC poles form a proxy for the \BCT of the exact $\resolventamp$ that the emulator approximates. This is the same as how the (near-)optimal rational approximation works. I now consider $\matrixform{\resolventamp}$ effectively having the discretized \BCT. 

As it turns out, choosing an exact $T$-matrix to be compared with the emulated one could be subtle in the 4th quadrant of the complex $E$ plane. Note in the current discussion, I define the Riemann sheet of the exact $\resolventamp$ according to the \BCT defined on the real axis (e.g., as in the analytic formulas in Appendix~\ref{app:twobody}). The exact result in the quadrant 1 to 3 is the one on the 1st sheet. In quadrant 4, above the new discretized \BCT, the exact result used for computing emulation errors is the $\resolventamp$ defined on the 2nd sheet; however, below the new \BCT, the exact result is the one on the 1st sheet. The ambiguity could arise for the Riemann sheet assignment in the region close to the \BCT poles. Here, I connect the adjacent points by a straight line to form a continuous \BCT curve, which determines the Riemann sheet of the exact $\resolventamp$. The well-behaved error plot near the new \BCT shows the simple procedure is sufficient in the current cases;  otherwise, sharp edges could show up in the region away from the discretized \BCT. 

The detailed comparisons between different panels in the left sub-figure indicate that the emulations with smaller $\Im E_\alpha$ (with the same accuracy in the training calculations) have smaller errors in the region around the real axis. Increasing $N_b$ also systematically reduces the extrapolation errors. Therefore, the applicability of the CEE method relies on a compromise: the training calculations need to be performed sufficiently above the real axis to obtain enough precision when a given amount of computing resources is fixed. Still, the training points must be close enough to the real axis to control the extrapolation error around the real axis. 

It is worth noting that when approaching the branch point, the emulation error increases quickly. However, the problematic region around the branching point shrinks if the approximation improves---here by reducing $\Im E_\alpha$ and increasing $N_b$. This is a general behavior of rational approximations of $\resolventamp$, as mentioned in Sec.~\ref{subsec:optimarational}.

In the right sub-figure, where the emulators are trained on both $E^\mathrm{tr}_\alpha$  and their complex conjugates, the dark orange dots, i.e., the discretized \BCT,  are back on the real axis in all the panels.  This reinforces the relation between the error pattern and the \BCT location. The error pattern shares the same mirror symmetry with respect to the real axis as the locations of the training points. Another possibility for the discretized \BCT, while respecting the same symmetry, is to have poles distributed symmetrically on both sides of the real axis. As argued in the Appendix~\ref{app:proof_pseduo_hermicity}, this case would have more significant emulation errors in general than the existing one with a single discretized \BCT on the real axis.

The difficulty in describing continuum physics with \BCT on the real axis is also exposed clearly in the right sub-figure. Reducing $\Im E^\mathrm{tr}_\alpha$ and increasing $N_b$ doesn't improve $\matrixform{\resolventamp}$ on the real axis, although it does improve in the region of the bound states. Therefore, one should follow the training point setup used in the right sub-figure to infer the discrete excited states from CEE; however, in the following parts of the paper focusing on continuum physics, I distribute the training points as in the left sub-figure. 

\begin{figure}
    \centering
    \includegraphics[width=0.48\textwidth]{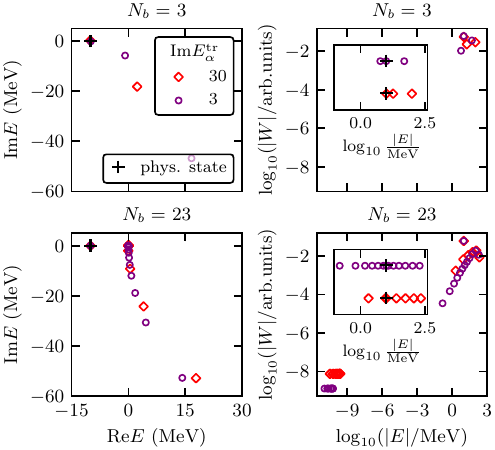}
    \caption{The left panels show the compressed spectra. The right ones plot the residues of the poles (see Eq.~\eqref{eq:emulated_amp_eigen_decomp}) vs. the absolute values of the corresponding eigenvalues. Their insets show the absolute values of the eigenenergies in the $\log$ scale. The physical (bound) state is marked with a black ``$+$'' in the left panels and the insets of the right panels. From the top and bottom rows, $N_b$ increases from $3$ to $23$. Each row compares two emulation results with different $\Im E^\mathrm{tr}_\alpha$.}
    \label{fig:Swave_BE_10_LAM2_200_poles}
\end{figure}

Figure~\ref{fig:Swave_BE_10_LAM2_200_poles} shows the information about the compressed spectra and the residues of the corresponding poles in $\matrixform{\resolventamp(E)}$s (see Eq.~\eqref{eq:emulated_amp_eigen_decomp}) from the emulators shown in the left sub-figure of  Fig.~\ref{fig:Swave_BE_10_LAM2_200_error_2dim}. From top to bottom, $N_b$ increases from 3 to 23. In each panel, the results from two emulators with different $\Im(E_\alpha^\mathrm{tr})$ are compared. 

The left panels again show the \BCT pattern of the DC state eigenenergies, while the bound state (not visible in Fig.~\ref{fig:Swave_BE_10_LAM2_200_error_2dim}) is separated from the DC states. This feature is similar to the eigenenergy distributions in the complex scaling calculations~\cite{Myo:2014ypa}, but the number of eigenvalues here is much smaller than that in a typical complex scaling calculation. Moreover, in the complex scaling method, the angle between the rotated {\BCT}s and the real axis is a control parameter, the maximum of which could be limited due to numeric difficulties. In the CEE, the moved {\BCT}s, which emerge from spectrum compression, could be affected by the $\Im E_\alpha^\mathrm{tr}$:  the smaller the $\Im E_\alpha^\mathrm{tr}$ gets, the further away the {\BCT}s are pushed from the real axis. Such behavior is reflected in the left panels.  

This could be understood intuitively: reducing $\Im E_\alpha^\mathrm{tr}$ decreases the emulation errors below the real axis, and thus effectively, the failure region, where the discretized {\BCT}s are located, is pushed further away from the real axis. Note the physical states, such as the bound state in the left panels, have less sensitivity numerically---supposedly no sensitivity at all in exact calculations. 

In fact, such sensitivity difference could be exploited to separate the physical states from the DC states, in addition to relying on the distribution pattern of the eigenenergies or poles. Both ways of separating states all trace back to the fact that the discretized \BCT poles represent the {\BCT}s of the exact $\resolventamp$ and thus can be ``redefined'' without changing the function behavior above both old and new {\BCT}s. This is also why one can use $-\Im \resolventamp$'s behavior on the real axis, specifically its decomposition into the smooth and peak components, to ``measure'' resonance properties (absent here but present in the $p$-wave case).  This point is further elaborated in Sec.~\ref{sec:three-body}.

The right panels in Fig.~\ref{fig:Swave_BE_10_LAM2_200_poles} plot the sizes of the pole residues  ($W_\beta$ in Eq.~\eqref{eq:emulated_amp_eigen_decomp}) against the absolute values of the eigenvalues. One can see that incredibly close to the branch point (threshold), there could be poles with tiny residues when $N_b$ becomes too large, such as in the $N_b =23$ case with over-sampled trainings. They could be the so-called Froissart doublets (or spurious poles, see Sec.~8.1 in Ref.~\cite{Trefethen2023review}), whose contributions are negligible unless $E$ gets exceptionally close to the threshold (much smaller than $10^{-10}$ MeV here). If the thresholds are inferred from subsystem mass computations, one can separate and eliminate these spurious poles in Eq.~\eqref{eq:emulated_amp_eigen_decomp}. Or, as mentioned in Sec.~\ref{sec:numerical_details}, I could apply the so-called greedy algorithm~\cite{sarkar2021selflearning, Quarteroni:218966} to eliminate redundancy, offering an effective regularization. For now, $N_b =3$ is under-spanning the subspace, while $N_b =23$ is over. Therefore, the effective dimension of the subspaces for both emulators is between $3$ and $23$.

Each inset plots the absolute values of the eigenenergies on a logarithmic scale, with the bound state marked in black $+$ (the number of symbols gives $\Neff$). The distribution of the DC states in terms of the distance between poles and the branch point is quasi-even on the log scale, with a gradual tapering (i.e., density reduction) towards both the threshold and the infinity, which is already visible in the left panels with linear scales. This distribution is similar to the tapered exponential clustering of poles in the (near-)optimal rational approximations discussed in Sec.~\ref{subsec:optimarational}. I thus identify the first similarity suggesting a close connection between CEE's dimension reduction efficiency and the optimality of the rational approximation. 

Also note that when reducing $\Im E_\alpha^\mathrm{tr}$, the smallest distance between the \BCT poles (or DC state eigenvalues)---excluding the spurious poles---and the threshold decreases; the distribution of poles or the eigenvalues also gets denser. The scale of that smallest separation controls to what proximity towards the threshold the emulator can be trusted because those close poles are responsible for the near-threshold behavior of $\matrixform{\resolventamp}$. The emulations and rational approximations eventually fail if $E$ gets too close to the threshold, as noted in the discussion of Fig.~\ref{fig:Swave_BE_10_LAM2_200_error_2dim} and in Sec.~\ref{subsec:optimarational}. 

\begin{figure}
    \centering
    \includegraphics[width=0.48\textwidth]{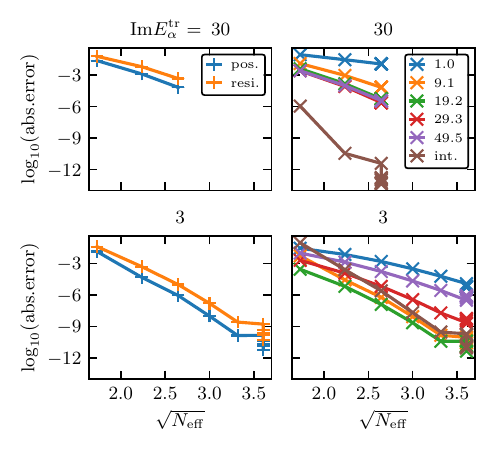}
    \caption{The emulation errors vs $\sqrt{\Neff}$. The left panels show the emulation errors for the bound state pole location and its residues. The right ones show the absolute errors for $\matrixform{\resolventamp}$ (1) on the real energy axis with different $E$ values marked in the legends, and (2) averaged over the interpolation region around the training energies (marked as ``int.''). From the top to bottom row, $\Im E^\mathrm{tr}_\alpha$ reduces from $30$ to $3$ MeV.}
    \label{fig:Swave_BE_10_LAM2_200_error_1dim}
\end{figure}

Of course, the emulation performance is also related to the number of effective poles (excluding the spurious poles), i.e., $\Neff$. As one can imagine, when $\Neff$ increases and approaches the dimension of the solution subspace, the quality of near-threshold approximation saturates, and the smallest distance between the pole and branch point becomes ``stuck''. (This saturation behavior is further illustrated later in the Fig.~\ref{fig:Swave_BE_10_LAM2_200_error_1dim}.) Therefore, by counting the number of symbols in the insets, i.e., the effective poles, in the $N_b = 23$ cases,  one sees that the $\Im E^\mathrm{tr}_\alpha =3 $ subspace is higher dimensional than that of $\Im E^\mathrm{tr}_\alpha =30$. This indicates $\Im E^\mathrm{tr}_\alpha$ could be viewed as a continuum resolution indicator for the solution manifold mapped out by varying the $\Re E$ parameter. Such resolution, an intrinsic property of the solution manifold, can only be ``measured'' by the emulator when its $\Neff$ saturates the subspace dimension. 

Fig.~\ref{fig:Swave_BE_10_LAM2_200_error_1dim} shows detailed information about the emulation errors vs $\sqrt{\Neff}$ for the pole properties, including its locations and residues\footnote{These residues could be related to the asymptotic normalization coefficients of the bound states~\cite{Zhang:2017yqc}.} in the left panels. The right panels show the same information for $\matrixform{\resolventamp(E)}$ in the interpolation region (marked as ``int'' in the legend) and along the real energy axis with locations marked in the legend (considered as extrapolations). $\Im E_\alpha^\mathrm{tr}$ decreases from the top to bottom panels. To get the data points for these plots, I run emulations with different $N_b$ and compute the corresponding $\Neff$ as already described. 

One can see the $e^{- c \sqrt{\Neff}}$ convergence behavior for all the quantities of my interests. Towards the largest $\Neff$, the employed SVD-based regularization could also contribute to the emulation errors. Moreover, the extrapolation errors are generally more significant than the interpolation errors but still satisfy the same scaling law. It is also clear that reducing $\Im E_\alpha$ increases the maximum $\Neff$ value that can be achieved, which is essential for lowering both interpolation and extrapolation errors. This supports the reasoning about $\Neff$ saturating the subspace dimension during the discussion of Fig.~\ref{fig:Swave_BE_10_LAM2_200_poles}. The convergence behavior is the second similarity that suggests the connection between the CEE and (near-)optimal rational approximations.

\begin{figure}
    \centering
    \includegraphics[width=0.48\textwidth]{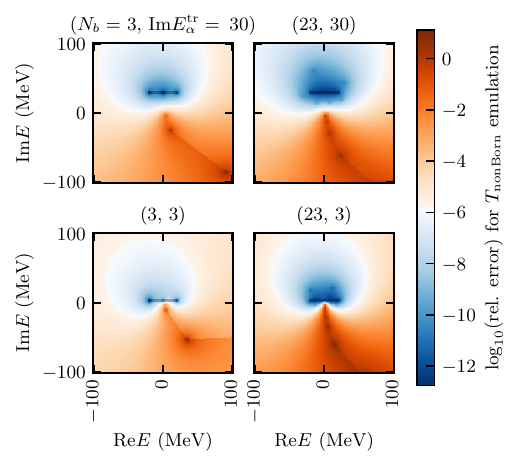}
    \caption{Similar to Fig.~\ref{fig:Swave_BE_10_LAM2_200_error_2dim}, but for the $p$-wave  case with $\LAMtwo = 200 $ MeV and $\lambda = 0.8$. The resonance pole is located at $ 3.9 - 2.6\,i$ MeV.}
    \label{fig:Pwave_lam2_mpt8_LAM2_200_error_2dim}
\end{figure}

\begin{figure}
    \centering
    \includegraphics[width=0.48\textwidth]{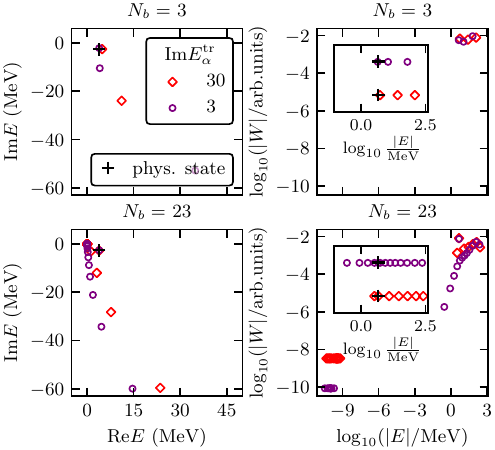}
    \caption{Similar to Fig.~\ref{fig:Swave_BE_10_LAM2_200_poles}, but for the $p$-wave case with $\LAMtwo = 200 $ MeV and $\lambda = 0.8$. The resonance pole is located at $ 3.9 - 2.6\,i$ MeV.}
    \label{fig:Pwave_lam2_mpt8_LAM2_200_poles}
\end{figure}

 Now, I look into the $p$-wave channel to investigate emulation performance with the presence of a resonance. Here, $\LAMtwo = 200$ MeV and $\lamtwo = 0.8$ with a resonance located at $E= 3.9 - 2.6\, i$ MeV. Figs.~\ref{fig:Pwave_lam2_mpt8_LAM2_200_error_2dim}
and~\ref{fig:Pwave_lam2_mpt8_LAM2_200_poles}
show the emulation error and spectrum in the complex-$E$ plane, in parallel to the  Figs.~\ref{fig:Swave_BE_10_LAM2_200_error_2dim} and~\ref{fig:Swave_BE_10_LAM2_200_poles} for the $s$-wave. Since the resonance pole is well reproduced by the emulator, no enhanced error (i.e., dark orange dot) can be found at the resonance location in Fig.~\ref{fig:Pwave_lam2_mpt8_LAM2_200_error_2dim}. The other results concerning emulation errors vs $\sqrt{\Neff}$  are also  qualitatively similar to those in the $s$-wave case, which  can be found in the SM~\cite{Zhang:2024gac_SM}.  The key difference, though, is that the physical pole is now located below the real axis and thus has bigger errors than in the previous $s$-wave results. However, all the other observations are the same as in the $s$-wave case.

\subsection{Emulation in ($\Re E$, $\Im E$, $\Ein$, $\lamtwo$, $\LAMtwo$) } \label{sec:two-body-manyH}

The CERPE is examined here by studying both $s$ and $p$ waves. The parameter space is now enlarged to five-dimensional: ($\Re E$, $\Im E$, $\Ein$, $\lamtwo$, $\LAMtwo$).  $\LAMtwo$ is a non-affine parameter, but I am not concerned with the emulation speed but the accuracy. Extra steps are needed for speed, which will be addressed elsewhere. 

As will be illustrated in the results of this subsection, the behaviors of the emulators, when the $H$ and sources are varied, are similar to the behaviors seen in the CEE case, except the clear indication of a higher-dimensional solution manifold mapped out by varying more parameters and a different $\Neff$ dependence of the error from $e^{- c \sqrt{N_\text{eff}}}$ scaling seen in the CEE case.

I start with the $s$-wave case. The training points are sampled using LHS in a 4-dimensional space: $\lamtwo \in [-3, -1.5 ]$, $\LAMtwo \in [150, 250]$ MeV,  $\Ein \in [0, 20]$ MeV, $\Re E^\mathrm{tr}_\alpha \in [-20, 20]$ MeV, with  $\Im E^\mathrm{tr}_\alpha$ fixed to either 3 or 30 MeV, as done previously. To test the emulators, $\lamtwo$, $\LAMtwo$, and $\Ein$ are sampled randomly 50 times in the same ranges as in sampling the training points, but the ranges for $E$ are more extensive. The $(\Re E, \Im E)$ values are on a  $30\times 30$ evenly-spaced grid spanning from $-50$ to $50$ MeV along both axis to study interpolation and extrapolation in the complex $E$ plane.

\begin{figure}
    \centering
    \includegraphics[width=0.48\textwidth]{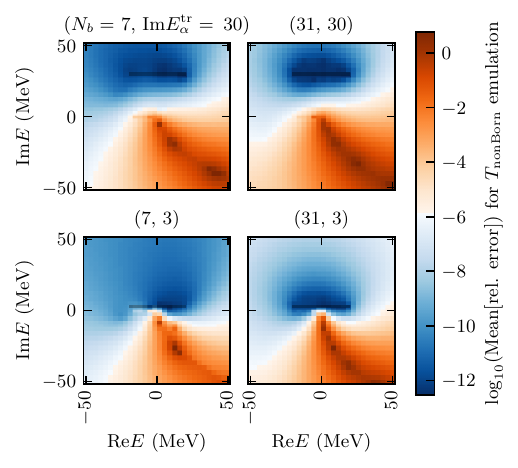}
    \caption{For the emulations in the $s$-wave channel in ($\Re E$, $\Im E$, $\Ein$, $\lamtwo$, $\LAMtwo$) space. The plots show the mean of the relative emulation errors of the emulations in the complex $E$ plane. See the text for more details. Note the range of the complex plane is smaller than those in, e.g., Fig.~\ref{fig:Swave_BE_10_LAM2_200_error_2dim}.}
    \label{fig:Swave_ave_error_2dim}
\end{figure}

Fig.~\ref{fig:Swave_ave_error_2dim} shows, at each $E$, the mean of the relative emulation errors, averaged over the sample with the same $E$ but different $(\Ein, \lambda, \LAMtwo)$ values. The error pattern is similar to the pattern, e.g., in the left sub-figure of Fig.~\ref{fig:Swave_BE_10_LAM2_200_error_2dim}, with a clear sign of moved \BCT. 

\begin{figure}
    \centering
    \includegraphics[width=0.48\textwidth]{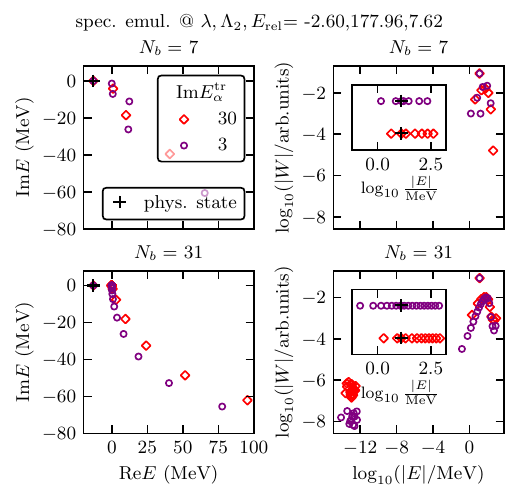}
    \caption{For the same emulators as shown in Fig.~\ref{fig:Swave_ave_error_2dim}. But, what is plotted are the distribution of the poles (or emulated spectra) in $\matrixform{\resolventamp}$s, at a random testing point with $\lambda, \LAMtwo, \Ein  = -2.60, 177.96, 7.62 $ (shown in the title).}
    \label{fig:Swave_poles_test1}
\end{figure}

Fig.~\ref{fig:Swave_poles_test1} shows the information about the emulated spectra (or the poles in the $\matrixform{\resolventamp}$) at a randomly chosen testing point (see the title), in the same way as being presented in Fig.~\ref{fig:Swave_BE_10_LAM2_200_poles}. Again, one notices similar emulator behaviors in the CEE cases, including the moved {\BCT}s, the tapered exponential clustering of the poles toward the branch points, how the physical state and the DC states respond differently to changing $\Im E^\mathrm{tr}_\alpha$ and $N_b$, the existence of the spurious poles when $N_b$ is too big, and the density increase of the non-spurious poles when lowering $\Im E_\alpha$ and increasing $N_b$. 

\begin{figure}
    \centering
    \includegraphics[width=0.48\textwidth]{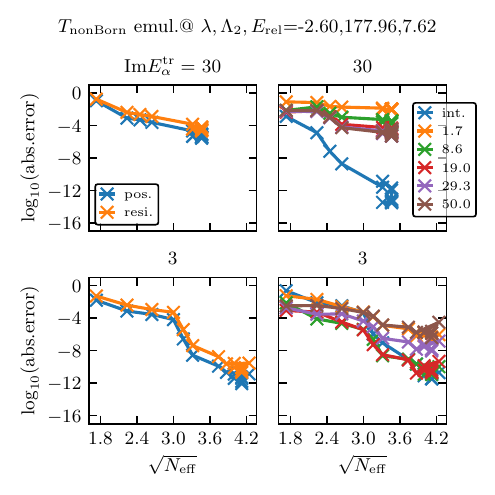}
    \caption{For the same emulators as shown in Fig.~\ref{fig:Swave_ave_error_2dim}. Similar to Fig.~\ref{fig:Swave_BE_10_LAM2_200_error_1dim}, the plots show the emulation errors vs $\sqrt{N_\text{eff}}$ at a testing point shown in the title (the same point as in Fig.~\ref{fig:Swave_poles_test1}). }
    \label{fig:Swave_error_1dim_test1}
\end{figure}

Figure~\ref{fig:Swave_error_1dim_test1} shows the emulation errors at the same testing point as in Fig.~\ref{fig:Swave_poles_test1}. In the same way as the presentation of Fig.~\ref{fig:Swave_BE_10_LAM2_200_error_1dim}, the panels show  the emulation errors vs. $\sqrt{\Neff}$ for the properties of the physical (bound) state pole and $\matrixform{\resolventamp}$ at the interpolation region and a few points on the real axis. The scaling of errors against $\sqrt{\Neff}$ differs from the scaling in the CEE cases. This reflects that I am working with multivariate approximations, different from the univariate cases with the (near-)optimal rational approximation. However, do note that the errors still decrease in a certain exponential fashion with $\sqrt{\Neff}$. As the other distinction, the maximum $\Neff$ value here is larger than that in Fig.~\ref{fig:Swave_BE_10_LAM2_200_error_1dim} with the same $\Im E^\mathrm{tr}_\alpha$ (see the symbol densities of the insets of the two figures). For example, when $\Im E^\mathrm{tr}_\alpha = 3$ MeV, $\Neff = 13$ and $19$ for the CEE and CERPE cases respectively. This is also expected, since the solution subspaces here have higher dimensionalities than those in the CEE case. 

\begin{figure}
    \centering
    \includegraphics[width=0.48\textwidth]{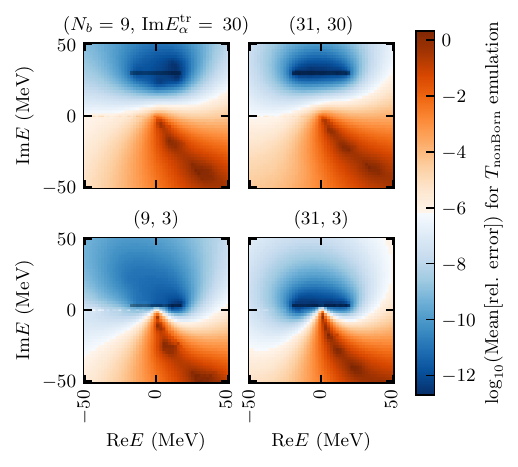}
    \caption{For the emulation of $p$-wave scattering in the ($\Re E$, $\Im E$, $\Ein$, $\lamtwo$, $\LAMtwo$) space. The plots show the mean of relative emulation errors in the complex $E$ plane in the same way as in Fig.~\ref{fig:Swave_ave_error_2dim}. See the text for more details. }
    \label{fig:Pwave_ave_error_2dim}
\end{figure}

The $p$-wave results, presented in the same fashion as the $s$-wave results, are collected in Fig.~\ref{fig:Pwave_ave_error_2dim} and additional figures in the SM~\cite{Zhang:2024gac_SM}. 
The 5-dimensional parameter space is defined as  $\lamtwo \in [-3, -0.5 ]$, $\LAMtwo \in [150, 250]$ MeV,  $\Ein \in [0, 20]$ MeV, and both $\Re E$ and $\Im E $ $\in [-50, 50]$ MeV. The training points are sampled using LHS with the same ranges for $\Ein$, $\lamtwo$, and $\LAMtwo$, but a narrower range for $\Re E \sim [-20, 20]$ MeV. $\Im E^\mathrm{tr}_\alpha$ is again fixed to $3$ or $30$ MeV. When testing the averaged emulation performance, $\lambda$, $\LAMtwo$, and $\Ein$ are sampled randomly 200 times in the parameter space for each $E$ in the complex $E$ plane. I provide results at two test points in the SM~\cite{Zhang:2024gac_SM} to probe two types of emulations: one with a bound state and the other with a resonance. The parameter values can be found in the title of each sub-figure therein. 

Again, the general behaviors are similar to those in the $s$-wave case. One difference is that the errors for the resonance pole positions and residues are generally larger than those for the bound states, as they are located further below the real axis.

\section{Three-body demonstrations} \label{sec:three-body}

In this section, I further study these emulators in a system of three identical bosons with $s$-wave pairwise and $s$-wave three-body interactions. Of course, the physics become more complex. There could be multiple thresholds, multiple bound states, and a near-threshold resonance state in the spectra, meaning the analytical behaviors of $\resolventamp$ and $\matrixform{\resolventamp}$ in the complex $E$ plane are closer to realistic systems (such as nucleus) than those in the two-body case. When exploring the $\vectheta$ parameter space, the system changes in nontrivial ways. However, the general observations about the properties and performances of the emulators are consistent with those seen in the two-body cases, lending further evidence about the robustness of the continuum extraction and emulation methods and the general understanding of the methods. 

\subsection{Three-body: basics} \label{sec:three-body-basics}

The Hamiltonian operator in the \schro equation is $H_{sr} = H_0 + V$, with $H_0$ as the kinetic energy operator in the center of mass frame, $V \equiv \sum_{i = 1}^4 V_i$,  $V_{i=1,2,3}$ as the pair-wise interaction with $i$th particle as the spectator, and $V_{4}$ as the three-body force. The notations in Ref.~\cite{Glockle:1983} are followed loosely here. All the interactions are separable, with Gaussian-like form factors. The two-body interaction is the same as the interaction discussed in Sec.~\ref{sec:two-body}. The system has been studied previously in Ref.~\cite{Zhang:2021jmi} for a different purpose. 

I work in the Faddeev framework to solve the \schro equation. Therefore, I deal with the Faddeev equations and their solutions, known as Faddeev components, in emulator development and high-fidelity benchmark calculations. It will be interesting to study the equivalence between the current emulation approach and the approach directly based on the \schro equation in the future.

\subsubsection{Three-body interaction}\label{subsec:three-body-force}

Similar to Refs.~\cite{Phillips:1966zza,Zhang:2021jmi}, I introduce separable potential for the three-body force with a coupling $\lamfour$: $V_4  = \lambda_4 | g_4 \rangle \langle g_4 |$. The corresponding $T$-matrix, $t_4$, defined with only the three-body interaction, is  

\begin{align}
    t_4(E) & = \tau_4(E) | g_4 \rangle \langle g_4 | \ , \\
    \tau_4(E)^{-1} & = \lambda_4^{-1}- \langle g_4|G_0(E)| g_4 \rangle \ , \ \text{with} \\
    G_0 (E) & \equiv \frac{1}{E - H_0} \ . 
\end{align}  
The form factor $| g_4 \rangle$ takes a Gaussian form~\cite{Zhang:2021jmi}: 

\begin{align}
    \langle  {P}_1,  {q}_1 | g_4 \rangle & = 
    \frac{4 \pi}{\sqrt{M \Lambda_4^4}} \exp\left[-\frac{M E_4}{2 \Lambda_4^2}\right] \ , \ \text{with}  \\
    E_4 & \equiv \frac{P_1^2}{2\mu^1} + \frac{q_1^2}{2\mu_1} \ . 
\end{align}
As in Ref.~\cite{Glockle:1983}, $\mu^1 = 2 M_N/3$ and $\mu_1 = M_N/2$ are the reduced mass between particle and dimer and between particle and particle, respectively; $P_1$ and $q_1$ are the corresponding relative momenta, with particle-1 as the spectator. The momentum pair with other spectators are $P_2, q_2$ and $P_3, q_3$.  Since the kinetic energy and  $E_4$ are invariant when changing the momenta sets, I have $ \langle {P}_2, {q}_2 | g_4 \rangle =  \langle {P}_3, {q}_3 | g_4 \rangle = \langle {P}_1, {q}_1 | g_4 \rangle $, which is a property I use to simplify the calculation.

\subsubsection{Faddeev equations} \label{subsubsec:Faddeev_eqs}

My interest is mainly the elastic scattering between particle and dimer, both below and above the dimer break-up threshold. As a preparation, suppose $|\phi_1 \rangle \equiv |\Pin \rangle |\varphi_B \rangle$, which is the product of the bound state of dimer-23 and a plane wave state for the relative motion between particle-1 and the dimer-23. Note $ |\Pin\rangle $ is a $s$-wave state. The associated many-body total energy of $|\phi_1 \rangle$ is $E = {\Pin^2}/{2\mu^1} - B_2$. A variable for the relative energy between a particle and a dimer $\Ein = {\Pin^2}/{2\mu^1}$ is also used later.  For the separable potential, the dimer bound state $|\varphi_B\rangle = \bsren G_{0,2b}(-B_2)|g\rangle$ with $G_{0,2b}(-B_2)$, the free two-body Green's function evaluated at two-body energy $e_2 = - B_2$ ($B_2$ as the dimer binding energy); $\xi$ is a factor properly normalizing the bound state. 

There exist various three-body Green's functions:

\begin{align}
    G_i(E) & \equiv \frac{1}{E- H_i}  \ , \\
    H_i & \equiv  H_0  + V_i \ , 
\end{align}
and $V^i  \equiv V - V_i$.  The Lippmann-Schwinger equation for the channel of scattering beween particle-1 and dimer-23 can be recast as~\cite{Glockle:1983}

\begin{align}
    | \wf_1^+ \rangle & = G_1 V^1 | \wf_1^+ \rangle + |\phi_1 \rangle  \notag \\ 
    & = G_i V^i | \wf_1^+ \rangle \ \text{with} \ i=2,3,4 \ . 
\end{align}
$| \wf_1^+ \rangle$ is the full scattering wave function of this channel labeled with subscript $1$.  Now, I introduce four different Faddeev components and specifically separate the incoming wave from these components:

\begin{align}
    | \psi_{1,1} \rangle & = G_0V_1 |\wf_1^+\rangle - | \phi_1 \rangle\,, \notag \\ 
    | \psi_{1,i=2,3,4} \rangle & = G_0V_i |\wf_1^+\rangle  \ . \label{eq:FC_def}  
\end{align}
I then arrive at the Faddeev equations for the channel-1 scattering: 

\begin{align}
    (E - H_i) |\psi_{1,i} \rangle = V_i \left(\bar{\delta}_{1,i} |\phi_1 \rangle + \sum_{j=1}^4 \bar{\delta}_{i,j} | \psi_{1, j} \rangle \right) \ , 
\end{align}
with  $i = 1,2,3,4$,  $\bar{\delta}_{i,j} \equiv 1 - \delta_{i,j}$ ($\delta_{i,j}$ as the Kronecker delta). 

The above equations are then symmetrized for the identical boson system by summing up the other two channels with particle-2 and -3 as the spectator, respectively. Four Faddeev components in the boson-dimer scattering channel then turn into

\begin{align}
    |\psi_j \rangle = \sum_{i=1}^3 | \psi_{i,j}\rangle  \ \text{for} \ j=1,2,3,4 \ . 
\end{align}
Note $|\psi_2\rangle$ and $|\psi_3\rangle$ can be computed by applying proper permutation operators on $|\psi_1\rangle$. For example, $|\psi_2\rangle + | \psi_3\rangle = \perm | \psi_1 \rangle$, with $\perm$ as the sum of two permutation operators~\cite{Glockle:1983}. Therefore, I focus on the $|\psi_{1} \rangle$ and $|\psi_{4} \rangle$ components. Their Faddeev equations become 

\begin{align}
    (E - H_1) | \psi_1 \rangle & = V_1\perm | \phi_1 \rangle + V_1\perm | \psi_1 \rangle + V_1 |\psi_4 \rangle \ ,   \label{eq:Faddeev1} \\ 
    (E - H_4) | \psi_4 \rangle & = V_4 \left( 1 + \perm \right)  \left( | \phi_1 \rangle + | \psi_1 \rangle\right)  \ .  \label{eq:Faddeev4} 
\end{align}
With these solutions, the symmetrized wave function can be expressed as  

\begin{equation}
|\Psi^+ \rangle = \sum_{i=1}^3|\Psi_i^+ \rangle =  (1 + \perm) \left(| \psi_1 \rangle + |\phi_1\rangle \right)+ | \psi_4 \rangle  \ . \label{eq:fullWF1}
\end{equation}

Eqs.~\eqref{eq:Faddeev1} and~\eqref{eq:Faddeev4} can be recast in the form of Eq.~\eqref{eqn:BasicEq} with 

\begin{align}
{M} \begin{pmatrix}
 | \psi_1 \rangle    \\
 | \psi_4 \rangle  
\end{pmatrix}  \equiv  M |\wf \rangle = |\so\rangle \ ,  \label{eq:three_body_eq1}
\end{align}

\begin{align}
{M} \equiv 
\begin{pmatrix}
 E - H_1 - V_1 \perm  & - V_1   \\
  -3 V_4 & E- H_4   
\end{pmatrix}  \equiv  E - H  \ ,  
\end{align}
and

\begin{align}
|\so \rangle \equiv 
\begin{pmatrix}
 V_1 \perm  | \phi_1 \rangle    \\
  3 V_4 | \phi_1 \rangle  
\end{pmatrix} 
\equiv \begin{pmatrix}
    \soket{1} \\ 
    \soket{4}
\end{pmatrix} \ .  \label{eq:so_3body}
\end{align}

Now $H$---not $H_{sr}$---is the effective Hamiltonian operator working in the space of $|\psi_1\rangle$ and $|\psi_4\rangle$; $|\wf \rangle $ in these equations is not the whole wave function $|\wf^+\rangle$.  To simplify the equations, I take advantage of the fact that $g_4(P,q)$ is invariant with respect to particle permutation (see Sec.~\ref{subsec:three-body-force}) so that $V_4 (1+\perm) = 3 V_4 $. 

Knowing the full scattering wave function $| \wf^+ \rangle$ and its expression in terms of Faddeev components via Eq.~\eqref{eq:fullWF1}, the corresponding scattering amplitude can be expressed as~\cite{Glockle:1983}

\begin{align}
T & = \langle \phi_1 | (V_2 + V_3 + V_4) |\Psi^+ \rangle  \notag \\ 
& = \langle \phi_1 | \perm V_1 (1 +  \perm) + 3 V_4 |\phi_1 \rangle + \langle \soadj| \begin{pmatrix} | \psi_1 \rangle \\  |\psi_4 \rangle  \end{pmatrix} \notag \\ 
& \equiv T_\mathrm{Born} + T_\mathrm{nonBorn}  \ , \label{eq:Tonshell_decomp}
\end{align}
with

\begin{align}
|\soadj \rangle \equiv 
\begin{pmatrix}
\bigg\{  ( 1 + \perm ) V_1 \perm  + 3 V_4 \bigg\} | \phi_1 \rangle    \\
 \left( V_1 \perm  +  V_4 \right ) | \phi_1 \rangle  
\end{pmatrix}  
\equiv \begin{pmatrix}
 \soadjket{1}    \\
 \soadjket{4}  
\end{pmatrix} \ .  \label{eq:soadj_3body}
\end{align}
The $T_\mathrm{Born}$ terms is explicitly presented in Appendix~\ref{app:threebody_formulas}. 

Since I am interested in the overlaps between $|\wf\rangle$ and the source $|\soadj \rangle$ for computing $T_\mathrm{nonBorn}$, I introduce the adjoint equation to develop emulators (see Sec.~\ref{sec:var_and_rbm})

\begin{align}
{M}^\dagger \begin{pmatrix}
 | \psiadj{1} \rangle    \\
 | \psiadj{4} \rangle  
\end{pmatrix} \equiv M^\dagger |\adj{\wf} \rangle = |\soadj \rangle  \label{eq:three_body_adjeq1}
\end{align}
The solutions of this equation, $|\psiadj{1,4}\rangle$, can be used in the variational approach to estimate $\langle \soadj | \psi\rangle $ with a second-order error. 

Moreover, both sources $| \so \rangle$ and $| \soadj \rangle$ are \emph{spatially localized}, which can be checked analytically using their definitions in Eqs.~\eqref{eq:so_3body} and~\eqref{eq:soadj_3body}. The explicit expressions of these sources in the momentum space can be found in Eqs.~\eqref{eq:Pq_so1_overlap}, \eqref{eq:Pq_so4_overlap}, \eqref{eq:Pq_soadj1_overlap}, and~\eqref{eq:Pq_soadj4_overlap}. The momentum dependence of these sources is smooth without any singular behaviors, meaning the sources are spatially localized.

\subsubsection{Training calculations, emulations, and benchmark calculations}

With the Faddeev equations developed in the Sec.~\ref{subsubsec:Faddeev_eqs}, the emulator developments follow the general discussion in Sec.~\ref{sec:interiorview}, as implemented in Sec.~\ref{sec:two-body} for the two-body system. Therefore, the emulation formalisms are not repeated here. However, some details about treating the parameters in $H$ and the sources are worth mentioning. The methods for performing training calculations at complex energies and benchmark calculations at real energies are also discussed.   

When computing or emulating \emph{on-shell} $T_\mathrm{nonBorn}$ amplitude, the $E$ variable in the resolvent operator and the real $\Ein$ parameter in the sources are related via $E = \Ein - B_2$; the potential parameters in the $H$ and sources are undoubtedly the same. However, when performing the training calculations in the form of Eqs.~\eqref{eq:training_eq} and~\eqref{eq:training_adjeq} to collect $| \wf^\mathrm{tr}_\alpha\rangle$ and $| \wfadj^\mathrm{tr}_\alpha\rangle$, the $E$ variable, being complex generally, is treated as independent of $\Ein$. 

In addition to the elastic scattering amplitude, I also study $\resolventamp$ with \emph{fixed} sources. In these cases, I assign particular values for $\Ein$ and the parameters of the $V_1$ and $V_4$ that are inside the sources (see Eqs.~\eqref{eq:so_3body} and~\eqref{eq:soadj_3body}) and fix them, while varying the $E$ and the $H$'s parameters in the resolvent operator. These calculations resemble computing response functions when $E$ is along the real axis with $\Im E = 0^+$. This type of emulator can also be used to access the compressed spectra and their dependence on the $H$ parameters because the spectrum, as a property of $H$, should be accessible by any sources (or ``probes'' in terms of response function terminology).

The training solutions are spatially localized and thus can be solved using an integrable basis, such as a Harmonic oscillator basis. However, I work in the momentum space in which the $G_0$ is a diagonal operator, and the Faddeev equations can be simplified so that one deals with only one momentum variable, $P$. The nonzero imaginary part of the complex $E$ smooths out the singular functions in solving the training equations, which would be present with real $E$ above the lowest threshold. The difficulty of dealing with moving singularities is mitigated in the training calculations.

These momentum-space equations are derived and simplified step by step in Appendix~\ref{app:threebody_formulas}. Note that in those equations, the parameters in the sources and those in the $H$ are labeled differently, as there are cases in which I need to treat some of them separately.  
To solve them, I apply the interpolation method~\cite{Baye:2015xoi} based on the Lagrange function of the Legendre polynomials to discretize the integral equations and obtain a linear equation system, which is solved numerically using a standard linear algebra package. I increase the mesh points to get the desired high accuracy, which could be improved by adapting the careful choice of mesh points and Gaussian quadratures developed in the CE calculations~\cite{Deltuva:2012fa}.  

I emphasize that the training equations in Eqs.~\eqref{eq:training_eq} and~\eqref{eq:training_adjeq} apply for any finite ranged interactions. In the model with separable potentials, however, those equations are simplified to the forms in Appendix~\ref{app:threebody_formulas}, which are then solved to obtain the training solutions.

For benchmarks, I perform two types of calculations: one is with $E$ in the region of training energies, i.e., the test $E$ is in a similar area of the complex plane as the training energies $E^\mathrm{tr}_\alpha$, while the other is on the real energy axis, corresponding to continuum physics of the interests. The conclusion for the first type of benchmark is the same as in the two-body sector: the emulators, as single or multivariate interpolants, have similar accuracy as that of the training calculations. 

The important benchmark is the second type. Note the form of my response-like calculation is new, but the scattering $T$-matrix calculation is well established. As checked analytically at the end of Appendix~\ref{app:three-body-highfidelty}, my equations and formulas provide the same results as the established calculations for on-shell  scattering $T$-matrix.   

When performing the second type of benchmark calculations, I apply the interpolation method based on the Lagrange function~\cite{Baye:2015xoi} of the Legendre polynomial to discretize the equations, including Eqs.~\eqref{eq:sol_psi1_1}--\eqref{eq:sol_psi4_1}, as I do in obtaining training solutions. However, the computation becomes significantly different.  When $E$ is larger than the lowest threshold, the kernel in the integral equation develops various singularities, and the solution itself has branch points, which require dedicated treatments. Those details can be found in Appendix~\ref{app:three-body-highfidelty}. With these real-$E$ benchmark calculations setup, I can also extract the information about the bound and resonance states. The procedures are discussed in Appendix~\ref{app:three-body-phys-states-highfidelty}. 

\subsubsection{Numerical values for the Hamitonians}

In this work, I fix $\LAMtwo$ and $\Lambda_4$ in the two and three-body interactions to $\LAMtwo = 200$ MeV and $\Lambda_4 = 300$ MeV. $\lambda_4$ is varied in  $[-0.5, 0.5]$. I vary $ B_2 \in [2,  10]$ MeV when looking at particle-dimer scattering. Note since $\LAMtwo$ is fixed, a $B_2$ value determines the $\lamtwo$ value, and vice versa. Associated with the $B_2$ range, $\lamtwo$ varies from $ -1.44 $ to $-2.15$. In addition, I vary $\lamtwo \in [-2.15, 1 ] $, a wider range  in which the resonances could appear in the three identical boson systems with unbound dimers. Bear in mind that these ranges define the parameter spaces where I will train emulators in the following demonstrations; I also test the extrapolation capability by choosing testing points in larger ranges, if explicitly stated.

\subsection{Emulation in ($\Re E$, $\Im E$) with fixed $H$ and sources} \label{subsec:three-body-fixedH-fixedsources}

The emulation is performed only in the complex $E$ plane, while the $H$ and sources are fixed. The source parameters are set as: $\Ein^s = 1$ MeV, $B_2^s = 10$ MeV, and $\lambda_4^s = -0.2$. They are specifically labeled with the ${}^s$ superscripts to differentiate them from those in the resolvent operator. I then study three cases with distinct $\lamtwo$ (and the associated $B_2$) and $\lambda_4$ values for $H$. I will show the compressed spectra and compare them with the ``spectra'' extracted using the AAA method mentioned in Sec.~\ref{subsec:optimarational}. The $\matrixform{\resolventamp}$ results are also presented. 

\begin{figure*}
    \centering
    \includegraphics[width=\textwidth]{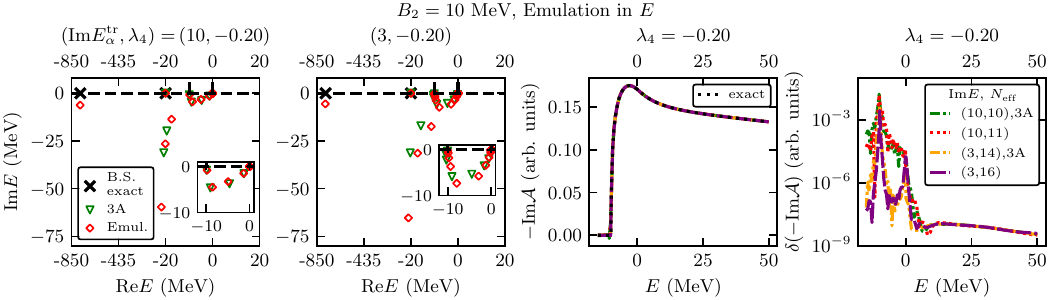}
    \caption{Emulations in $E$ using the RBM emulator and the AAA method. The $H$ is fixed by setting $B_2 = 10$ MeV and $\lambda_4 = -0.2$, while the information for the fixed sources is given in the text. The first two panels, sharing the same legend, show the eigenvalues in the complex plane, extracted from the $\Im(E_\alpha^\mathrm{tr}) = 10$ MeV emulators (1st) and the $\Im(E_\alpha^\mathrm{tr}) = 3$ MeV emulators (2nd). Note that in these two panels, the left and right half of the real axis have different scales to accommodate the deep bound state location. The spectra extracted from the same training data sets but using the AAA method (``3A'') are shown for comparisons.
    The exact results for the bound states are marked as black $\times$, and exact thresholds are 
    indicated by the short vertical black lines. 
    The insets zoom in to the region close to the branch points. The 3rd panel compares various $-\Im\matrixform{\resolventamp}$ to the exact results on the real energy axis, while their differences, indicating the emulation errors, are shown in the last panel. Both panels share the same legend showing the information about $\Im(E^\mathrm{tr}_\alpha)$ and $\Neff$.}
    \label{fig:spectra_fixSfixpot_BE2_10_lam4_-0.2}
\end{figure*}

For Fig.~\ref{fig:spectra_fixSfixpot_BE2_10_lam4_-0.2}, I work with a $H$, whose $B_2 = 10$ MeV and $\lambda_4 = -0.2 $. I trained two emulators with $(\Im E^\mathrm{tr}_\alpha, N_b )= (3, 48)$ and $(10, 30)$ respectively. Similar to the two-body studies, $\Re E^\mathrm{tr}_\alpha$ are evenly sampled in the $[-20, 50]$ MeV interval. Note that the effective dimension $\Neff$ (see the legend of the 4th panel), i.e., the number of training points needed, is smaller than $N_b$ in each case. Furthermore, I performed analytical continuations of the two training data sets of $\resolventamp(E^\mathrm{tr}_\alpha)$ using the AAA method and collected the corresponding spectra and $\matrixform{\resolventamp}$ results. 

The 1st and 2nd panels, sharing the same legend, show the spectra of the two emulators separately, together with the corresponding AAA results (``3A''). For benchmarking, the exact bound states are marked with black ``$\times$'' and the exact thresholds with short vertical lines. To accommodate the deep bound state in the plots, the left and right half of the $x$-axis have different scales. The insets in these plots zoom in on the regions near the two branch points. The corresponding $\Neff$ of both emulations and the AAA calculations can be found in the legend of the 4th panel. The AAA $\Neff$ is computed similarly to the emulator $\Neff$ by counting the number of effective poles in the corresponding $\matrixform{\resolventamp}$ excluding those extremely close to the branch points and with tiny residues. The 3rd panel compares th emulation results of $-\Im \matrixform{\resolventamp}$ with the exact ones on the real energy axis below and above the particle-dimer threshold; their differences, i.e., the emulation errors, are plotted in the last panel. Note the two panels share a legend. 

As can be seen in the first two panels, the basic features of the spectra are correctly reproduced by the emulators, including the bound states, the thresholds indicated by the crossings between the real axis and the discretized {\BCT}s, and the exponential clustering of the DC states towards the thresholds. The details here are more complex than in the two-body case. There are now multiple bound states. Two thresholds exist, one corresponding to particle-dimer and the other to three-particle. However, the general characteristics of the spectra are the same as in the two-body case.  

Note that in this specific system, no resonance exists when the dimer is bound. This is consistent with the fact that no $s$-wave resonance exists in a two-body scattering. It is satisfying to see no isolated resonance state exists in the spectra. 
 
Meanwhile, the amplitude, specifically $-\Im \resolventamp$ on the real energy axis, is also well reproduced by the emulators across the particle-dimer threshold in the 3rd panel. 

The comparisons between the two emulation results reinforce my understanding of how $\Im E^\mathrm{tr}_\alpha$ impacts the emulator performance. When reducing $\Im E^\mathrm{tr}_\alpha$  from $10$  to $3$ MeV, $\Neff$ increases from $11$ to $16$; the density of the DC states increases; and the emulation errors for $-\Im \matrixform{\resolventamp}$ decreases in the challenging region around the two thresholds. Although it is analytically known that below the particle-dimer threshold, $-\Im \resolventamp$ should be zero, and above that, it starts increasing, the numerical reproduction of this sharp turn behavior is nontrivial. Around that point, the function can not be approximated by Taylor expansions but by rational approximations. 

It is worth pointing out that the two emulators in the last panel have similar errors in the region above those thresholds, indicating the method's robustness for dealing with that energy region. On the other hand, more elaborate training calculations are needed for better results in the lower energy region. This is the general behavior throughout this work. 

The similarities between the emulations and the AAA results, including the spectra and amplitudes, are striking. Note that the two are independent calculations; the only information shared is the training data. I.e., no fine-tuning is invoked to achieve such agreement. Therefore, this three-body case further indicates the close connection between the CEE emulation and the near-optimal rational approximation.  

\begin{figure*}
    \centering
    \includegraphics[width=\textwidth]{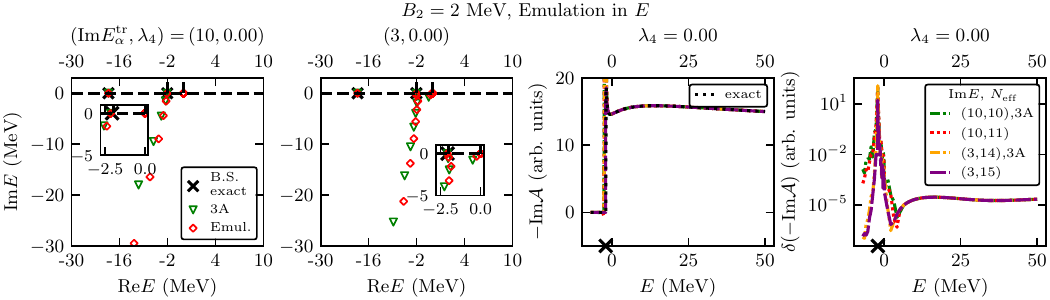}
    \caption{Emulations in $E$ using the RBM emulator and the AAA method. The $H$ is fixed by setting $B_2 = 2$ MeV and $\lambda_4 = 0$, while the sources are the same as in Fig.~\ref{fig:spectra_fixSfixpot_BE2_10_lam4_-0.2}. The presentation follows that in Fig.~\ref{fig:spectra_fixSfixpot_BE2_10_lam4_-0.2}, and thus, see the later caption for the explanation of the current plots.}
    \label{fig:spectra_fixSfixpot_BE2_2_lam4_0}
\end{figure*}

Fig.~\ref{fig:spectra_fixSfixpot_BE2_2_lam4_0} shows similar results but for a more numerically challenging case: $B_2 = 2$ MeV (i.e., a shallow dimer) and $\lambda_4 = 0$ in $H$, while the sources are the same as used before.  There are two three-body bound states; the deeper one has its binding energy $B_3$ about 20 MeV, while the other one is loosely bound, with its $B_3 = 2.06$ MeV---only $0.06$ MeV below the particle-dimer threshold. The inset shows the latter bound state more clearly. The emulators and AAA can reproduce these bound states: with $\Im E^\mathrm{tr} =10$, both emulator and AAA give $B_3 = 2.15$, while with $\Im E^\mathrm{tr} =3$, they give $B_3 = 2.08$ MeV.  In addition, the emulators and AAA also produce two correct thresholds and discretized {\BCT}s. However, for $\Im E^\mathrm{tr}=10$, the \BCT starting at the $E=0$ threshold is severely discretized. Interestingly, the $\Neff$s here have the same dependence on $\Im E^\mathrm{tr}_\alpha$ and almost the same values as in Fig.~\ref{fig:spectra_fixSfixpot_BE2_10_lam4_-0.2}. 

From the 3rd panel, one can see a sharp peak in the exact calculation caused by two competing factors: when moving toward the particle-dimer threshold from its right side, the fast increase of the amplitudes due to the nearby bound state pole and the rapid decrease of the phase space\footnote{This feature was found in other studies with shallow bound states (see, e.g., Ref.~\cite{Hammer:2011ye}).}. Although the approximations can reproduce that behavior qualitatively, the errors are significant in that region, relatively speaking, according to the last panel. This is expected based on the low density of the DC states near the thresholds. In the 4th panel, the errors do not peak at the peak location of the $-\Im \resolventamp$, but rather to its left. Again, the emulation and AAA errors increase towards the threshold but are further amplified by the shallow three-body bound state, as will be encountered in similar cases. 

\begin{figure*}
    \centering
    \includegraphics[width=\textwidth]{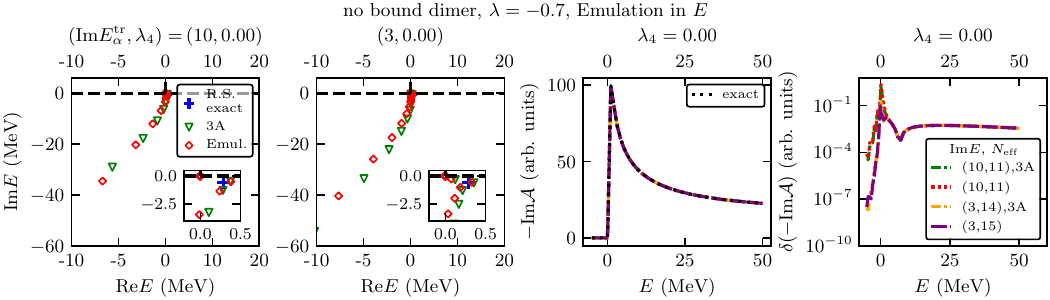}
    \caption{Emulations in $E$ using the RBM emulator and the AAA method. Here, $\lamtwo = -0.7$ (corresponding to unbound dimer) and $\lambda_4 = 0$, but the sources are the same as in Fig.~\ref{fig:spectra_fixSfixpot_BE2_10_lam4_-0.2}. Also, see the caption of Fig.~\ref{fig:spectra_fixSfixpot_BE2_10_lam4_-0.2} for explaining the plots.}
    \label{fig:spectra_fixSfixpot_lam2_-0.7_lam4_0}
\end{figure*}

Figure~\ref{fig:spectra_fixSfixpot_lam2_-0.7_lam4_0} shows another interesting case with $\lamtwo=-0.7$ and $\lambda_4 = 0$ in $H$ and again the same sources as before. 

The general features of the compressed spectra, their dependence on $\Im E^\mathrm{tr}_\alpha$, and the similarities between the emulators and the AAA results are the same as discussed in Figs.~\ref{fig:spectra_fixSfixpot_BE2_10_lam4_-0.2} and \ref{fig:spectra_fixSfixpot_BE2_2_lam4_0}. The differences in the details include that (1) the two-body interaction is too weak to support a bound dimer, and thus only a three-particle threshold ($E=0$) exists, which is correctly indicated by the single discretized \BCT; (2) no three-body bound state exists, but a three-body resonance, as the only physical state, emerges at the location marked by a blue ``$+$'' in the first two panels. Its exact eigenenergy is $E_R = 0.29-0.58\,i$ MeV. The resonance is close to the $E=0$ threshold and, at the same time, broad in the sense that the $|\Im E_R/\Re E_R|$ ratio is large; and (3) although one of the eigenstates from the emulations and AAA calculations should be identified as the resonance state, there are states in its proximity. This makes resonance identification more challenging than it is for bound states. Again, one can rely on the separation between the DC and physical states regarding their distributions in the complex plane and their sensitivity to changing $\Im E^\mathrm{tr}_\alpha$. However, in the current case, a high density of the DC states (or the discretized \BCT poles) is needed to sharpen these separations, which requires a small $\Im E^\mathrm{tr}_\alpha$ value. 

The $-\Im \resolventamp$ (or the response-like function) results on the real axis, as shown in the 3rd panel, can also be used to identify the resonance, similar to how one identifies them from the experimental data. However, this procedure faces the same difficulty as just mentioned: the resonance is near the threshold, which calls for precise control of the threshold physics and, thus, again, a high density of the DC states or the discretized \BCT poles. 

In the 3rd panel, the emulators reproduce the exact results well, including in the region close to the threshold. The errors' dependence on $\Im E^\mathrm{tr}_\alpha$ is similar to what I have seen previously. Moreover, the similarity between the CEE and AAA results is again seen here. The $\Neff$ quantities are also very close to those in Figs.~\ref{fig:spectra_fixSfixpot_BE2_10_lam4_-0.2} and \ref{fig:spectra_fixSfixpot_BE2_2_lam4_0}. 

\begin{figure*}
    \centering
    \includegraphics[width=\textwidth]{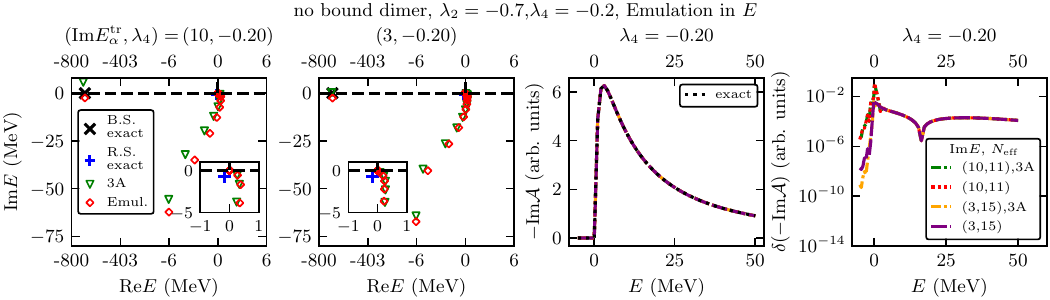}
    \caption{Emulations in $E$ using the RBM emulator and the AAA method. Here, $\lamtwo = -0.7$ (corresponding to unbound dimer) and $\lambda_4 = -0.2$, but the sources are the same as in Fig.~\ref{fig:spectra_fixSfixpot_BE2_10_lam4_-0.2}. Also, see the caption of Fig.~\ref{fig:spectra_fixSfixpot_BE2_10_lam4_-0.2} for explaining the plots.}
    \label{fig:spectra_fixSfixpot_lam2_-0.7_lam4_-0.2}
\end{figure*}

\begin{figure*}
    \centering
    \includegraphics[width=\textwidth]{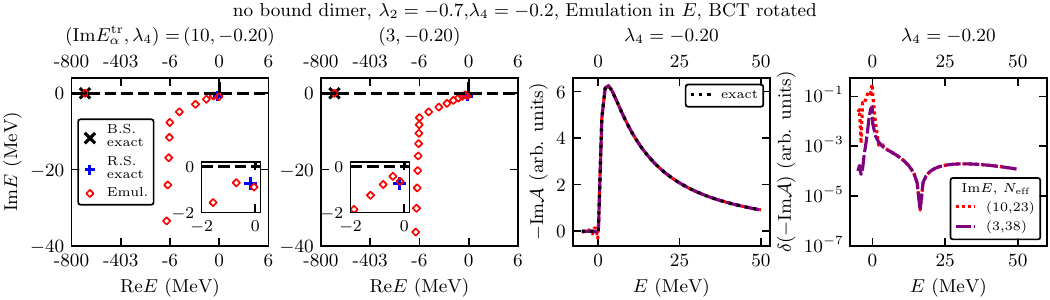}
    \caption{The results from the reanalysis of the same training data sets used in generating Fig.~\ref{fig:spectra_fixSfixpot_lam2_-0.7_lam4_-0.2}. In the new analysis, the {\BCT}s are rotated to a large angle by hand, as discussed in the text. The {\BCT}s seem bent but are on straight lines on a linear scale.  No AAA-related results are shown here, as I can't redefine the {\BCT}s inside the algorithm.} \label{fig:spectra_fixSfixpot_lam2_-0.7_lam4_-0.2_wBCrotated}
\end{figure*}

Fig.~\ref{fig:spectra_fixSfixpot_lam2_-0.7_lam4_-0.2} shows a case with $\lamtwo=-0.7$ and $\lambda_4 = -0.2$ in $H$ and the same sources as before. According to the exact calculations, the system has one deep bound state with $B_3 = 691$ MeV and a near-threshold and broad resonance ($E_R = -0.15 - 0.70$ MeV). Interestingly, the resonance is located to the \emph{left} of the $E=0$ threshold---since $\Re E_R < 0$. 

All the compressed spectra in the first two panels are pretty interesting. They reproduce the bound state and a discretized \BCT with an exponentially clustering behavior as in the previous cases and pointing to the correct threshold. However, no resonance state exists in those results. Meanwhile, for $-\Im \resolventamp$ on the real axis, both emulators and the AAA method perform as well as before, which shows the peak feature but is extremely close to the threshold. 

This suggests that the {\BCT}s from the emulators and AAA calculations are not ``rotated'' away enough to expose the resonance---the exact resonance is to the left of the discretized {\BCT}s, as can be seen in the insets. Thus, the resonance is still on a different Riemann sheet from the bound state and can't be identified in the compressed spectra. 

To check this explanation, I apply a (near-)optimal rational approximant to analytically continue the same training data.  The approximant takes the form of Eq.~\eqref{eq:pole_expansion}. The \BCT poles are exponentially distributed according to Eq.~\eqref{eq:BCT_distribution}, but different $\theta_\text\BCT$ angles are explored here. The branch point location, as needed for placing the {\BCT}s, is taken to be $E=0$. As for the bound state pole locations needed in Eq.~\eqref{eq:pole_expansion}, I use the $E_B$ result from a given emulator and plug $\Re E_B$ as the bound state location in the approximant---$\Im E_B$ is tiny. 
Eventually, the free parameters, including the pole residues and the location of the resonance $E_R$, are fitted against the training data $\resolventamp(E^\mathrm{tr}_\alpha)$.  

Fig.~\ref{fig:spectra_fixSfixpot_lam2_-0.7_lam4_-0.2_wBCrotated} shows a set of results with $\theta_\text\BCT=0.75\pi$. In the first two panels, due to the different horizontal scales on the left and right sides of the plots, the discretized {\BCT}s seem bent, but they are on straight lines. With this extra step and a large $\theta_{\text{\BCT}}$, the resonances, extracted based on the training data and partial information from the previous emulators, now agree with the exact result. This also shows that the resonance can be identified as an isolated pole and, thus, part of the eigenstates. The 3rd and the 4th panels show that the fit has similar performance, regarding $-\Im \resolventamp$ on the real energy axis, as those indicated in Fig.~\ref{fig:spectra_fixSfixpot_lam2_-0.7_lam4_-0.2}.

It is worth pointing out that although I use the $\Im E^\mathrm{tr}_\alpha$ parameter to vary the discretized {\BCT}s\footnote{This is different from the \BCT rotation in the complex scaling method, in which the angle is a controlling parameter.}, a thorough study of other $E^\mathrm{tr}_\alpha$ distributions and their impacts on the discretized {\BCT}s could reveal more effective controls and thus provide a better solution to the resonance identification task, which could substitute the extra step of forcefully rotating the {\BCT}s in Fig.~\ref{fig:spectra_fixSfixpot_lam2_-0.7_lam4_-0.2_wBCrotated}. The so-called greedy algorithm~\cite{sarkar2021selflearning,Quarteroni:218966} could help achieve this goal.  

\subsection{Emulation for parameterized $H$ and sources} \label{sec:three-body-manyHs}

This section examines emulations in $E$, the Hamiltonian parameters, and/or those in the sources. 

First, I fix the sources and study spectrum emulations. When sampling the training points, $\Im E^\mathrm{tr}_\alpha$ is fixed, while $\Re E^\mathrm{tr}_\alpha$, $\lambda_4$ and/or $\lamtwo$ are sampled using LHS in the corresponding space.  

If the systems at the emulation and training points have the same thresholds, as done in Sec.~\ref{subsubsec:three-body_emul_E_lam4_source_fixed}, the emulated spectra follow the characteristics seen in Sec.~\ref{subsec:three-body-fixedH-fixedsources}. However, when I vary $\lamtwo$ and thus the particle-dimer threshold in Sec.~\ref{subsubsec:three-body_emul_E_lam2_lam4_source_fixed}, the emulated spectra are somewhat distorted in the region away from the threshold(s) at the emulation point. In both sections, $-\Im\resolventamp$ is accurately reproduced in the emulations, even though system behavior varies dramatically in the parameter space.  

Sec.~\ref{subsubsec:three-body_emul_full} studies the emulations of particle-dimer scattering amplitudes, which require emulations in the source parameters. One source parameter is the scattering energy between the particle and dimer ($\Ein$) in the incoming channel. Moreover, the sources depend on $\lamtwo$ and $\lambda_4$. Therefore, the emulation parameter space is five-dimensional in $(\Re E, \Im E, \Ein, \lamtwo, \lambda_4) $. Again, when sampling the training calculations using LHS, $\Im E^\mathrm{tr}_\alpha$ is fixed.  

By examining the structure of the sources in Eqs.~\eqref{eq:so_3body} and~\eqref{eq:soadj_3body}, one can see that $\lambda_4$ is an affine parameter, but $\Ein$ and $\lamtwo$ are not---$\lamtwo$ is affine in $H$. Besides the affine $\lamtwo$ dependence in $V_1$ inside the sources, the dimer (as a subsystem) bound state varies with $\lamtwo$ in a nonlinear way, which, however, can be emulated using RBM-based bound state emulators~\cite{Konig:2019adq, Ekstrom:2019lss, Duguet:2023wuh, Drischler:2022ipa, Melendez:2022kid}. The $\Ein$ dependence could be emulated using data-driven methods or approximated by affine structures. These extra considerations will be investigated in the future. 

For now, I study the accuracy of the scattering amplitude emulations. By probing individual cases and collecting a large testing sample, one can see good emulation performance and its variations against the two-body binding energy.

\subsubsection{Emulation in $(\Re E, \Im E, \lambda_4)$ with fixed sources}
\label{subsubsec:three-body_emul_E_lam4_source_fixed}

Here, the sources are fixed as in Sec.~\ref{subsec:three-body-fixedH-fixedsources}. $\lamtwo$ is fixed in $H$ so that the dimer binding energy $B_2$ and thus the particle-dimer threshold are fixed. $\Im E^\mathrm{tr}_\alpha$ is fixed to either $3$ or $10$ MeV, when sampling the training points in $(\Re E, \lambda_4)$ space. In all the results presented here, I sampled $N_b = 60$ training points, but as will be seen in the results, $\Neff$s are smaller.  

\begin{figure*}
    \centering
    \includegraphics[width=0.9\textwidth]{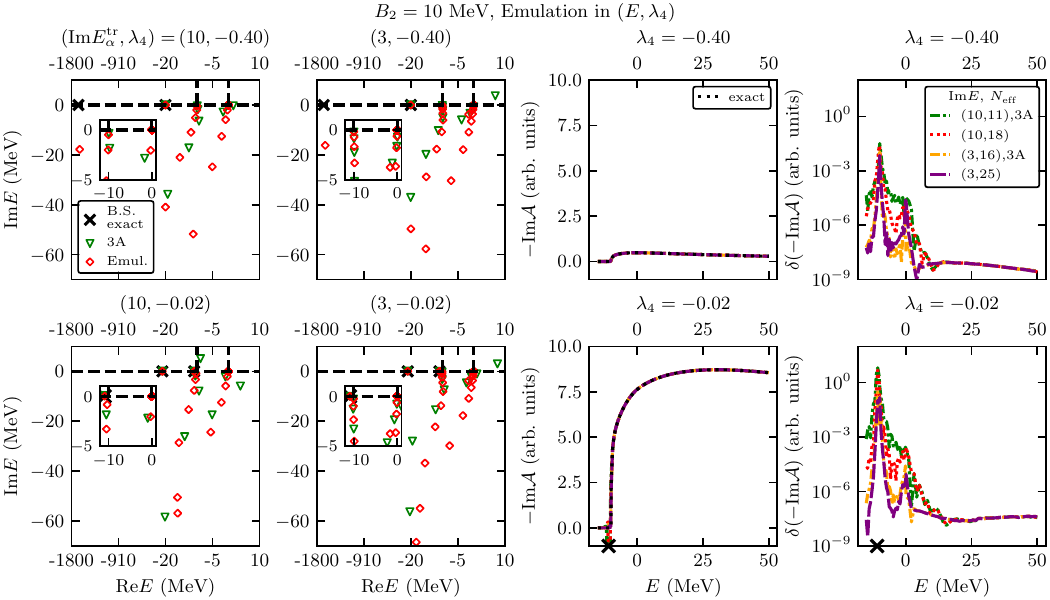}
    \caption{Emulations in the $(\Re E, \Im E, \lambda_4)$ space. The sources are the same as in Fig.~\ref{fig:spectra_fixSfixpot_BE2_10_lam4_-0.2}. I choose $B_2 =10$, which fixes $\lamtwo$ accordingly. Each row shows the same information  as in Fig.~\ref{fig:spectra_fixSfixpot_BE2_10_lam4_-0.2}, but the rows differ by the $\lambda_4$ value at the emulation point.}
    \label{fig:spectra_fixSvaryV4_BE2_10}
\end{figure*}

Figure~\ref{fig:spectra_fixSvaryV4_BE2_10} follows the presentation, for example, in Fig.~\ref{fig:spectra_fixSfixpot_BE2_10_lam4_-0.2}, except that the results at different $\lamfour$ values are stacked together vertically (more results in the SM~\cite{Zhang:2024gac_SM}). Here, $\lamtwo$ is tuned so that $B_2 = 10$ MeV. The $\Im E^\mathrm{tr}_\alpha$ and $\lamfour$ values are shown in the title of each panel. 

The compressed spectra from emulations have similar essential characteristics as those in Sec.~\ref{subsec:three-body-fixedH-fixedsources}, including the physical states, the discretized {\BCT}s, and the thresholds. However, a significant difference emerges: the value of $\Neff$ is significantly bigger than the value in Sec.~\ref{subsec:three-body-fixedH-fixedsources}, and in terms of spectra, the two {\BCT}s merge at $\Im E \sim -50$ MeV, a location much deeper than the merging point ($\Im E \sim -10$ MeV) in Fig.~\ref{fig:spectra_fixSfixpot_BE2_10_lam4_-0.2}. This is also expected since the solution subspace's dimension increases with the parameter space's dimension. For the same reason, the differences observed here are also seen in the later plots. 

The results extracted using the AAA algorithm are also plotted. Since the algorithm only works for univariate functions, it is not feasible to fit the rational form in Eq.~\eqref{eq:AAAdef} to the training data in a multiple-dimensional parameter space. Instead, I fit it to the \emph{emulation} results at the fixed $\lamfour$ but different $E^\mathrm{tr}_\alpha$ values. Essentially, AAA is used to analytically continue these emulation results in the complex $E$ plane. Thus, spectra and $\Neff$ from this procedure (still labeled as ``3A'') are more similar to those in Sec.~\ref{subsec:three-body-fixedH-fixedsources} than the CERPE results are. This difference also suggests that the analytical continuation in CEREP is not the same as the (near-)optimal rational approximation.  

In addition, emulators and the AAA procedure reproduce the exact $-\Im \resolventamp$ very well on the real axis, as shown in the 3rd and 4th columns. Again, reducing $\Im E^\mathrm{tr}_\alpha$ is needed to better predict the near-threshold behavior. Note the errors in the 2nd and 3rd rows are enhanced due to the near-threshold three-body bound states, as seen in Fig.~\ref{fig:spectra_fixSfixpot_BE2_2_lam4_0}.

In the plot, $\lamfour$ (including those in the SM~\cite{Zhang:2024gac_SM}) is varied to trace the movements of the two three-body bound states, which are accurately reproduced by the emulators. Therefore, although my primary interest is in continuum physics, these emulators can also be used to compute the discrete bound states. For that, Sec.~\ref{subsec:two-body-givenHandSources} suggest a better training-point setup.

\begin{figure*}
    \centering
    \includegraphics[width=\textwidth]{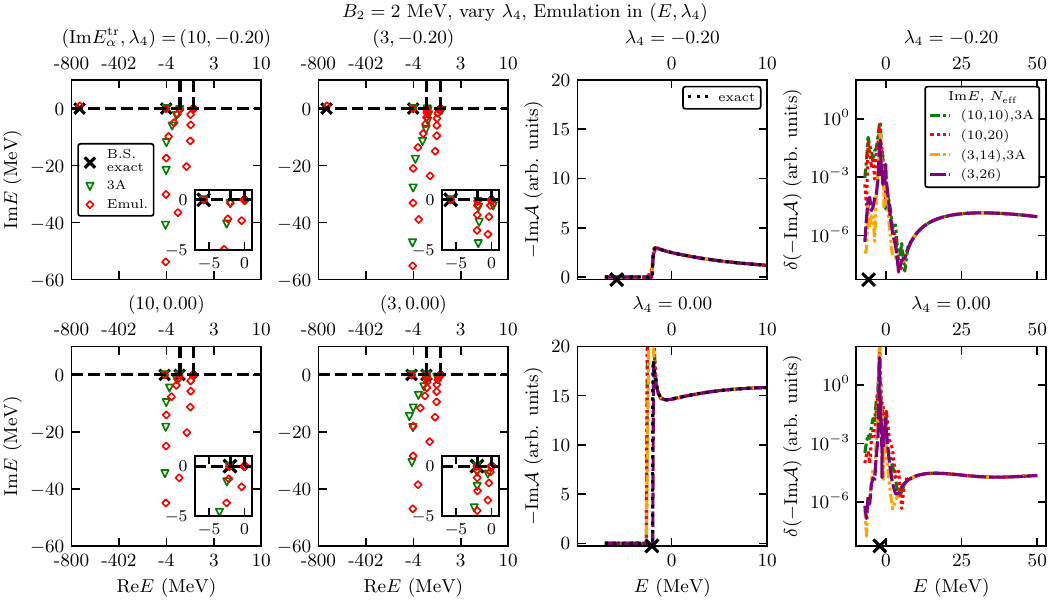}
    \caption{Similar to Fig.~\ref{fig:spectra_fixSvaryV4_BE2_10}. The sources are the same as in Fig.~\ref{fig:spectra_fixSvaryV4_BE2_10}, but $B_2 = 2$ MeV in $H$. }
    \label{fig:spectra_fixSvaryV4_BE2_2}
\end{figure*}

Fig.~\ref{fig:spectra_fixSvaryV4_BE2_2} shows the same results for a smaller $B_2$ value ($B_2 = 2$ MeV) with its extended version in the SM~\cite{Zhang:2024gac_SM}. The emulator spectra are similar to those in the $B_2 = 10$ MeV case in  Fig.~\ref{fig:spectra_fixSvaryV4_BE2_10} but again different from those in Fig.~\ref{fig:spectra_fixSfixpot_BE2_2_lam4_0}. The AAA results, however, only show one discretized \BCT, corresponding to the particle-dimer threshold. Remarkably, the emulators and the AAA procedure can reproduce the spectra and $\resolventamp$ on the real energy axis, although they change dramatically when varying $\lamfour$. 

\begin{figure}
    \centering
    \includegraphics[width=0.48\textwidth]{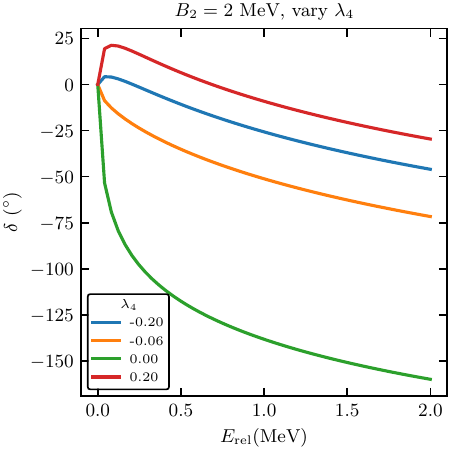}
    \caption{Full calculations of the particle-dimer scattering phase shifts below the breakup thresholds, with the parameters corresponding to those shown in Fig.~\ref{fig:spectra_fixSvaryV4_BE2_2}.} \label{fig:elastic_phase_shift_varyV4_BE2_2_lam4list}
\end{figure}

Note that a bound state could be extremely close to the particle-dimer threshold in some plotted cases. There are no resonances here, but virtual states could exist, for example, in the first row of Fig.~\ref{fig:spectra_fixSvaryV4_BE2_2}, based on the near-threshold behavior of $-\Im\resolventamp$ on the real axis. The particle-dimer scattering phase shifts in  Fig.~\ref{fig:elastic_phase_shift_varyV4_BE2_2_lam4list} also points to the presence of virtual states for $\lamfour = -0.2$ and $0.2$ (see the SM~\cite{Zhang:2024gac_SM} for the $\lamfour = 0.2$ case). It is worth mentioning that the sources in the phase shift scattering vary with $\Ein$, while they are fixed in the emulations performed here. 
However, such a difference is irrelevant when identifying virtual states. Also note that with the {\BCT}s as given, the virtual states are still on the 2nd Riemann sheet and thus can not be identified as an eigenstate; however, the phase shifts can be emulated, as shown in Sec.~\ref{subsubsec:three-body_emul_full}.

\begin{figure*}
    \centering
    \includegraphics[width=\textwidth]{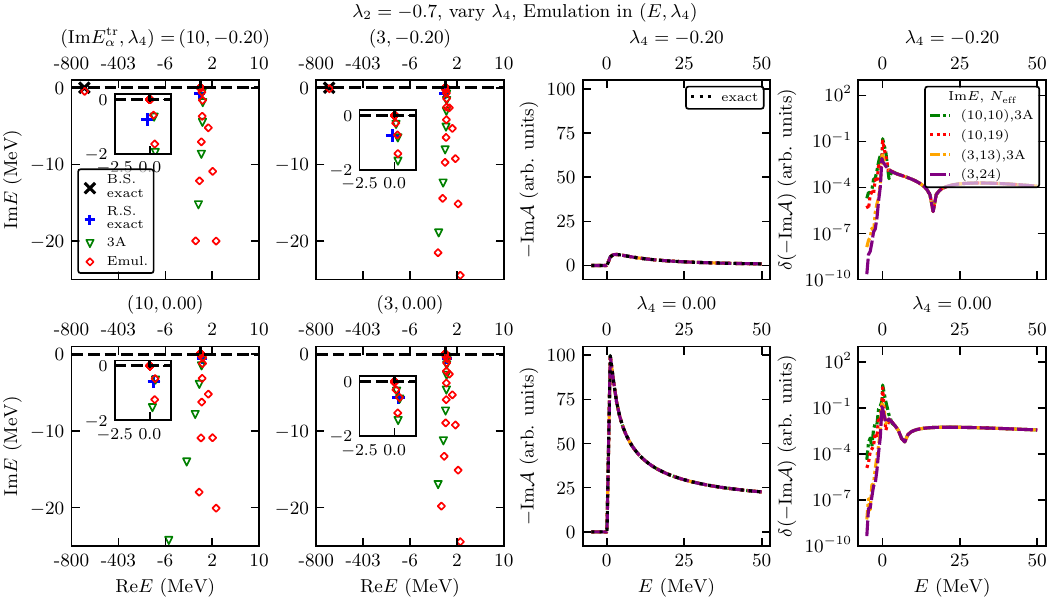}
    \caption{Similar to Fig.~\ref{fig:spectra_fixSvaryV4_BE2_10}, but now with unbound dimer ($\lamtwo = -0.7$). The sources are the same as in Fig.~\ref{fig:spectra_fixSvaryV4_BE2_10}.}
    \label{fig:spectra_fixSvaryV4_lam2_-0.7}
\end{figure*}

I further reexamine the case without a bound dimer in the current situation. Fig.~\ref{fig:spectra_fixSvaryV4_lam2_-0.7} (and its extension in the SM~\cite{Zhang:2024gac_SM}) shows the examples with $\lamtwo = -0.7$ and varying $\lamfour$.  For each parameter set, there is a near-threshold broad three-body resonance. Both emulators and AAA with $\Im E^\mathrm{tr} =3$ MeV can reproduce the resonance state, but separating the state from the DC states turns tricky. As discussed around Fig.~\ref{fig:spectra_fixSfixpot_lam2_-0.7_lam4_0}, smaller $\Im E^\mathrm{tr}_\alpha$ is needed to increase the density of the DC states to sharpen the separation. 

As related, the peak feature of $-\Im\resolventamp$ on the real axis also sits extremely close to the three-particle threshold. As the result, it is difficult to extract the resonance information based on $-\Im\resolventamp$ unless the threshold behavior can be separated from the resonance. This again asks for the high DC state density. Do note that both emulators and the AAA procedure reproduce $-\Im\resolventamp$ on the real axis.   

The discretized \BCT from the AAA procedure is similar to that in Fig.~\ref{fig:spectra_fixSfixpot_lam2_-0.7_lam4_0}, but the single \BCT pattern from the emulators is distorted when $\Im E < -5$ MeV. Interestingly, it looks like a single \BCT started bifurcating into two around $\Im E = -5$ MeV. Such distortion is not problematic for identifying narrow resonances. 

\subsubsection{Emulation in $(\Re E, \Im E, \lamtwo, \lambda_4)$ with fixed sources} \label{subsubsec:three-body_emul_E_lam2_lam4_source_fixed}

I expand the emulations in Sec.~\ref{subsubsec:three-body_emul_E_lam4_source_fixed} by varying both $\lamtwo$ and $\lamfour$ in $H$ but still using the same sources as in Sec.~\ref{subsec:three-body-fixedH-fixedsources}. 
$\Im E^\mathrm{tr}_\alpha$ is fixed to either $3$ or $10$ MeV, when I sample the training points in the $(\Re E, \lamtwo, \lamfour)$ space.  For all the results presented here, a $N_b = 60$ number of training points are sampled. But, as before, $\Neff$ is smaller. The results based on the AAA algorithm are not shown, as the results are qualitatively similar to the AAA results in Sec.~\ref{subsubsec:three-body_emul_E_lam4_source_fixed}. 

\begin{figure*}
    \centering
    \includegraphics[width=\textwidth]{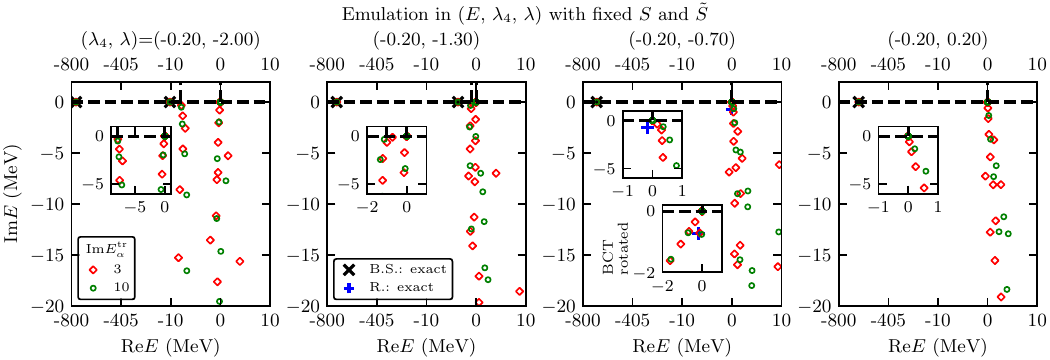}
    \caption{Emulation in the $(\Re E, \Im E, \lamtwo, \lamfour)$ space. The emulated spectra are plotted for different $\lamtwo$ and $\lamfour$ values. The spectrum plot in each panel follows the previous spectrum presentation, e.g., in Fig.~\ref{fig:spectra_fixSfixpot_BE2_10_lam4_-0.2}. $\lamtwo$ increases from left to right, while from top to bottom, $\lamfour$ increases.}
    \label{fig:spectra_fixSvaryallpot}
\end{figure*}

\begin{figure*}
    \centering
    \includegraphics[width=0.9\textwidth]{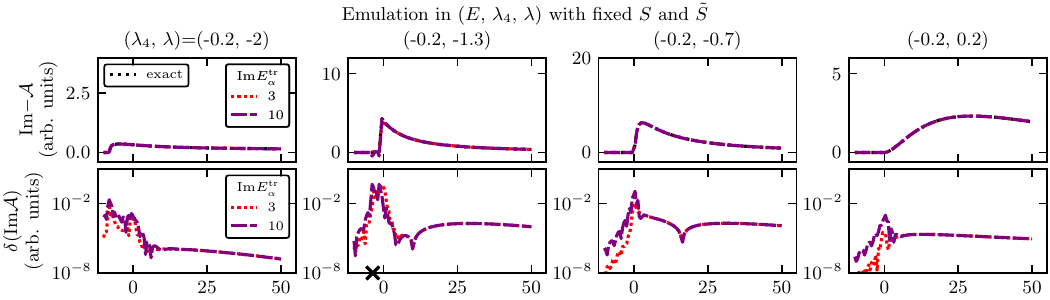}
    \caption{Emulation in the $(\Re E, \Im E, \lamtwo, \lamfour)$ space. The emulated $-\Im\resolventamp$ on the real energy axis is compared to the exact results for every $\lamfour$ and $\lamtwo$ combination whose spectrum is shown in previous Fig.~\ref{fig:spectra_fixSvaryallpot}. The emulator errors are also shown explicitly.}
    \label{fig:response_fixSvaryallpot}
\end{figure*}

Figure~\ref{fig:spectra_fixSvaryallpot} and its extension in the SM~\cite{Zhang:2024gac_SM} plot emulated spectra at different $(\lamfour,\lamtwo)$ values (provided in each panel title). The spectra from $\Im E^\mathrm{tr}_\alpha = 3$ and $10$ MeV emulators are contrasted in each panel. From left to right, $\lamtwo$ increases from $-2$ to $0.2$. Accordingly, the dimer becomes less bound in the first two columns and then becomes unbound in the last two; the particle-dimer threshold moves to the right and eventually merges with the three-particle threshold.  See the SM~\cite{Zhang:2024gac_SM} for more results with increasing $\lamfour$, where the three-body bound states become less bound and eventually disappear. 

The emulators reproduce the noted variations of the spectrum properties well. However, in the 3rd column, where a near-threshold resonance exists---to the left of the threshold, it is still on the 2nd Riemann sheet according to the discretized {\BCT}s and thus does not show up in the emulator spectra (see one of the two insets). I then perform a second analysis in each panel, which is the same as the analysis done in Fig.~\ref{fig:spectra_fixSfixpot_lam2_-0.7_lam4_-0.2_wBCrotated}, to ``rotate'' the {\BCT}s even further away from the real axis ($\theta_{\mathrm{\BCT}} = 3\pi/4$). As shown in the other inset in each panel, the resonance state emerges and agrees well with the exact result. 

In parallel to Fig.~\ref{fig:spectra_fixSvaryallpot}, Fig.~\ref{fig:response_fixSvaryallpot} shows the corresponding $-\Im \resolventamp$ emulations on the real energy axis and the emulations errors.  These tests (including the extended figure in the SM~\cite{Zhang:2024gac_SM}) show good emulation capabilities at different parameters, although $-\Im \resolventamp$ varies significantly in terms of magnitude and threshold behavior. The emulation errors behave as before: when reducing $\Im E^\mathrm{tr}$, the errors around the thresholds generally decrease; however, they would persist if there is a near-threshold three-body bound state (see the 2nd column). For the latter, one could eliminate the bound state contribution in  $\resolventamp$ by projecting out the bound state in the sources, as applied in the LIT studies~\cite{Efros:2007nq}. Also, note that the two emulators perform almost the same at higher energies. 

\begin{figure*}
    \centering
    \includegraphics[width=\textwidth]{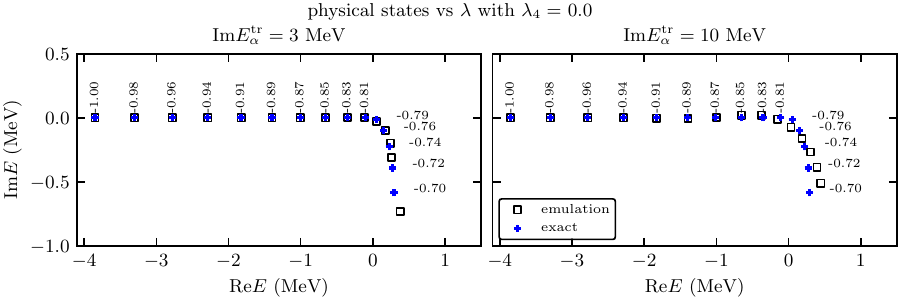}
    \caption{Emulation in the $(E, \lamtwo, \lamfour)$ space. Here, I test the spectrum emulations by fixing $\lamfour = 0$ and varying $\lamtwo$ in $[-1, -0.7]$. The corresponding trajectory of the emulated resonance in the complex $E$ plane is shown for both emulators. The exact results are also plotted for comparison. }
    \label{fig:physical_state_vs_lam2_w_lam4_0}
\end{figure*}

\begin{figure*}
    \centering
    \includegraphics[width=\textwidth]{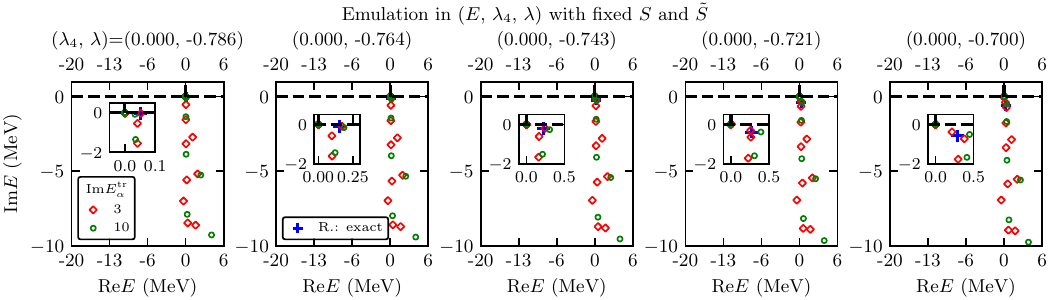}
    \caption{Emulation in the $(E, \lamtwo, \lamfour)$ space. The panels plot the emulated spectra along the trajectory shown in Fig.~\ref{fig:physical_state_vs_lam2_w_lam4_0}. Note the spectra without resonance are not shown.}
    \label{fig:spectra_fixSvaryV2_lam4_0}
\end{figure*}

A particular application of the spectrum emulation is quickly mapping out the bound and resonance states' dependence on the $H$ parameters. To demonstrate that, each panel in  Fig.~\ref{fig:physical_state_vs_lam2_w_lam4_0} plots an emulated trajectory of a physical state in the complex plane when $\lamtwo $ is varied from $-1$ to $-0.7$ with $\lamfour = 0$. 

According to the exact results shown there,  the three-body bound state becomes increasingly less bound, turns into a resonance at $\lamtwo \sim -0.8$, and moves deeper into the complex plan in the 4th quadrant. These resonances are incredibly close to the three-particle threshold. 

One can see that the bound state emulations are generally better than the resonance state emulations. The resonance emulation error increases when the state moves further away from the real axis. This is consistent with the increase of extrapolation errors when moving away from the training points in the complex plane. 
For reader's information, Fig.~\ref{fig:spectra_fixSvaryV2_lam4_0} shows the emulated spectra, from which I identify the resonance states plotted in  Fig.~\ref{fig:physical_state_vs_lam2_w_lam4_0}.

\subsubsection{Emulation in $(\Re E, \Im E, \Ein, \lamtwo, \lambda_4)$} \label{subsubsec:three-body_emul_full}

In contrast to the previous emulations in Sec.~\ref{subsubsec:three-body_emul_E_lam4_source_fixed} and~\ref{subsubsec:three-body_emul_E_lam2_lam4_source_fixed}, I need to vary the sources to emulate the particle-dimer elastic scattering amplitudes. Recall that, for each on-shell scattering calculation or emulation, the $\Ein$ parameter in the sources is related to the $E$ variable via $E = \Ein - B_2$. Note $B_2$ is a function of $\lamtwo$. However, in training calculations, I only compute at complex $E$s. Therefore, the $\Ein$ and $E$ parameters are treated separately at the emulator off-line training stage. Meanwhile, the  $\lamtwo$ and $\lamfour$ parameters in the $H$ and sources are treated as the same.  

In short, I sample the training points in the $(\Re E, \Ein, \lamtwo, \lamfour)$ space using LHS while fixing $\Im E$ to $3$ or $10$ MeV. Emulations are then performed in the entire $(\Re E, \Im E, \Ein, \lamtwo, \lamfour)$ space, including extrapolating training results at complex $E$s to the real energies.   

\begin{figure*}
    \centering
    \includegraphics[width=0.9\textwidth]{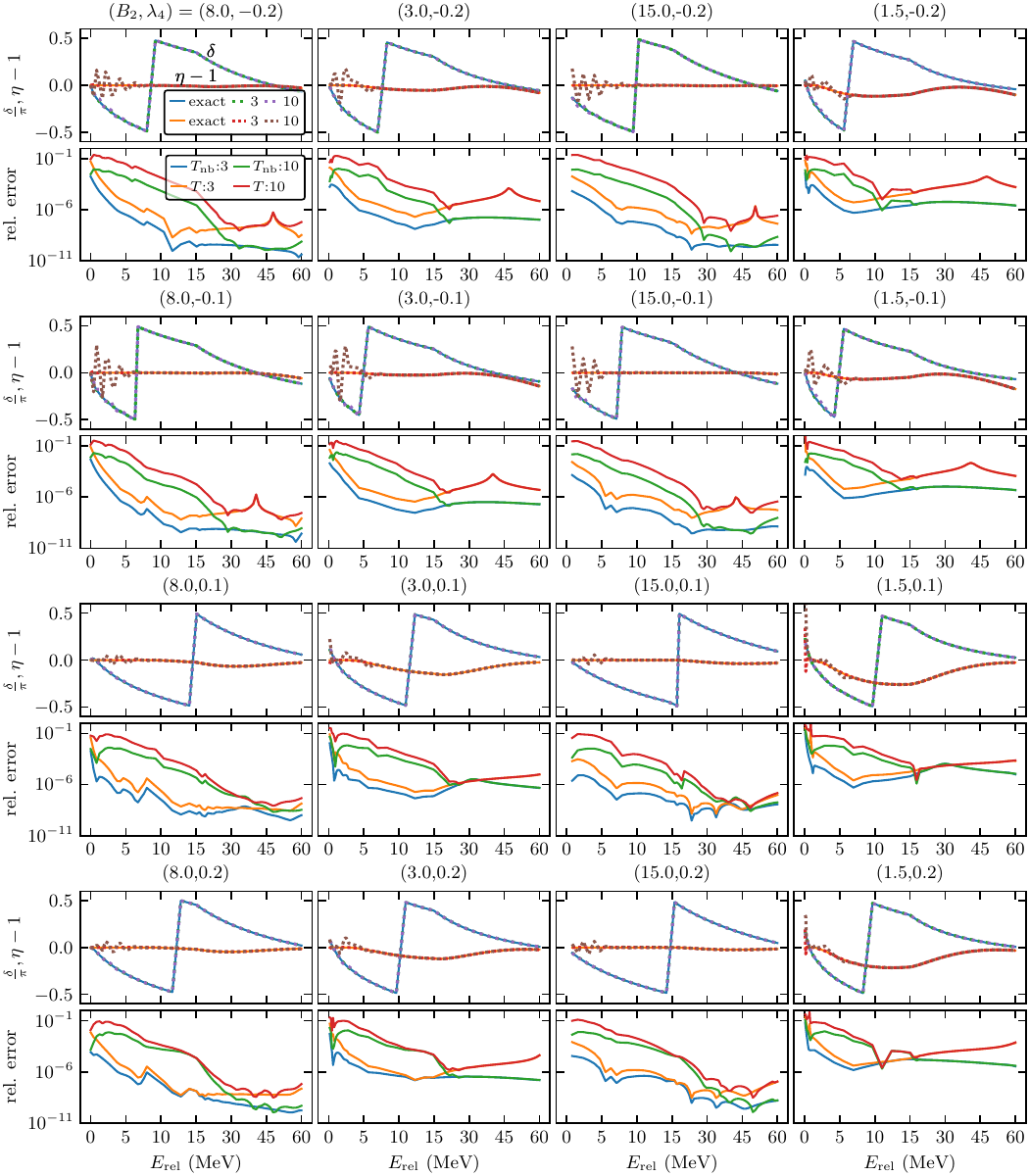}
    \caption{Emulation in the $(E, \Ein, \lamtwo, \lamfour)$ space for the partilce-dimer scattering amplitude. I show the tests of the emulations at 16 cases with different $(B_2, \lamfour)$ values. From left to right, the emulation interpolates for the $\lamtwo$ variable in the first two columns while it extrapolates in the last two columns. There are four rows, with each row having two sub-rows. From top to bottom, $\lamfour$ values increase. The top sub rows plot elastic scattering observables, including the elastic scattering phase shift, normalized to $\pi$, and the modulus of the $S$ matrix offset by $-1$. The emulation results are compared to the exact results. The lower sub-row shows the relative errors for the full $T$-matrix, and its non-Born piece $T_\mathrm{nb}$.}
    \label{fig:Emul_error_varySallpot_LAM2_200_LAM3_300_small_sample}
\end{figure*}

\begin{figure}
    \centering
    \includegraphics[width=0.48\textwidth]{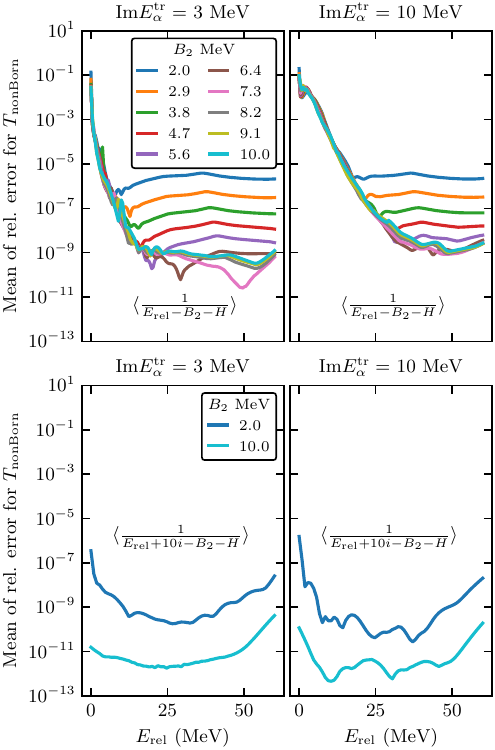}
    \caption{Emulation in the $(E, \Ein, \lamtwo, \lamfour)$ space for the particle-dimer scattering amplitude. I test the emulation performance in a grid in the ($\lamtwo, \lamfour$) space with 250 points; for each ($\lamtwo, \lamfour$), an even mesh with 128 points is picked for the $\Ein$ variable. The top two panels show the relative errors for emulating $T_{NB}$ vs $\Ein$ for different $B_2$ values but then averaged over $\lamfour$. $E = \Ein -B_2$ is set there. Similar errors are plotted in the bottom panels but with $E = \Ein -B_2 + 10\, i$.  }
    \label{fig:Emul_error_varySallpot_LAM2_200_LAM3_300}
\end{figure}

Figure~\ref{fig:Emul_error_varySallpot_LAM2_200_LAM3_300_small_sample}  shows the detailed emulation results for a sample of 16 test points with different combinations of $B_2$ and $\lamfour$. There are four rows. Within each row, the panels in the top sub-row show rescaled scattering phase shifts ($\delta/\pi$) and the shifted elasticities ($\eta - 1$) vs. particle-dimer scattering energy $\Ein$. The elastic scattering phase shift $\delta$ and the inelasticity $\eta$ are derived from the $S$ matrix via $\log(S)/2i = \delta + i \delta_I$ and $\eta \equiv e^{-2\delta_I} -1$. The emulation errors, including for the non-Born $T$-matrix and full $T$-matrix, are plotted in the bottom sub-row panels. I use two different meshes for the horizontal axis, one between $0$ and $15$ and the other above $15$, to better illustrate the details around the particle-dimer threshold and, at the same time, present the trends up to 60 MeV. 

Again, the errors are generally tiny but grow toward the near-threshold region. At small $\Ein$, the $\eta$ emulation errors can reach $10^{-2}$ to $10^{-1}$, when $\Im E^\mathrm{tr}_\alpha = 10$ MeV. With $\Im E^\mathrm{tr} = 3$ MeV, those errors are suppressed significantly, as inferred in the sub rows. As a side note, in the last column with $B_2 = 1.5$ MeV (an extrapolation case), one can see the signs of the virtual states can be seen from emulated phase shift results. 

In the sub rows, I notice enhancements of emulation errors for the $T$-matrix compared to the errors for the non-Born term due to strong cancellations between the Born and non-Born terms in their contribution to the full $T$-matrix. Across the table, the shallow dimer scattering errors are more significant than those for other dimers. 

To have a better understanding of the emulation performance, I systematically checked the two emulators by sampling a large number of test points. In the sampling, $\lamtwo$ and $\lamfour$ values are distributed in an even grid in their two-dimensional space ($250$ grid points in total), and  $\Ein$ is sampled independently from an even mesh ($128$ points).

In the top row plots of Fig.~\ref{fig:Emul_error_varySallpot_LAM2_200_LAM3_300}, I look at the emulations of the on-shell $T$-matrix, and thus $E$ is determined by $\Ein$ according to $E = \Ein - B_2$.  The panels show the relative emulation errors for the emulated scattering amplitudes vs $\Ein$ for a series of $B_2$ values. Each curve corresponds to  (with a fixed $B_2$)  the mean of a result sample with different $\lamfour$ values. One sees dependence of the errors on $B_2$ value and $\Ein$, as was noticed in Fig.~\ref{fig:Emul_error_varySallpot_LAM2_200_LAM3_300_small_sample}. The errors increase when approaching the particle-dimer threshold or reducing the $B_2$ value. Again, reducing $\Im E^\mathrm{tr}_\alpha$ decreases the emulation errors in the region close to the particle-dimer threshold for all the $B_2$ values.

In the bottom panels, the value of $E$ is set according to $E = \Ein -B_2 + 10\, i$ so that I can probe the emulation performance for interpolating complex-$E$ $\resolventamp$ in the space of ($\lamtwo$, $\lamfour$). Only two curves with the smallest and largest $B_2$ values are shown; the error curves fall between the two for the other $B_2$ values. As expected, these errors are systematically smaller than the extrapolation errors shown in the top panels. The errors increase slightly towards the particle-dimer threshold and higher energies outside the training-energy range. The implication from the bottom panel is further discussed in Sec.~\ref{sec:couplingwithothers}.

\section{Potential couplings with existing methods} \label{sec:couplingwithothers}

The CEE developed here can analytically continue $\resolventamp$ from one region of the complex $E$ plane to the other. Therefore, when a particular \schro equation solver or method is coupled with CEE---via applying the solver to obtain the training solutions required in CEE, the reach of the solver is expanded in the complex $E$ plane.

Specifically, for solvers based on Hermitian $H$ matrices (e.g., Harmonic oscillator basis methods), the new region they can access via CEE includes the real axis and the region below. In this way, the continuum physics can be computed using such solvers. As for the existing NHQM methods (e.g., the complex scaling method), they can solve the training equations with $E$ on the real axis or even below it as long as $E$ is above their rotated {\BCT}s. The CEE can help these NHQM methods reach the region to the left of these {\BCT}s, effectively rotating their {\BCT}s even further away from the real axis. 

The CERPE tool can further expand the functionality of all these solvers, by providing potential users easy access to the solvers to explore their predictions in the input-parameter spaces. The applications have been mentioned in Sec.~\ref{sec:introduction} and are thus not repeated here. 

It is worth mentioning that Ref.~\cite{Liu:2024pqp} recently developed an emulator specifically for two-body coupled-channel scattering calculations based on the complex scaling method. However, the emulation in the complex $E$ plane and the spectrum emulation were not studied there.  

Another group of continuum physics methods, the CE and LIT methods, can also be coupled with the CERPE.
Like the CEE, these methods make use of the \schro equation solutions at complex energies and extrapolate the results to real energy axis using their specific procedures. It is the complex-energy calculation component that can be emulated using the CERPE method, significantly reducing the computing cost of this component when exploring the parameter space. The excellent performance of the CERPE emulations at \emph{complex energies} has been demonstrated in Fig.~\ref{fig:Emul_error_varySallpot_LAM2_200_LAM3_300}. See the discussion about this point around the figure.

The CE method has been primarily applied in the few-body studies~\cite{Schlessinger:1966zz, Schlessinger:1968vsk, Schlessinger:1968zz, McDonald:1969zza, Uzu:2003ms, Deltuva:2012fa, Deltuva:2013qf, Deltuva:2013mda, Deltuva:2014pda}. Although they have been focused on computing scattering amplitudes by solving integral equations, the emulations developed in this work can still apply because the wave functions or Faddeev components can be computed from these amplitudes and vice versa. Or, one can use the CERPE directly to emulate the integral equations, as a type of linear equations. Through either of these two routes, one can emulate the complex energy amplitudes in $\vectheta$. These complex energy amplitudes can then be extrapolated to the real energy axis using the continuous fraction method, a rational approximation used extensively in the CE calculations. 

The LIT method~\cite{efros1985computation, Efros:1994iq, Efros:2007nq, Orlandini:2013eya, Sobczyk:2021dwm, Sobczyk:2023sxh, Bonaiti:2024fft}, which has been applied mostly for computing response functions\footnote{Note  Ref.~\cite{Efros:2007nq} pointed out ways to compute other amplitudes beyond response functions using the LIT type method, and reviewed various amplitude calculations of \emph{exclusive} electromagnetic-current-induced disintegration of nuclei.}, connect $\Im \resolventamp (E)$ at complex energy $E$ to the same function evaluated just above the real-$E$ axis (i.e., at $E_0 + i 0^+$ with $E_0$ being real): 

\begin{align}
   \frac{  \Im \resolventamp(E)}{\Im E} &  = \frac{1}{\pi}\int dE_0  \frac{ \Im \resolventamp(E_0 + i 0^+)}{|E - E_0|^2} \ , \label{eq:LIT_1} 
\end{align}
which is conventionally expressed as:

\begin{align}
    \langle \wf | \wf \rangle & = \int dE_0  \frac{R(E_0)}{|E - E_0|^2} \,. \label{eq:LIT_2}
\end{align}
To see the equivalence of these two expressions, one note that (1) $\langle \wf | \wf \rangle$ (with $| \wf \rangle$ as the solution of Eq.~\eqref{eq:inhomeeqn_1} at complex $E$) is the same as $-\Im\resolventamp(E)/\Im E$ with  $ | \so \rangle = | \soadj \rangle $ and (2) the response function $R(E_0) =  - \Im \resolventamp(E_0 + i 0^+)/\pi $.  

The response function $R(E_0)$ as a function of $E_0$ can then be extracted from $ \langle \wf | \wf \rangle$ via solving a typical inverse problem~\cite{Efros:2007nq}. Existing studies computed the $\langle \wf | \wf \rangle$ data using nuclear structure methods, such as the Hyperspherical harmonics basis and coupled cluster methods~\cite{Efros:2007nq,Orlandini:2013eya, Sobczyk:2021dwm, Sobczyk:2023sxh, Bonaiti:2024fft}. These expensive calculations can now be emulated using CERPE. By solving the inverse problem, one then obtains emulations of $R(E_0)$ in $\vectheta$. Moreover, the same strategy can be used to emulate Green's Function-based optical potential calculations~\cite{Rotureau:2016jpf, Burrows:2023ygq}.

\section{Summary}\label{sec:summary}

This work arrives at a remarkable conclusion concerning a parametrized inhomogeneous \schro equation for a finite quantum system. That is, one can efficiently emulate the continuum spectra and the associated $\resolventamp$ by adopting the RBM and treating the complex $E$ variable and other parameters in the equation in a whole parameter space. The spectra, as projected in a low-dimensional subspace and thus compressed, indicate that the method belongs to the NHQM category if the non-Hermitization is carried out by properly setting up the complex training energies.  Therefore, this study adds a new NHQM method for studying continuum physics. Moreover, it offers a new understanding of related LIT and CE methods from the perspective of the NHQM framework. It also unifies them with other NHQM methods, such as the complex scaling method.

I also see a close connection between the emulation in the complex $E$ plane (i.e., CEE) and the recent new understanding of the (near-)optimal rational approximation of univariate function with branch points, in terms of pole distributions (i.e., spectrum pattern) and the error scaling with the number of poles (i.e., the effective dimension of the emulator subspace, $\Neff$). 

On the computational advantage, the CEE method allows people to extract continuum physics based on bound-state-like training calculations. For example, one could, in principle, use the harmonic oscillator basis method~\footnote{Whether or not harmonic oscillator basais is optimal for this purpose is a separate issue.} to compute resonance without modifying the underlying $H$, which is needed in the other NHQM methods. 
The CERPE expands the CEE further by interpolating (and extrapolating) CEE in the space of other parameters in the inhomogeneous \schro equation. The potential applications of the CERPE are immense, as discussed in Secs.~\ref{sec:introduction} and~\ref{sec:couplingwithothers}. 

I also suggest a rudimentary test of a NHQM calculation, as discussed at the end of Sec.~\ref{subsec:optimarational}, by inspecting its eigenenergy distribution (spectrum) in the complex energy plane. The physical and DC states must be well separated so that the observable calculation, such as $\resolventamp$, makes physical sense. 

However, the puzzle concerning the bias of the emulators developed here, as mentioned at the end of Sec.~\ref{sec:introduction}, is still lingering. How does a small-matrix-based emulation work better than a large-Hermitian-matrix-based high-fidelity calculation for continuum physics, even though the emulators are trained by high-fidelity calculations? To answer this question, one can go back to  Fig.~\ref{fig:Tnb_in_CE_plane} and focus on the region above the positive real axis, i.e., $E_0 + i\eta$ with $E_0$ being real and $\eta$ as a positive but small number. The difference between the finite $\RIR$ results and the $\RIR\to \infty$ limit shows at $\RIR\to \infty$, $\resolventamp$ changes in a much smoother way than the variations at finite $\RIR$. That is, the finite $\RIR$ boundary condition changes $\resolventamp$ in a dramatic way~\footnote{In the time-dependent picture, the dynamics of a wave packet scattering off a potential center and then into infinity is simpler than a wave packet scattering back and forth inside a finite spatial volume.}. From the perspective of rational approximations, to describe that smooth transition, one needs to move those \BCT poles away from the real axis. The discussion of (near-)optimal rational approximation in Sec.~\ref{subsec:optimarational} shows the DC states can be efficiently discretized while the approximation still reproduces $\resolventamp$, including in the near-threshold region. This explains why my emulators are well suited for describing the physical $\resolventamp$ instead of its approximation based on a high-dimensional Hermitian $H$ matrix, when $E$ is real and above kinematic thresholds. 

Undoubtedly, additional investigations are required. The version of the RBM method, specifically the component relying on the variational approach for solving linear equations, could be further improved by employing other projection approaches or test function space. The regularization, for which I use a truncation on the SVD singular values, could become more sophisticated if  relevant prior information is incorporated~\cite{Hicks:2022ovs}. Greedy algorithms can be implemented to reduce the computing costs of trainings and provide another avenue of regularization. Based on these improvements, CEE and CERPE could tolerate significant numerical errors in the training calculations. 

One also needs to explore avenues to deal with non-affine parameters, such as the $\Ein$ variable, to further minimize the computational costs of the emulations. Moreover, it is necessary to expand the framework developed here to treat systems with the presence of significant Coulomb effects in their continuum physics, likely by following the procedures outlined in the existing LIT studies~\cite{Efros:2007nq}. 

Lastly, these emulators could  be transformed into a data-driven type by calibrating them  directly to the training data $\resolventamp(E^\mathrm{tr}_\alpha, \vectheta^\mathrm{tr}_\alpha)$ without explicitly projecting the underlying equation system. This would generalize the so-called parameterized matrix model (PMM)~\cite{Cook:2024toj}, a recently proposed machine learning platform based on Hermitian or Unitary matrices, to deal with problems with continuous spectra. Such data-driven emulators would also generalize the (near-)optimal rational approximations for a univariate function  to a method for multivariate functions.

\begin{acknowledgments}
 Discussions with Dean Lee, Dick Furnstahl, Nobuo Hinohara, Chong Qi, Simin Wang, Alex Gnech, Bijaya Acharya, and Ante Ravlic during this work are greatly appreciated. The author also thanks Kevin Fossez,  Sebastian K\"{o}nig, Simin Wang and Witold Nazarewicz for valuable comments on the manuscript. This material is based on work supported by the U.S. Department of Energy, Office of Science, Office of Nuclear Physics, under the FRIB Theory Alliance Award No. DE-SC0013617 and under the  STREAMLINE Collaboration Award No. DE-SC0024586.
\end{acknowledgments}

\appendix 

\section{On the Hermiticity of the generalized eigenvalue problem in a special type of CEE} \label{app:proof_pseduo_hermicity}

If the training calculations fulfill the requirements mentioned at the end of Sec.~\ref{subsec:spectra}, i.e., the source states $| \so_\alpha \rangle  = | \soadj_\alpha \rangle$ and  are invariant under time-reversal transformation, and both $E_\alpha^\mathrm{tr}$ and its complex conjugate $E_\alpha^{\mathrm{tr}\ast}$ are included in the training set, $\matrixform{H}$ and $\matrixform{N}$ in fact have the same generic structure and are Hermitian. For example, 
\begin{align}
    \matrixform{N} = \begin{pmatrix}
            M_{hm}, M_{cs} \\ 
            M_{cs}^\ast, M_{hm}^\ast 
    \end{pmatrix} \ , \label{eq:HNsubspace_general_structure_special_case}
\end{align}
with $M_{cs}$ as a $N_b \times N_b$ complex symmetric  matrix and $M_{hm}$ a Hermitian matrix of the same size. 

The choice of the basis  for this representation is 
\begin{align}
\bigg\{ & |\wf(E^\mathrm{tr}_1,\vectheta^\mathrm{tr}_1 )\rangle,,,|\wf(E^\mathrm{tr}_{N_b},\vectheta^\mathrm{tr}_{N_b})\rangle, \notag \\ 
 & |\wf(E^{\mathrm{tr}\ast}_1, ,\vectheta^\mathrm{tr}_1)\rangle,,,|\wf(E^{\mathrm{tr}\ast}_{N_b}, ,\vectheta^\mathrm{tr}_{N_b})\rangle\bigg\}
\end{align}
and 
\begin{align}
\bigg\{& \langle \wfadj(E^{\mathrm{tr}\ast}_1, \vectheta^\mathrm{tr}_1)|,,,\langle\wfadj(E^{\mathrm{tr}\ast}_{N_b}, \vectheta^\mathrm{tr}_{N_b})|, \notag \\ 
& \langle \wfadj(E^\mathrm{tr}_1, \vectheta^\mathrm{tr}_1)|,,,\langle \wfadj(E^\mathrm{tr}_{N_b}, \vectheta^\mathrm{tr}_{N_b})|\bigg\}
\end{align} 
 
To see the Hermiticity,  one should note that with $| \so(\vectheta) \rangle  = | \soadj(\vectheta) \rangle$, the solutions of Eqs.~\eqref{eq:training_wf} and~\eqref{eq:training_wfadj} have the following connections:
\begin{align}
 | \wfadj(E, \vectheta) \rangle & = | \wf(E^{\ast},\vectheta) \rangle  \ , \\
 | \wf(E, \vectheta) \rangle & = | \wfadj(E^{\ast}, \vectheta) \rangle  \ ,
\end{align}
as they are analytical functions of $E$ away from the real-$E$ axis.  
Therefore, the basis sets are, in fact, the same as 
\begin{align}
\bigg\{ & |\wf(E^\mathrm{tr}_1,\vectheta^\mathrm{tr}_1 )\rangle,,,|\wf(E^\mathrm{tr}_{N_b},\vectheta^\mathrm{tr}_{N_b})\rangle, \notag \\ 
 & |\wfadj(E^\mathrm{tr}_1, ,\vectheta^\mathrm{tr}_1)\rangle,,,|\wfadj(E^\mathrm{tr}_{N_b}, ,\vectheta^\mathrm{tr}_{N_b})\rangle\bigg\}
\end{align}
and 
\begin{align}
\bigg\{& \langle \wf(E^\mathrm{tr}_1, \vectheta^\mathrm{tr}_1)|,,,\langle\wf(E^\mathrm{tr}_{N_b}, \vectheta^\mathrm{tr}_{N_b})|, \notag \\ 
& \langle \wfadj(E^\mathrm{tr}_1, \vectheta^\mathrm{tr}_1)|,,,\langle \wfadj(E^\mathrm{tr}_{N_b}, \vectheta^\mathrm{tr}_{N_b})|\bigg\} \ . 
\end{align} 
Indeed, both $\matrixform{H}$ and $\matrixform{N}$ are Hermitian. 

One can also conclude that the diagonal blocks in Eq.~\eqref{eq:HNsubspace_general_structure_special_case} which are Hermitian are  the complex conjugate of each other, considering that via  Eqs.~\eqref{eq:solution_connection_T_1_special_case} and~\eqref{eq:solution_connection_T_2_special_case},  
\begin{gather}
\langle \wfadj (E_\alpha^\mathrm{tr}, \vectheta_\alpha^\mathrm{tr}) | \wfadj (E_\beta^\mathrm{tr}, \vectheta_\beta^\mathrm{tr}) \rangle  = \langle \Treversal\wf (E_\alpha^\mathrm{tr}, \vectheta_\alpha^\mathrm{tr}) | \Treversal\wf (E_\beta^\mathrm{tr}, \vectheta_\beta^\mathrm{tr}) \rangle  \notag \\ 
 =  \langle \wf (E_\alpha^\mathrm{tr}, \vectheta_\alpha^\mathrm{tr}) | \wf (E_\beta^\mathrm{tr}, \vectheta_\beta^\mathrm{tr}) \rangle^\ast \ .
\end{gather} 
Similarly by using $\Treversal$ transformation, I have 
\begin{gather}
\langle \wfadj (E_\alpha^\mathrm{tr}, \vectheta_\alpha^\mathrm{tr}) | \wf (E_\beta^\mathrm{tr}, \vectheta_\beta^\mathrm{tr}) \rangle  =  \langle \wf (E_\alpha^\mathrm{tr}, \vectheta_\alpha^\mathrm{tr}) | \wfadj (E_\beta^\mathrm{tr}, \vectheta_\beta^\mathrm{tr}) \rangle^\ast \notag \\ 
 = \langle \wfadj (E_\beta^\mathrm{tr}, \vectheta_\beta^\mathrm{tr}) | \wf (E_\alpha^\mathrm{tr}, \vectheta_\alpha^\mathrm{tr}) \rangle \ ,
\end{gather}
i.e. the off-diagonal blocks in Eq.~\eqref{eq:HNsubspace_general_structure_special_case} are the complex conjugate of each other (from the first equal sign) and each block is complex symmetric (from the second equal sign). Based on the same argument, one can see that the same block structure for $\matrixform{H}$.

\section{Two-body}\label{app:twobody}

For a separable potential, the analytical results for the $\resolventamp(E)$ in the complex $E$ plane exist.  
Suppose that in the $\ell$-partial wave channel, the potential takes the form of
\begin{align}
    V & =  \lambda | g \rangle \langle g | \ , \text{with} \ 
     \langle q, \ell | g \rangle = g_\ell(q)  \ . 
\end{align}
The $T$-matrix at complex $e_2$ can be expressed as 
\begin{align}
    t(e_2) & =  \tau(e_2)  | g \rangle \langle g | \\
    \tau^{-1} & = \lambda^{-1} - \langle g | G_{0}(e_2) | g \rangle \notag \\
    & =  \lambda^{-1} - \int dq\, q^2 \frac{|g_\ell(q) |^2}{e_2-\frac{q^2}{2 \mu_1}} \ , 
\end{align}
with $\mu_1$ as the reduced mass, and $G_0(e_2)$ the free Green's function in the two-body sector. This $G_0$ should be differentiated from the free propagator in the three-body sector. At the location of a bound state (with binding energy $B_2$), $\tau(e_2)$ should have a pole, i.e., 
\begin{align}
     \tau^{-1}(e_2 = -B_2)  = 0   \ . 
   \label{eq:two_body_sep_pot_BS_cond1}
\end{align}
The corresponding bound state is analytically known, 
\begin{align}
|\varphi_B\rangle = \bsren G_{0}(-B_2)|g\rangle \ ,   \label{eq:dimer_bs}
\end{align}
with $\bsren$ properly normalizing the bound state. 

For the $s$-wave channel, I employ a Gaussian form factor, 
\begin{align}
    g_0(q)  = \frac{C}{\Lambda_2} \exp\left[-\frac{q^2}{2 \Lambda_2^2}\right] \ \text{with} \ C^2 \equiv \frac{\LAMtwo}{\sqrt{\pi}\mu_1  }  \ ,   \label{eq:gauFF2def}  
\end{align} 
The following formula can be used to compute the $T$-matrix analytically: 
\begin{multline}
 \langle g | G_0(e_2) | g \rangle  =   \\ 
  \left(\frac{C}{\LAMtwo}\right)^2 (-\LAMtwo \mu_1)\left[\sqrt{\pi} - \pi \sqrt{D}\, e^{D}\, \erfc\left(\sqrt{D}\right)\right]   \ , 
\end{multline}
with $D = - \frac{2\mu_1 e_2^+}{\LAMtwo^2}$. 

For the $p$-wave case, I have 
\begin{align}
    g_1(q) & = \frac{C q }{\Lambda_2^2} \exp\left[-\frac{q^2}{2 \Lambda_2^2}\right] \ \text{with} \ C^2 \equiv \frac{2\LAMtwo}{\sqrt{\pi}\mu_1  }  \ ,   \label{eq:gauFF2defpwave}   \\ 
    \langle g | G_0(e_2) | g \rangle &  = \left(\frac{C}{\LAMtwo}\right)^2 (-\LAMtwo \mu_1)\Big[\sqrt{\pi} \left(\frac{1}{2}-D \right)  \notag \\ 
    & \quad + \pi  D^{\frac{3}{2}}\, e^{D}\, \erfc\left(\sqrt{D}\right)\Big] 
\end{align} 

To get the $T$-matrix in the 2nd Riemann sheet (with the positive real-$E$ axis defined as the branch cut), one must use the correct branch for the  $\sqrt{D}$ value.

\section{Three-body}\label{app:threebody}
\begin{widetext}
\subsection{Analytical formulas and equations}\label{app:threebody_formulas}

I first introduce important variables, such as $Z_{12}$ and $Z_{14}$, which will appear in the following formulas and equations. Here, $Z$s are defined as\footnote{Here, the orderings of the subscripts in $Z_{12}$ and $Z_{14}$ are not relevant. Their definitions differ from those in a previous work~\cite{Zhang:2021jmi} by factors of $\bsren^2$ and proper integer powers of $\sqrt{4\pi}$. This difference is due to the partial wave basis implemented here, while in Ref.~\cite{Zhang:2021jmi}, they were defined as averaged over the solid angle of the momentum vector.} 
\begin{gather}
  Z_{21}({P}_2, {P}_1, E)   =  \langle {P}_{2}, g | G_0(E) |{P}_{1}, g  \rangle  = (-)(2\pi)\frac{M}{P_1 P_2}\left(\frac{C}{\LAMtwo}\right)^2e^{-\frac{5\left(P_1^2 + P_2^2\right)}{8\LAMtwo^2} } \left\{e^{\tilde{D}} \left[\Gamma\left(0, \tilde{D}-\tilde{d}\right)- \Gamma\left(0, \tilde{D}+\tilde{d}\right) \right]\right\}  \ , \\ 
 \tilde{D}  = \frac{P_1^2 + P_2^2 - ME}{\LAMtwo^2}, \quad \tilde{d}=\frac{P_1 P_2}{\LAMtwo^2} \ , 
\end{gather}
and 
\begin{gather}
Z_{14}(P,E)   =   \langle P,g | G_0(E) | g_4 \rangle = - (4\pi)^{3/2} M  \frac{C}{\LAMtwo} \frac{ e^{-\frac{3 P^2}{8\Lambda_4^2}} }{2\sqrt{t} \sqrt{M}\Lambda_4^2}  \left[\sqrt{\pi} - 
    \pi e^{D\,t} \sqrt{D\,t}\erfc\left(\sqrt{D\,t}\right) \right] \\ 
    t  \equiv \frac{1}{2}\left(\frac{1}{\LAMtwo^2} +\frac{1}{\Lambda_4^2} \right)  ,  \ 
    D  \equiv \frac{3}{4} P^2 - M E  \ .
\end{gather}
Here, $G_0$ is the free Green's function in the three-body sector, defined in Sec.~\ref{sec:three-body-basics}.

With these definitions, the Born term in Eq.~\eqref{eq:Tonshell_decomp} can be expressed as 
\begin{equation}
T_\mathrm{Born}   =   2 \bsren^2 \lamtwo Z_{21}(P_{in}, P_{in}, E_{in})  
  + 3\lambda_4 \bsren^2 Z_{41}(P_{in}, E_{in}) Z_{14}(P_{in}, E_{in})  
 + \int dP P^2 \lamtwo \bsren^2 2 Z_{12}(P_{in}, P, E_{in}) 2 Z_{21}(P_{in}, P, E_{in}) \ . 
\end{equation}
Here, $\bsren$ is the dimer bound state normalization parameter in the $|\phi_1 \rangle \equiv |\Pin \rangle |\varphi_B \rangle$ (see Eq.~\eqref{eq:dimer_bs}). 

The sources $|\so_{1,4}\rangle$, expressed in momentum space, are 
\begin{align}
 \langle P, q | \so_1 \rangle  & =  \langle P, q | V_1^s \perm | \phi_1 \rangle  = \bsren_s  \lamtwo^s g(q) 2 Z_{12}(P, P_{in}, E_{in}^s) \label{eq:Pq_so1_overlap} \\ 
 \langle P, q | \so_4 \rangle  & =  \langle P, q | 3 V_4^s | \phi_1 \rangle  =  3 \bsren_s \lambda_4^s g_4(P,q)Z_{14}(P_{in}, E_{in}^s)   \label{eq:Pq_so4_overlap}   
\end{align}
Note the sub and super-scripts $s$ are intended to explicitly show how the sources depend on the couplings and $\Ein^s \equiv \Pin^2/(2\mu^1) - B_2^s$ depending on the dimer binding energy $B_2^s$. In this work, I will sometimes fix the parameters inside the source, whose values can differ from those in the $H$. 

For the sources $|\soadj_{1,4}\rangle$, 
\begin{gather}
 \langle P, q | \soadj_1 \rangle   =  \langle P, q | (1+ \perm ) V_1^s \perm + 3 V_4^s| \phi_1 \rangle   = \langle P, q | \so_1 \rangle + \langle P, q | \so_4 \rangle   + \lamtwo^s \bsren_s \int d P' P'^2 \langle P,q | \perm|P',{g}\rangle  2 Z_{12}(P', P_{in}, \Ein^s) \ ,  \label{eq:Pq_soadj1_overlap}         \\ 
\langle P, q | \soadj_4 \rangle   = \langle P, q | \so_1 \rangle + \frac{1}{3} \langle P, q | \so_4 \rangle  \ ,  \label{eq:Pq_soadj4_overlap} \\
\langle Pq | \perm|P' {g}\rangle = \frac{2 {g}(\sqrt{q^2-\frac{3}{4}(P'^2-P^2)})}{P P' q} \Theta\left( |P'-\frac{P}{2}| \leq q \leq P'+\frac{P}{2}\right) \ .
\end{gather}
Here $\Theta(a \leq x \leq b)$ is $1$ when $ x\in [a,  b]$, and $0$ otherwise. 

In the current three-body model, I take advantage of the potentials' separable nature to reduce the dimensionality of the Faddeev equations. 
I first define
\begin{align}
     t_1(E) \equiv V_1 + V_1 G_1(E) V_1 \ , \\
 t_4(E) \equiv V_4 + V_4 G_4(E) V_4 \   . 
\end{align}
To solve $|\psi_1\rangle$, based on Eq.~\eqref{eq:three_body_eq1}, I can express $|\psi_4\rangle$ in terms of $|\psi_1\rangle$:
\begin{align}
| \psi_4 \rangle & = 3 G_0(E) t_4(E) \left(\frac{\lambda_4^s}{\lambda_4} | \phi_1 \rangle + | \psi_1 \rangle\right) \ . \label{eq:three_body_psi4_1} 
\end{align}
Then, the equation for $|\psi_1\rangle$ can be derived:    
\begin{gather}
 \left(E - H_0 - V_1 - V_1  \left(\perm + 3  G_0(E) t_4(E) \right)   \right) | \psi_1 \rangle  =  \soket{1} + V_1 G_4(E) \soket{4}  \\ 
 |\psi_1 \rangle  = G_0(E) t_1(E) \left(\perm + 3  G_0(E) t_4(E) \right) | \psi_1 \rangle  + G_1(E)\soket{1} + G_1(E)V_1 G_4(E) \soket{4} \ . \label{eq:three_body_psi1_eq_1}
\end{gather}
Note, in the three-body Fock space, $V_1 = \int d P P^2 \lamtwo | P, {g} \rangle\langle P, {g} |$. Similarly, $t_1(E) = \int d P P^2 \tau(e_2) | P, {g} \rangle\langle P, {g} |$, with the full three-body total energy  $E$ as its argument, $e_2 = E - P^2/(2\mu^1)$, and $P$ as the spectator's momentum. $\tau(e_2)$ is discussed in Appendix~\ref{app:twobody}. $t_4$ and the related $\tau_4$ have been discussed in Sec.~\ref{subsec:three-body-force}. Its analytical form is  
\begin{align}
    \frac{1}{\tau_4(E)} 
    = & \frac{1}{\lambda_4} + \frac{4\pi^3}{3\sqrt{3}} \frac{M^2}{\Lambda_4^4} \left[\frac{1}{t^2} + \frac{E^+}{t} + e^{- t E } E^2 \Gamma\left(0, - t E^+\right) \right], \text{with} \ t \equiv \frac{M}{\Lambda_4^2} \ . 
\end{align}
Note $\Gamma\left(0, - t E^+\right) = E_1(-t E^+)$ \cite[Eq.~8.19.1]{NIST:DLMF} which cares about the $i 0^+$ in $E^+ \equiv E + i 0^+$. 

Considering that in Eq.~\eqref{eq:three_body_psi1_eq_1},  the three terms in the right side are proportional to $G_0 t_1$, I introduce a new state $| F_1 \rangle$,
\begin{align}
|\psi_1 \rangle & \equiv   G_0(E) t_1(E)  |F_1 \rangle \equiv G_0 t_1 | F_1 \rangle  \ .  \label{eq:sol_psi1_1}  
\end{align}
It satisfies a new equation:
\begin{align}
|F_1\rangle & = V_1^{-1}\soket{1} + G_4 \soket{4} +  \left(\perm + 3  G_0 t_4 \right) G_0 t_1 | F_1 \rangle \  . 
\end{align}
Eq.~\eqref{eq:sol_psi1_1} suggests that one only needs to know $\langle P, g | F_1 \rangle$ to compute $| \psi_1 \rangle$. I first define $\ F_1(P)  \equiv \langle P,  {g}|F_1 \rangle$, and then express $|\psi_1\rangle$ in terms of $F_1$: 
\begin{align}
\langle P, q |\psi_1 \rangle & = \frac{ {g}(q) {\tau}_1(E-P^2/(2\mu^1))}{E - P^2/(2\mu^1) - q^2/(2\mu_1)} F_{1}(P) \ . 
\end{align}
The equation for $F_1(P)$ is 
\begin{align}
F_1(P) & = \frac{\lamtwo^s}{\lamtwo}  \langle P,g| \perm | \phi_1 \rangle +  3 \frac{\lambda_4^s}{\lambda_4} \langle P, g| G_0 t_4 |\phi_1 \rangle  +   \langle P, g | \left(\perm + 3  G_0 t_4 \right) G_0 t_1 | F_1 \rangle \notag  \\ 
& = \bsren_s 2 \bar{Z}(P, \Pin, \Ein, \Ein, \frac{\lamtwo^s}{\lamtwo} , \frac{\lambda_4^s}{\lambda_4})   + \int d P' P'^2  2 \bar{Z}(P, P', E, E, 1, 1)  {\tau_1}(E-P'^2/(2\mu^1))F_1(P') \ ,  \label{eq:sol_psi1_2}  
\end{align}
with 
\begin{align}
 2 \bar{Z}(P, P', E, E', c, c') & \equiv   2 c\, Z_{12}(P, P', E') + 3 c'\, \tau_4(E) Z_{14}(P, E) Z_{14}(P', E')  \ .    \label{eq:sol_psi1_3}
\end{align}
The inhomogeneous equation~\eqref{eq:three_body_psi1_eq_1}, which has two variables $P$ and $q$, now reduces to Eq.~\eqref{eq:sol_psi1_2} with just one variable $P$. Interestingly, the equation for $F_1(P)$ is similar to the equation for scattering amplitude in Ref.~\cite{Zhang:2021jmi}, but $E$ could be complex here. 

Knowing $| \psi_1 \rangle$, $|\psi_4\rangle$ can be computed via Eq.~\eqref{eq:three_body_psi4_1}: 
\begin{align}
\langle P q   | \psi_4 \rangle & =  \frac{3 \tau_4(E) g_4(P, q)}{E - \frac{P^2}{2\mu^1} -\frac{q^2}{2\mu_1} } \left( \bsren_s \frac{\lambda_4^s}{\lambda_4} Z_{14}(P_{in}, E_{in})  + \langle g_4 | \psi_1 \rangle \right) \label{eq:sol_psi4_1}
\end{align}

The adjoint equations in Eq.~\eqref{eq:three_body_adjeq1} can also be simplified. $ | \psiadj{4} \rangle $ depends on $ | \psiadj{1} \rangle$ via  
\begin{align}
| \psiadj{4} \rangle & = G_4^\ast \left\{|\soadj_4\rangle + V_1 | \psiadj{1} \rangle \right\}  \ , \label{eq:three_body_adjpsi4_1} \\ 
G_4^\ast & \equiv G_4(E^\ast) = [G_4(E)]^\ast  \ . 
\end{align}
Plugging this expression back into Eq.~\eqref{eq:three_body_adjeq1}, I get 
\begin{align}
 &  \left(E^\ast - H_0 - V_1 -  \left(\perm + 3  t_4^\ast G_0^\ast \right)   V_1  \right) | \psiadj{1} \rangle 
 =   | \soadj_1 \rangle + 3 V_4 G_4^\ast |\soadj_4 \rangle   \ \text{i.e.,} \\ 
 & | \psiadj{1} \rangle  = G_1^\ast \left(\perm + 3  t_4^\ast G_0^\ast \right) V_1  | \psiadj{1} \rangle  +  G_1^\ast \left( | \soadj_1 \rangle + 3 t_4^\ast G_0^\ast| \soadj_4 \rangle  \right)     \label{eq:three_body_adjpsi1_2} \  .
\end{align}
Eq.~\eqref{eq:three_body_adjpsi1_2} shows that if $\langle P, {g}| \psiadj{1} \rangle \equiv \tilde{F}_1(P)$ is known, I can plug it on the right side and get $\langle P, q | \psiadj{1}\rangle$. Therefore, one can first look at the equation for $\tilde{F}_1(P)$:
\begin{align} 
\tilde{F}_1(P)  =   \langle P, g|  G_1^\ast \left( | \soadj_1 \rangle + 3 t_4^\ast G_0^\ast| \soadj_4 \rangle  \right) + \int dP' P'^2  \langle P, {g}|  G_1^\ast \left(\perm + 3  t_4^\ast G_0^\ast \right) | P', g\rangle \lamtwo   \tilde{F}_1(P') \ .   
\end{align}
Knowing $\tilde{F}_1(P)$, I can compute 
\begin{align}
\langle P, q| \psiadj{1} \rangle & = \langle P, q| G_1^\ast \left( | \soadj_1 \rangle + 3 t_4^\ast G_0^\ast| \soadj_4 \rangle  \right)   + \int d P' P'^2  {}_1\langle P, q| G_1^\ast \left(\perm + 3  t_4^\ast G_0^\ast \right) | P', g\rangle \lamtwo \tilde{F}_1(P') \ .
\end{align}
The $|\psiadj{4}\rangle$ can then be computed as 
\begin{multline}
 \langle P, q| \psiadj{4} \rangle =  \langle P, q| G_4^\ast \left[ |\soadj_4 \rangle + V_1 | \psiadj{1} \rangle \right] \\
 =  \frac{ 1 }{E^\ast - \frac{P^2}{2\mu^1} - \frac{q^2}{2\mu_1} } \bigg\{\langle P,q |\soadj_4 \rangle + g_4(P,q)\tau_4^\ast \langle g_4 | G_0^\ast |\soadj_4\rangle + \lamtwo g(q) \tilde{F}_1(P) + \lamtwo g_4(P,q)\tau_4^\ast \int dP P^2 Z_{14}(P, E^\ast) 
\tilde{F}_1(P)\bigg\} \ . 
\end{multline}
Recall that the sources in the momentum space representations can be found in Eqs.~\eqref{eq:Pq_so1_overlap}--\eqref{eq:Pq_soadj4_overlap}.

In these integral equations, I need to have the explicit expression for the kernels, including $\langle P, {g}| G_1^\ast \left(\perm + 3  t_4^\ast G_0^\ast \right) |P', g\rangle$ and $\langle P, q| G_1^\ast \left(\perm + 3  t_4^\ast G_0^\ast \right) | P', g\rangle$. 
By expanding $ G_1 = G_0 + G_0 t_1 G_0 $, I get 
\begin{align}
& \langle P, {g}| G_1^\ast \left(\perm + 3  t_4^\ast G_0^\ast \right) |P', g\rangle  = \left(1  + \langle {g}|G_{0,2b}(E^\ast - \frac{P^2}{2\mu^1})|{g}\rangle {\tau}_1(E^\ast - \frac{P^2}{2\mu^1})\right) 2 \bar{Z}(P, P', E^\ast, E^\ast, 1, 1) \notag \\    
& = \frac{\tau_1(E^\ast - \frac{P^2}{2\mu^1})}{\lamtwo} 2 \bar{Z}(P, P', E^\ast, E^\ast, 1, 1) =  \frac{\tau_1(E^\ast - \frac{P^2}{2\mu^1})}{\lamtwo} 
     \langle P, {g}| G_0^\ast \left(\perm + 3 t_4^\ast G_0^\ast \right) |P', {g}\rangle   \ , 
\end{align}
and 
\begin{align}
& \langle P, q| G_1^\ast\left(\perm + 3 t_4^\ast G_0^\ast \right) |P', g\rangle \notag \\ 
& = \frac{1}{E^\ast - \frac{P^2}{2\mu^1} -\frac{q^2}{2\mu_1} } \bigg( \langle Pq | \perm + 3 t_4^\ast G_0^\ast |P' g\rangle  + g(q) {\tau}_1(E^\ast - \frac{P^2}{2\mu^1}) \langle P, g|  G_0^\ast \left(\perm + 3 t_4^\ast G_0^\ast \right) |P', g\rangle     \bigg) \\
& = \frac{1}{E^\ast - \frac{P^2}{2\mu^1} -\frac{q^2}{2\mu_1} } \bigg( \langle Pq | \perm|P' {g}\rangle  + 3 \tau_4^\ast g_4(P,q)  Z_{14}(P', E^\ast) + {g}(q) {\tau}_1(E^\ast - \frac{P^2}{2\mu^1}) 2 \bar{Z}(P, P', E^\ast, E^\ast, 1, 1)  \bigg)
\end{align}
In the derivations, I make use of the fact that 
\begin{align}
1  + \langle {g}|G_{2b,0}(e_2)|{g}\rangle {\tau}_1(e_2) = 1  + \langle {g}|G_{2b,0}(e_2)|{g}\rangle  {\tau}_1(e_2) = 1+ (\lamtwo^{-1} -\tau^{-1}_1(e_2)) \tau_1(e_2) = \frac{\tau_1(e_2)}{\lamtwo}  . 
\end{align}

I also need to know the driving terms explicitly in the integral equations. First, 
\begin{align}
& \langle P, {g}|  G_1^\ast  |\soadj_1 \rangle  = \frac{\tau_1(E^\ast - \frac{P^2}{2\mu^1})}{\lamtwo} \langle P, {g}|  G_0^\ast  |\soadj_1 \rangle \, \label{eq:pg_G1star_soadj_1} \\
& = \frac{\tau_1(E^\ast - \frac{P^2}{2\mu^1})}{\lamtwo} \int dq q^2 \frac{g(q)}{E^\ast - P^2/(2\mu^1) - q^2/(2\mu_1)}  \langle P,q |\soadj_1\rangle \ , 
\end{align}
and
\begin{align}
& \langle P, {g}|  G_1^\ast  3 t_4^\ast G_0^\ast  |\soadj_4 \rangle  = 3 \tau_4^\ast \frac{\tau_1(E^\ast - \frac{P^2}{2\mu^1})}{\lamtwo} \langle P, {g}|  G_0^\ast |g_4 \rangle \langle g_4 | G_0^\ast   |\soadj_4 \rangle \label{eq:pg_G1start4starG0star_soadj_4}\\
& =  3 \tau_4^\ast  \frac{\tau_1(E^\ast - \frac{P^2}{2\mu^1})}{\lamtwo} \left[\int dq q^2 \frac{g(q) g_4(P,q)}{E^\ast - P^2/(2\mu^1) - q^2/(2\mu_1)} \right]
\left[\int dP dq P^2 q^2 \frac{\langle P,q |\soadj_4\rangle g_4(P,q)}{E^\ast - P^2/(2\mu^1) - q^2/(2\mu_1)}\right] 
\end{align}

Moreover, the overlaps between sources and $|P, q \rangle$ can be computed in similar ways:
\begin{align}
& \langle P, q|  G_1^\ast  |\soadj_1 \rangle  =  \langle P, q|  G_0^\ast + G_0^\ast t_1^\ast G_0^\ast|\soadj_1 \rangle\\
& =  \frac{1}{E^\ast - P^2/(2\mu^1) - q^2/(2\mu_1)}  \left[\langle P,q |\soadj_1\rangle + \tau_1(E^\ast - P^2/(2\mu^1)) g(q) \langle P,g |G_0^\ast | \soadj_1 \rangle  \right] \ , 
\end{align}
and
\begin{align}
& \langle P, q|  G_1^\ast  3 t_4^\ast G_0^\ast  |\soadj_4 \rangle  = 3 \tau_4^\ast \langle P, q|  G_1^\ast |g_4 \rangle \langle g_4 | G_0^\ast   |\soadj_4 \rangle\\
& =  \frac{3 \tau_4^\ast }{E^\ast - P^2/(2\mu^1) - q^2/(2\mu_1)} \left[g_4(P,q) + \tau_1(E^\ast - \frac{P^2}{2\mu^1}) g(q) \langle P,g |G_0|g_4\rangle  \right]  \langle g_4 | G_0^\ast   |\soadj_4 \rangle
\end{align}
Note the $\langle P,g| G_0^\ast |g_4\rangle$ and $\langle g_4 | G_0^\ast   |\soadj_4 \rangle$ have been computed in Eq.~\eqref{eq:pg_G1start4starG0star_soadj_4}. 

It is worth noting that when $\Im E \neq 0$,  the functions in the integral equations and those in computing integrals are smooth functions without any singular behaviors. However, the singularities appear when taking $\Im E \to 0^+$, as in the high-fidelity benchmark calculations. Specific treatments are needed to deal with these singular functions, which are discussed in Appendix~\ref{app:three-body-highfidelty}. 

As a partial consistency check of the formulas derived so far, I note first that $\xi F_1(\Pin)$, as the solution of Eq.~\eqref{eq:sol_psi1_2} with $E = E_0 $ and $E_0 \equiv \Ein -B_2 + i 0^+$, is the \emph{full} on-shell scattering amplitude as defined according to Eq.~\eqref{eq:Tonshell_decomp}. To prove this, one goes back to  Eq.~\eqref{eq:Tonshell_decomp} and makes the following derivations:
\begin{align}
    T & = \langle \phi_1(E_0) | V_2 + V_3 + V_4 | \wf^+(E_0) \rangle  \\ 
      & =  \big[\langle \phi_1(E_0) | V_1 G_0(E_0)\big] \big[ (V_2 + V_3 + V_4) | \wf^+(E_0) \rangle \big] \label{eq:T_as_F1_tricky3}  \\
    & = \langle \phi_1(E_0) | V_1  | \perm (\psi_1(E_0) + \phi_1(E_0)) + \psi_4(E_0) \rangle  \label{eq:T_as_F1_tricky2} \\ 
    & = \langle \phi_1(E_0) | \big[(E_0 - H_1)  |  \psi_1(E_0) \rangle\big] \label{eq:T_as_F1_tricky} \\ 
    & = \langle \phi_1(E_0) | \big[(E_0 - H_1)  G_1(E_0) V_1 | F_1(E_0) \rangle\big] 
    =  \xi \langle \Pin, g | G_0(E_0) V_1 | F_1(E_0) \rangle \\ 
    & = \xi \lamtwo \langle g|G_{2b,0}(-B_2)|g\rangle F_1(\Pin) = \xi F_1(\Pin)
\end{align}
The first step comes from that $ (H_0 + V_1) |\phi_1(E_0)\rangle = E_0 |\phi_1(E_0)\rangle $. From Eq.~\eqref{eq:T_as_F1_tricky3} to~\eqref{eq:T_as_F1_tricky2}, I utilize the definition of the Faddeev components in Eq.~\eqref{eq:FC_def}. The step from Eq.~\eqref{eq:T_as_F1_tricky2} to~\eqref{eq:T_as_F1_tricky} is based on Eq.~\eqref{eq:Faddeev1}. Moreover, in Eq.~\eqref{eq:T_as_F1_tricky}, one would conclude that the matrix element becomes zero if  $H_1$ is applied to the left. However, $H_1$ can not be applied to $\langle \phi_1 |$ via integration by parts because both $\langle \phi_1 |$ and $| \psi_1 \rangle$ have oscillating asymptotics in the coordinate space when particle-1 is far away from dimer-23. Finally, in the last step, I use the fact that $\langle g|G_{2b,0}(-B_2)|g\rangle = 1/\lamtwo$ because of the bound-state pole in the $\tau$ function. 

Meanwhile, the other route to verify $\xi F_1(\Pin)$ as the on-shell scattering amplitude, when $E = E_0$, is to inspect its Eq.~\eqref{eq:sol_psi1_2} directly. It is easy to see the equation is the same as the one for on-shell scattering amplitude computed in Ref.~\cite{Zhang:2021jmi}. Therefore, one sees the consistency of the two routes and, thus, the consistency of the formula derivation in this section. 

\end{widetext}

\subsection{Benchmark calculations of $\resolventamp$ at real energies} \label{app:three-body-highfidelty}

In Eq.~\eqref{eq:sol_psi1_2}, the inhomogeneous term is always smooth in momentum space and spatially localized, but the kernel in the integration term is singular when $\Im(E)=0^+$, which requires dedicated treatments. Such treatment should apply to the equation with smooth sources, such as in computing on-shell $T$-matrix and the response-function-like calculations I set up by fixing the $\Ein$ parameter in the inhomogeneous term. 

To deal with singular kernel in Eq.~\eqref{eq:sol_psi1_2}, I follow the procedure of Ref.~\cite{Matsuyama:2006rp, Larson:1974zza} by carefully choosing the momentum mesh points for both $P$ and $P'$ variable and by proper variable transformations around the singularities of the $\bar{Z}$ function. The other factor in the kernel, $\tau_1(E-P'^2/(2\mu^1))$, has a simple pole located at $E-P'^2/(2\mu^1) \to -B_2$. This pole can be dealt with by a simple pole subtraction method commonly used in solving two-body Lippmann-Schwinger equation (see, e.g., Sec.~18.3 in Ref.~\cite{Landau:1996}). 

Specifically, in the integral equation, treating $f(P';P,E) \equiv \bar{Z}(P, P', E, E, 1, 1) $ as a function of $P'$ variable with a fixed $P$ and $E$, requires careful analysis. For $E<0$, the $f(P';P,E)$ is smooth. However, when $E>0$, $f(P';P,E)$ becomes singular near two points (collectively labeled as $P_{ls} \equiv | \sqrt{ME -3 P^2/4} \pm P/2 |$):  $\Re f$ diverges logarithmically and $\Im f$ becomes discontinuous. Therefore, I first decompose the mesh for $P'$ (and $P$) into several intervals~\cite{Matsuyama:2006rp,Larson:1974zza}: $[0, P_1]$, $[P_1, P_2]$, $[P_2, P_3]$, $[P_3, \Pin]$, and $[\Pin, P_{max}]$ with $P_1 \equiv \sqrt{M_N E/3} $, $P_2 \equiv \sqrt{M_N E}$, $P_3 \equiv \sqrt{4M_N E/3}$, $\Pin = \sqrt{4M_N (E+B_2)/3}$, and $P_{max}$ as the UV cut off. When there is no bound dimer, $B_2 = 0$ is assumed in this setup of the intervals. Note that these two singularities at $P_{ls}$ are always in two different intervals. In the intervals without the two singularities, I apply the Lagrange function-based interpolant, which assigns momentum mesh in each interval according to the Gaussian quadrature~\cite{Baye:2015xoi}. To deal with the intervals with the two singularities, I make a variable transformation with $P' - P_{ls} = t^3$ around the singularities, with which the kernel becomes smooth in terms of $t$. Then, the Gaussian quadrature mesh in terms of the $t$ variable informs the momentum mesh. 

Moreover, as the solution to the Eq.~\eqref{eq:sol_psi1_2}, $F(P)$ has a branch point at $P_3$, when $E\geq 0$. Near that point, $F(P) \sim \sqrt{P_3 - P}$~\cite{Matsuyama:2006rp,Larson:1974zza}. I apply another variable transformation~\cite{Matsuyama:2006rp} so that when $P < P_3$, $t \sim  \sqrt{P_3 - P}$ and when $P> P_3$, $t \sim \sqrt{P-P_3}$ with proper normalizations. Again, the Gaussian quadrature mesh for $t$ provides the corresponding momentum mesh.  

It is worth pointing out that in Refs.~\cite{Matsuyama:2006rp, Larson:1974zza}, there were no consideration of three-body interactions. However, adding this interaction does not add new singularities in $f(P')$ and no new branch points in the solution $F(P)$ to the integral equation. 

Eventually, by setting up the proper momentum mesh and the pole subtraction, the integral equation changes to an inhomogeneous linear equation, which can be solved directly. The proper overlaps provide the benchmark calculations at real $E$ below or above the break-up threshold. Note that the $H$ spectrum dictates the analytical structure of the integration kernel, while the sources are irrelevant as long as they are smooth in momentum space. 

\subsection{Benchmark calculations for bound and resonance states} \label{app:three-body-phys-states-highfidelty}

Here, I discuss the benchmark calculations of the locations of physical states. In order to compute bound state eigenenergies, I look for the zeros of the kernel, $\bar{Z}(P, P', E, E, 1, 1)$, as a function of $E$ with $E$ restricted to be real and below the particle-dimer threshold. The kernel, numerically, is a matrix in two-dimensional momentum mesh for $P$ and $P'$. A simple Gaussian quadrature mesh suffices since the kernel varies smoothly in the bound state region when changing $P$ and $P'$. I then look for the zeros of the matrix's determinant. 

To compute the resonance location, I could have followed the established method using contour deformation or complex-scaling method. However, for simplicity, I apply the rational approximation as an analytical continuation method to continue $\resolventamp$ from the benchmark calculation results on the \emph{real} energy axis ($\Im E \to 0^+$) to the region below the real axis in the complex $E$ plane. The rational approximation is  
\begin{align}
    f(E) =  \frac{w_r}{E - E_r} + \sum_{i\in \mathrm{B.S.}}\frac{w_i}{ E - E_i} +  \sum_{i\in \mathrm{\BCT}} \frac{w_i}{ E - E_i}  \ . \label{eq:pole_expansion}
\end{align}

In this expression, the first term is the resonance pole contribution---assuming  the existence of a single resonance, the second is the pole contribution from possible three-body bound states, and the last term summing up the \BCT poles. For a given continuous \BCT that starts from 
the corresponding branching point $E = E_{th}$ and has a relative angle ($\theta_\mathrm{\BCT}$) from the positive real axis, the discretized \BCT poles are distributed on that line according to the tapered exponential distribution~\cite{Trefethen2021NodeClustering}. The location distribution is parameterized with $ E_{th}$, $c_0$, $c_1$, $N_\mathrm{\BCT}$, and  $\theta_\mathrm{\BCT}$: 
\begin{align}
    E_i = E_{th} + c_0\, e^{c_1 \sqrt{i} } \times e^{i \theta_\mathrm{\BCT}} \ ,  \label{eq:BCT_distribution}
\end{align}
with  $i = 0,,,N_\mathrm{\BCT}$. Equivalently, the \BCT pole locations are controlled by the smallest and the largest absolute values of separation between the poles and the branch point,  $ E_{th}$, $N_\mathrm{\BCT}$, and $\theta_\mathrm{\BCT}$. 

Then, I approximate $\resolventamp(E) \approx f(E)$ and look for the best fits that minimize the difference between the two at chosen $E^\mathrm{tr}_\alpha$. It should be emphasized that here $E^\mathrm{tr}_\alpha$ are on the upper rim of the real energy axis ($\Im E^\mathrm{tr}_\alpha \to 0^+$) and spread out over large energy intervals. 

The minimization is performed in two steps. For a given $E_r$, the residues $w_i$ and $w_r$ can be solved by applying the pseudo-inverse method to solve the over-constrained linear equation system: $f(E^\mathrm{tr}_\alpha) = \resolventamp(E^\mathrm{tr}_\alpha)$. Then, I compute the loss function defined as $\sum_{\alpha} | f(E^\mathrm{tr}_\alpha) - \resolventamp(E^\mathrm{tr}_\alpha) |$. The best $E_r$ is identified if it minimizes the loss function. The robustness of the minimization is checked by varying the parameters for controlling the distribution of the \BCT poles and looking for the consistency between these different fits.


\begin{thebibliography}{120}%
\makeatletter
\providecommand \@ifxundefined [1]{%
 \@ifx{#1\undefined}
}%
\providecommand \@ifnum [1]{%
 \ifnum #1\expandafter \@firstoftwo
 \else \expandafter \@secondoftwo
 \fi
}%
\providecommand \@ifx [1]{%
 \ifx #1\expandafter \@firstoftwo
 \else \expandafter \@secondoftwo
 \fi
}%
\providecommand \natexlab [1]{#1}%
\providecommand \enquote  [1]{``#1''}%
\providecommand \bibnamefont  [1]{#1}%
\providecommand \bibfnamefont [1]{#1}%
\providecommand \citenamefont [1]{#1}%
\providecommand \href@noop [0]{\@secondoftwo}%
\providecommand \href [0]{\begingroup \@sanitize@url \@href}%
\providecommand \@href[1]{\@@startlink{#1}\@@href}%
\providecommand \@@href[1]{\endgroup#1\@@endlink}%
\providecommand \@sanitize@url [0]{\catcode `\\12\catcode `\$12\catcode
  `\&12\catcode `\#12\catcode `\^12\catcode `\_12\catcode `\%12\relax}%
\providecommand \@@startlink[1]{}%
\providecommand \@@endlink[0]{}%
\providecommand \url  [0]{\begingroup\@sanitize@url \@url }%
\providecommand \@url [1]{\endgroup\@href {#1}{\urlprefix }}%
\providecommand \urlprefix  [0]{URL }%
\providecommand \Eprint [0]{\href }%
\providecommand \doibase [0]{https://doi.org/}%
\providecommand \selectlanguage [0]{\@gobble}%
\providecommand \bibinfo  [0]{\@secondoftwo}%
\providecommand \bibfield  [0]{\@secondoftwo}%
\providecommand \translation [1]{[#1]}%
\providecommand \BibitemOpen [0]{}%
\providecommand \bibitemStop [0]{}%
\providecommand \bibitemNoStop [0]{.\EOS\space}%
\providecommand \EOS [0]{\spacefactor3000\relax}%
\providecommand \BibitemShut  [1]{\csname bibitem#1\endcsname}%
\let\auto@bib@innerbib\@empty
\bibitem [{\citenamefont {Zhang}(2024)}]{Zhang:2024ril}%
  \BibitemOpen
  \bibfield  {author} {\bibinfo {author} {\bibfnamefont {X.}~\bibnamefont
  {Zhang}},\ }\bibfield  {title} {\bibinfo {title} {{Non-Hermitian Quantum
  Mechanics Approach for Extracting and Emulating Continuum Physics Based on
  Bound-State-Like Calculations}}\ }\href {https://doi.org/10.1103/5frj-w5xh}
  {10.1103/5frj-w5xh} (\bibinfo {year} {2024}),\ \Eprint
  {https://arxiv.org/abs/2408.03309} {arXiv:2408.03309 [nucl-th]} \BibitemShut
  {NoStop}%
\bibitem [{\citenamefont {Duguet}\ \emph {et~al.}(2024)\citenamefont {Duguet},
  \citenamefont {Ekstr\"om}, \citenamefont {Furnstahl}, \citenamefont
  {K\"onig},\ and\ \citenamefont {Lee}}]{Duguet:2023wuh}%
  \BibitemOpen
  \bibfield  {author} {\bibinfo {author} {\bibfnamefont {T.}~\bibnamefont
  {Duguet}}, \bibinfo {author} {\bibfnamefont {A.}~\bibnamefont {Ekstr\"om}},
  \bibinfo {author} {\bibfnamefont {R.~J.}\ \bibnamefont {Furnstahl}}, \bibinfo
  {author} {\bibfnamefont {S.}~\bibnamefont {K\"onig}},\ and\ \bibinfo {author}
  {\bibfnamefont {D.}~\bibnamefont {Lee}},\ }\bibfield  {title} {\bibinfo
  {title} {{Colloquium: Eigenvector continuation and projection-based
  emulators}},\ }\href {https://doi.org/10.1103/RevModPhys.96.031002}
  {\bibfield  {journal} {\bibinfo  {journal} {Rev. Mod. Phys.}\ }\textbf
  {\bibinfo {volume} {96}},\ \bibinfo {pages} {031002} (\bibinfo {year}
  {2024})},\ \Eprint {https://arxiv.org/abs/2310.19419} {arXiv:2310.19419
  [nucl-th]} \BibitemShut {NoStop}%
\bibitem [{\citenamefont {Drischler}\ \emph {et~al.}(2023)\citenamefont
  {Drischler}, \citenamefont {Melendez}, \citenamefont {Furnstahl},
  \citenamefont {Garcia},\ and\ \citenamefont {Zhang}}]{Drischler:2022ipa}%
  \BibitemOpen
  \bibfield  {author} {\bibinfo {author} {\bibfnamefont {C.}~\bibnamefont
  {Drischler}}, \bibinfo {author} {\bibfnamefont {J.~A.}\ \bibnamefont
  {Melendez}}, \bibinfo {author} {\bibfnamefont {R.~J.}\ \bibnamefont
  {Furnstahl}}, \bibinfo {author} {\bibfnamefont {A.~J.}\ \bibnamefont
  {Garcia}},\ and\ \bibinfo {author} {\bibfnamefont {X.}~\bibnamefont
  {Zhang}},\ }\bibfield  {title} {\bibinfo {title} {{BUQEYE Guide to
  Projection-Based Emulators in Nuclear Physics}},\ }\href
  {https://doi.org/10.3389/fphy.2022.1092931} {\bibfield  {journal} {\bibinfo
  {journal} {Front. Phys.}\ }\textbf {\bibinfo {volume} {10}},\ \bibinfo
  {pages} {92931} (\bibinfo {year} {2023})},\ \bibinfo {note} {supplemental,
  interactive Python code can be found on the companion
  website~\url{https://github.com/buqeye/frontiers-emulator-review}},\ \Eprint
  {https://arxiv.org/abs/2212.04912} {arXiv:2212.04912} \BibitemShut {NoStop}%
\bibitem [{\citenamefont {Melendez}\ \emph {et~al.}(2022)\citenamefont
  {Melendez}, \citenamefont {Drischler}, \citenamefont {Furnstahl},
  \citenamefont {Garcia},\ and\ \citenamefont {Zhang}}]{Melendez:2022kid}%
  \BibitemOpen
  \bibfield  {author} {\bibinfo {author} {\bibfnamefont {J.~A.}\ \bibnamefont
  {Melendez}}, \bibinfo {author} {\bibfnamefont {C.}~\bibnamefont {Drischler}},
  \bibinfo {author} {\bibfnamefont {R.~J.}\ \bibnamefont {Furnstahl}}, \bibinfo
  {author} {\bibfnamefont {A.~J.}\ \bibnamefont {Garcia}},\ and\ \bibinfo
  {author} {\bibfnamefont {X.}~\bibnamefont {Zhang}},\ }\bibfield  {title}
  {\bibinfo {title} {{Model reduction methods for nuclear emulators}},\ }\href
  {https://doi.org/10.1088/1361-6471/ac83dd} {\bibfield  {journal} {\bibinfo
  {journal} {J. Phys. G}\ }\textbf {\bibinfo {volume} {49}},\ \bibinfo {pages}
  {102001} (\bibinfo {year} {2022})},\ \Eprint
  {https://arxiv.org/abs/2203.05528} {arXiv:2203.05528 [nucl-th]} \BibitemShut
  {NoStop}%
\bibitem [{\citenamefont {Newton}(2002)}]{newton2002scattering}%
  \BibitemOpen
  \bibfield  {author} {\bibinfo {author} {\bibfnamefont {R.~G.}\ \bibnamefont
  {Newton}},\ }\href@noop {} {\emph {\bibinfo {title} {Scattering theory of
  waves and particles}}}\ (\bibinfo  {publisher} {Dover},\ \bibinfo {address}
  {Mineola, New York},\ \bibinfo {year} {2002})\BibitemShut {NoStop}%
\bibitem [{\citenamefont {Humblet}\ and\ \citenamefont
  {Rosenfeld}(1961)}]{HUMBLET1961529}%
  \BibitemOpen
  \bibfield  {author} {\bibinfo {author} {\bibfnamefont {J.}~\bibnamefont
  {Humblet}}\ and\ \bibinfo {author} {\bibfnamefont {L.}~\bibnamefont
  {Rosenfeld}},\ }\bibfield  {title} {\bibinfo {title} {Theory of nuclear
  reactions: I. resonant states and collision matrix},\ }\href
  {https://doi.org/https://doi.org/10.1016/0029-5582(61)90207-3} {\bibfield
  {journal} {\bibinfo  {journal} {Nuclear Physics}\ }\textbf {\bibinfo {volume}
  {26}},\ \bibinfo {pages} {529} (\bibinfo {year} {1961})}\BibitemShut
  {NoStop}%
\bibitem [{\citenamefont {Joachain}(1975)}]{JoachainQCT1975}%
  \BibitemOpen
  \bibfield  {author} {\bibinfo {author} {\bibfnamefont {C.~J.}\ \bibnamefont
  {Joachain}},\ }\href@noop {} {\emph {\bibinfo {title} {Quantum Collision
  Theory}}}\ (\bibinfo  {publisher} {North-Holland},\ \bibinfo {address}
  {Amsterdam},\ \bibinfo {year} {1975})\BibitemShut {NoStop}%
\bibitem [{\citenamefont {Ram\'\i{}rez~Jim\'enez}\ and\ \citenamefont
  {Kelkar}(2018)}]{RamirezJimenez:2018dge}%
  \BibitemOpen
  \bibfield  {author} {\bibinfo {author} {\bibfnamefont {D.~F.}\ \bibnamefont
  {Ram\'\i{}rez~Jim\'enez}}\ and\ \bibinfo {author} {\bibfnamefont {N.~G.}\
  \bibnamefont {Kelkar}},\ }\bibfield  {title} {\bibinfo {title} {{Different
  manifestations of S-matrix poles}},\ }\href
  {https://doi.org/10.1016/j.aop.2018.07.001} {\bibfield  {journal} {\bibinfo
  {journal} {Annals Phys.}\ }\textbf {\bibinfo {volume} {396}},\ \bibinfo
  {pages} {18} (\bibinfo {year} {2018})},\ \Eprint
  {https://arxiv.org/abs/1802.09467} {arXiv:1802.09467 [hep-ph]} \BibitemShut
  {NoStop}%
\bibitem [{\citenamefont {Yamada}\ \emph {et~al.}(2022)\citenamefont {Yamada},
  \citenamefont {Morimatsu},\ and\ \citenamefont {Sato}}]{Yamada:2022xam}%
  \BibitemOpen
  \bibfield  {author} {\bibinfo {author} {\bibfnamefont {W.~A.}\ \bibnamefont
  {Yamada}}, \bibinfo {author} {\bibfnamefont {O.}~\bibnamefont {Morimatsu}},\
  and\ \bibinfo {author} {\bibfnamefont {T.}~\bibnamefont {Sato}},\ }\bibfield
  {title} {\bibinfo {title} {{Analytic Map of Three-Channel S Matrix:
  Generalized Uniformization and Mittag-Leffler Expansion}},\ }\href
  {https://doi.org/10.1103/PhysRevLett.129.192001} {\bibfield  {journal}
  {\bibinfo  {journal} {Phys. Rev. Lett.}\ }\textbf {\bibinfo {volume} {129}},\
  \bibinfo {pages} {192001} (\bibinfo {year} {2022})},\ \Eprint
  {https://arxiv.org/abs/2203.17069} {arXiv:2203.17069 [hep-ph]} \BibitemShut
  {NoStop}%
\bibitem [{\citenamefont {{Reinhardt}}(2007)}]{ReinhardtComlexScaling2007}%
  \BibitemOpen
  \bibfield  {author} {\bibinfo {author} {\bibfnamefont {W.~P.}\ \bibnamefont
  {{Reinhardt}}},\ }\bibfield  {title} {\bibinfo {title} {{Complex Scaling in
  Atomic Physics: A Staging Ground for Experimental Mathematics and for
  Extracting Physics from Otherwise Impossible Computations}},\ }in\ \href@noop
  {} {\emph {\bibinfo {booktitle} {Spectral Theory and Mathematical
  Physics}}},\ Vol.~\bibinfo {volume} {76},\ \bibinfo {editor} {edited by\
  \bibinfo {editor} {\bibfnamefont {F.}~\bibnamefont {{Gesztesy}}}, \bibinfo
  {editor} {\bibfnamefont {P.}~\bibnamefont {{Deift}}}, \bibinfo {editor}
  {\bibfnamefont {C.}~\bibnamefont {{Galvez}}}, \bibinfo {editor}
  {\bibfnamefont {P.}~\bibnamefont {{Perry}}},\ and\ \bibinfo {editor}
  {\bibfnamefont {W.}~\bibnamefont {{Schlag}}}}\ (\bibinfo {year} {2007})\ p.\
  \bibinfo {pages} {357}\BibitemShut {NoStop}%
\bibitem [{\citenamefont {Reinhardt}(1982)}]{reinhardt1982complex}%
  \BibitemOpen
  \bibfield  {author} {\bibinfo {author} {\bibfnamefont {W.~P.}\ \bibnamefont
  {Reinhardt}},\ }\bibfield  {title} {\bibinfo {title} {Complex coordinates in
  the theory of atomic and molecular structure and dynamics},\ }\href@noop {}
  {\bibfield  {journal} {\bibinfo  {journal} {Annual Review of Physical
  Chemistry}\ }\textbf {\bibinfo {volume} {33}},\ \bibinfo {pages} {223}
  (\bibinfo {year} {1982})}\BibitemShut {NoStop}%
\bibitem [{\citenamefont {Moiseyev}(2011)}]{Moiseyev_2011}%
  \BibitemOpen
  \bibfield  {author} {\bibinfo {author} {\bibfnamefont {N.}~\bibnamefont
  {Moiseyev}},\ }\href
  {https://doi.org/https://doi.org/10.1017/CBO9780511976186} {\emph {\bibinfo
  {title} {Non-Hermitian Quantum Mechanics}}}\ (\bibinfo  {publisher}
  {Cambridge University Press},\ \bibinfo {year} {2011})\BibitemShut {NoStop}%
\bibitem [{\citenamefont {Myo}\ \emph {et~al.}(2014)\citenamefont {Myo},
  \citenamefont {Kikuchi}, \citenamefont {Masui},\ and\ \citenamefont
  {Kat\={o}}}]{Myo:2014ypa}%
  \BibitemOpen
  \bibfield  {author} {\bibinfo {author} {\bibfnamefont {T.}~\bibnamefont
  {Myo}}, \bibinfo {author} {\bibfnamefont {Y.}~\bibnamefont {Kikuchi}},
  \bibinfo {author} {\bibfnamefont {H.}~\bibnamefont {Masui}},\ and\ \bibinfo
  {author} {\bibfnamefont {K.}~\bibnamefont {Kat\={o}}},\ }\bibfield  {title}
  {\bibinfo {title} {{Recent development of complex scaling method for
  many-body resonances and continua in light nuclei}},\ }\href
  {https://doi.org/10.1016/j.ppnp.2014.08.001} {\bibfield  {journal} {\bibinfo
  {journal} {Prog. Part. Nucl. Phys.}\ }\textbf {\bibinfo {volume} {79}},\
  \bibinfo {pages} {1} (\bibinfo {year} {2014})},\ \Eprint
  {https://arxiv.org/abs/1410.4356} {arXiv:1410.4356 [nucl-th]} \BibitemShut
  {NoStop}%
\bibitem [{\citenamefont {Afnan}(1991)}]{afnan1991resonances}%
  \BibitemOpen
  \bibfield  {author} {\bibinfo {author} {\bibfnamefont {I.}~\bibnamefont
  {Afnan}},\ }\bibfield  {title} {\bibinfo {title} {Resonances in few-body
  systems},\ }\href@noop {} {\bibfield  {journal} {\bibinfo  {journal}
  {Australian journal of physics}\ }\textbf {\bibinfo {volume} {44}},\ \bibinfo
  {pages} {201} (\bibinfo {year} {1991})}\BibitemShut {NoStop}%
\bibitem [{\citenamefont
  {Trefethen}(2019)}]{Trefethen_approximation_theory_book}%
  \BibitemOpen
  \bibfield  {author} {\bibinfo {author} {\bibfnamefont {L.~N.}\ \bibnamefont
  {Trefethen}},\ }\href {https://doi.org/10.1137/1.9781611975949} {\emph
  {\bibinfo {title} {Approximation Theory and Approximation Practice, Extended
  Edition}}}\ (\bibinfo  {publisher} {Society for Industrial and Applied
  Mathematics},\ \bibinfo {address} {Philadelphia, PA},\ \bibinfo {year}
  {2019})\ \Eprint
  {https://arxiv.org/abs/https://epubs.siam.org/doi/pdf/10.1137/1.9781611975949}
  {https://epubs.siam.org/doi/pdf/10.1137/1.9781611975949} \BibitemShut
  {NoStop}%
\bibitem [{\citenamefont {Trefethen}(2023)}]{Trefethen2023review}%
  \BibitemOpen
  \bibfield  {author} {\bibinfo {author} {\bibfnamefont {L.~N.}\ \bibnamefont
  {Trefethen}},\ }\bibfield  {title} {\bibinfo {title} {Numerical analytic
  continuation},\ }\href {https://doi.org/10.1007/s13160-023-00599-2}
  {\bibfield  {journal} {\bibinfo  {journal} {Japan Journal of Industrial and
  Applied Mathematics}\ }\textbf {\bibinfo {volume} {40}},\ \bibinfo {pages}
  {1587} (\bibinfo {year} {2023})}\BibitemShut {NoStop}%
\bibitem [{\citenamefont {Schlessinger}\ and\ \citenamefont
  {Schwartz}(1966)}]{Schlessinger:1966zz}%
  \BibitemOpen
  \bibfield  {author} {\bibinfo {author} {\bibfnamefont {L.}~\bibnamefont
  {Schlessinger}}\ and\ \bibinfo {author} {\bibfnamefont {C.}~\bibnamefont
  {Schwartz}},\ }\bibfield  {title} {\bibinfo {title} {{Analyticity as a Useful
  Computation Tool}},\ }\href {https://doi.org/10.1103/PhysRevLett.16.1173}
  {\bibfield  {journal} {\bibinfo  {journal} {Phys. Rev. Lett.}\ }\textbf
  {\bibinfo {volume} {16}},\ \bibinfo {pages} {1173} (\bibinfo {year}
  {1966})}\BibitemShut {NoStop}%
\bibitem [{\citenamefont {Efros}(1985)}]{efros1985computation}%
  \BibitemOpen
  \bibfield  {author} {\bibinfo {author} {\bibfnamefont {V.}~\bibnamefont
  {Efros}},\ }\bibfield  {title} {\bibinfo {title} {{Computation of inclusive
  transition spectra and reaction cross sections without use of the continuum
  wave functions}},\ }\href@noop {} {\bibfield  {journal} {\bibinfo  {journal}
  {Sov. J. Nucl. Phys.(Engl. Transl.);(United States)}\ }\textbf {\bibinfo
  {volume} {41}},\ \bibinfo {pages} {949} (\bibinfo {year} {1985})}\BibitemShut
  {NoStop}%
\bibitem [{\citenamefont {Hesthaven}\ \emph {et~al.}(2015)\citenamefont
  {Hesthaven}, \citenamefont {Rozza},\ and\ \citenamefont
  {Stamm}}]{hesthaven2015certified}%
  \BibitemOpen
  \bibfield  {author} {\bibinfo {author} {\bibfnamefont {J.}~\bibnamefont
  {Hesthaven}}, \bibinfo {author} {\bibfnamefont {G.}~\bibnamefont {Rozza}},\
  and\ \bibinfo {author} {\bibfnamefont {B.}~\bibnamefont {Stamm}},\ }\href
  {https://doi.org/10.1007/978-3-319-22470-1} {\emph {\bibinfo {title}
  {Certified Reduced Basis Methods for Parametrized Partial Differential
  Equations}}},\ SpringerBriefs in Mathematics\ (\bibinfo  {publisher}
  {Springer International Publishing},\ \bibinfo {year} {2015})\BibitemShut
  {NoStop}%
\bibitem [{\citenamefont {Quarteroni}\ \emph {et~al.}(2016)\citenamefont
  {Quarteroni}, \citenamefont {Manzoni},\ and\ \citenamefont
  {Negri}}]{Quarteroni:218966}%
  \BibitemOpen
  \bibfield  {author} {\bibinfo {author} {\bibfnamefont {A.}~\bibnamefont
  {Quarteroni}}, \bibinfo {author} {\bibfnamefont {A.}~\bibnamefont
  {Manzoni}},\ and\ \bibinfo {author} {\bibfnamefont {F.}~\bibnamefont
  {Negri}},\ }\href {https://doi.org/10.1007/978-3-319-15431-2} {\emph
  {\bibinfo {title} {Reduced Basis Methods for Partial Differential Equations.
  An Introduction}}},\ La Matematica per il 3+2. 92\ (\bibinfo  {publisher}
  {Springer International Publishing},\ \bibinfo {year} {2016})\BibitemShut
  {NoStop}%
\bibitem [{\citenamefont {Benner}\ \emph
  {et~al.}(2017{\natexlab{a}})\citenamefont {Benner}, \citenamefont
  {Ohlberger}, \citenamefont {Patera}, \citenamefont {Rozza},\ and\
  \citenamefont {Urban}}]{Benner_2017aa}%
  \BibitemOpen
  \bibinfo {editor} {\bibfnamefont {P.}~\bibnamefont {Benner}}, \bibinfo
  {editor} {\bibfnamefont {M.}~\bibnamefont {Ohlberger}}, \bibinfo {editor}
  {\bibfnamefont {A.}~\bibnamefont {Patera}}, \bibinfo {editor} {\bibfnamefont
  {G.}~\bibnamefont {Rozza}},\ and\ \bibinfo {editor} {\bibfnamefont
  {K.}~\bibnamefont {Urban}},\ eds.,\ \href
  {https://doi.org/10.1007/978-3-319-58786-8} {\emph {\bibinfo {title} {Model
  Reduction of Parametrized Systems}}}\ (\bibinfo  {publisher} {Springer},\
  \bibinfo {year} {2017})\BibitemShut {NoStop}%
\bibitem [{\citenamefont {Benner}\ \emph
  {et~al.}(2017{\natexlab{b}})\citenamefont {Benner}, \citenamefont {Cohen},
  \citenamefont {Ohlberger},\ and\ \citenamefont
  {Willcox}}]{Benner2017modelRedApprox}%
  \BibitemOpen
  \bibfield  {author} {\bibinfo {author} {\bibfnamefont {P.}~\bibnamefont
  {Benner}}, \bibinfo {author} {\bibfnamefont {A.}~\bibnamefont {Cohen}},
  \bibinfo {author} {\bibfnamefont {M.}~\bibnamefont {Ohlberger}},\ and\
  \bibinfo {author} {\bibfnamefont {K.}~\bibnamefont {Willcox}},\ }\href
  {https://doi.org/doi:10.1137/1.9781611974829} {\emph {\bibinfo {title} {Model
  Reduction and Approximation}}}\ (\bibinfo  {publisher} {Society for
  Industrial and Applied Mathematics: Computational Science \& Engineering},\
  \bibinfo {year} {2017})\BibitemShut {NoStop}%
\bibitem [{\citenamefont {Benner}\ \emph {et~al.}(2015)\citenamefont {Benner},
  \citenamefont {Gugercin},\ and\ \citenamefont {Willcox}}]{benner2015survey}%
  \BibitemOpen
  \bibfield  {author} {\bibinfo {author} {\bibfnamefont {P.}~\bibnamefont
  {Benner}}, \bibinfo {author} {\bibfnamefont {S.}~\bibnamefont {Gugercin}},\
  and\ \bibinfo {author} {\bibfnamefont {K.}~\bibnamefont {Willcox}},\
  }\bibfield  {title} {\bibinfo {title} {A survey of projection-based model
  reduction methods for parametric dynamical systems},\ }\href
  {https://doi.org/10.1137/130932715} {\bibfield  {journal} {\bibinfo
  {journal} {SIAM Review}\ }\textbf {\bibinfo {volume} {57}},\ \bibinfo {pages}
  {483} (\bibinfo {year} {2015})}\BibitemShut {NoStop}%
\bibitem [{\citenamefont {Frame}\ \emph {et~al.}(2018)\citenamefont {Frame},
  \citenamefont {He}, \citenamefont {Ipsen}, \citenamefont {Lee}, \citenamefont
  {Lee},\ and\ \citenamefont {Rrapaj}}]{Frame:2017fah}%
  \BibitemOpen
  \bibfield  {author} {\bibinfo {author} {\bibfnamefont {D.}~\bibnamefont
  {Frame}}, \bibinfo {author} {\bibfnamefont {R.}~\bibnamefont {He}}, \bibinfo
  {author} {\bibfnamefont {I.}~\bibnamefont {Ipsen}}, \bibinfo {author}
  {\bibfnamefont {D.}~\bibnamefont {Lee}}, \bibinfo {author} {\bibfnamefont
  {D.}~\bibnamefont {Lee}},\ and\ \bibinfo {author} {\bibfnamefont
  {E.}~\bibnamefont {Rrapaj}},\ }\bibfield  {title} {\bibinfo {title}
  {{Eigenvector continuation with subspace learning}},\ }\href
  {https://doi.org/10.1103/PhysRevLett.121.032501} {\bibfield  {journal}
  {\bibinfo  {journal} {Phys. Rev. Lett.}\ }\textbf {\bibinfo {volume} {121}},\
  \bibinfo {pages} {032501} (\bibinfo {year} {2018})},\ \Eprint
  {https://arxiv.org/abs/1711.07090} {arXiv:1711.07090} \BibitemShut {NoStop}%
\bibitem [{\citenamefont {Sarkar}\ and\ \citenamefont
  {Lee}(2021{\natexlab{a}})}]{Sarkar:2020mad}%
  \BibitemOpen
  \bibfield  {author} {\bibinfo {author} {\bibfnamefont {A.}~\bibnamefont
  {Sarkar}}\ and\ \bibinfo {author} {\bibfnamefont {D.}~\bibnamefont {Lee}},\
  }\bibfield  {title} {\bibinfo {title} {{Convergence of Eigenvector
  Continuation}},\ }\href {https://doi.org/10.1103/PhysRevLett.126.032501}
  {\bibfield  {journal} {\bibinfo  {journal} {Phys. Rev. Lett.}\ }\textbf
  {\bibinfo {volume} {126}},\ \bibinfo {pages} {032501} (\bibinfo {year}
  {2021}{\natexlab{a}})},\ \Eprint {https://arxiv.org/abs/2004.07651}
  {arXiv:2004.07651 [nucl-th]} \BibitemShut {NoStop}%
\bibitem [{\citenamefont {Sarkar}\ and\ \citenamefont
  {Lee}(2022)}]{Sarkar:2021fpz}%
  \BibitemOpen
  \bibfield  {author} {\bibinfo {author} {\bibfnamefont {A.}~\bibnamefont
  {Sarkar}}\ and\ \bibinfo {author} {\bibfnamefont {D.}~\bibnamefont {Lee}},\
  }\bibfield  {title} {\bibinfo {title} {{Self-learning emulators and
  eigenvector continuation}},\ }\href
  {https://doi.org/10.1103/PhysRevResearch.4.023214} {\bibfield  {journal}
  {\bibinfo  {journal} {Phys. Rev. Res.}\ }\textbf {\bibinfo {volume} {4}},\
  \bibinfo {pages} {023214} (\bibinfo {year} {2022})},\ \Eprint
  {https://arxiv.org/abs/2107.13449} {arXiv:2107.13449 [nucl-th]} \BibitemShut
  {NoStop}%
\bibitem [{\citenamefont {K\"onig}\ \emph {et~al.}(2020)\citenamefont
  {K\"onig}, \citenamefont {Ekstr\"om}, \citenamefont {Hebeler}, \citenamefont
  {Lee},\ and\ \citenamefont {Schwenk}}]{Konig:2019adq}%
  \BibitemOpen
  \bibfield  {author} {\bibinfo {author} {\bibfnamefont {S.}~\bibnamefont
  {K\"onig}}, \bibinfo {author} {\bibfnamefont {A.}~\bibnamefont {Ekstr\"om}},
  \bibinfo {author} {\bibfnamefont {K.}~\bibnamefont {Hebeler}}, \bibinfo
  {author} {\bibfnamefont {D.}~\bibnamefont {Lee}},\ and\ \bibinfo {author}
  {\bibfnamefont {A.}~\bibnamefont {Schwenk}},\ }\bibfield  {title} {\bibinfo
  {title} {{Eigenvector Continuation as an Efficient and Accurate Emulator for
  Uncertainty Quantification}},\ }\href
  {https://doi.org/10.1016/j.physletb.2020.135814} {\bibfield  {journal}
  {\bibinfo  {journal} {Phys. Lett. B}\ }\textbf {\bibinfo {volume} {810}},\
  \bibinfo {pages} {135814} (\bibinfo {year} {2020})},\ \Eprint
  {https://arxiv.org/abs/1909.08446} {arXiv:1909.08446 [nucl-th]} \BibitemShut
  {NoStop}%
\bibitem [{\citenamefont {Demol}\ \emph {et~al.}(2020)\citenamefont {Demol},
  \citenamefont {Duguet}, \citenamefont {Ekstr\"om}, \citenamefont {Frosini},
  \citenamefont {Hebeler}, \citenamefont {K\"onig}, \citenamefont {Lee},
  \citenamefont {Schwenk}, \citenamefont {Som\`a},\ and\ \citenamefont
  {Tichai}}]{Demol:2019yjt}%
  \BibitemOpen
  \bibfield  {author} {\bibinfo {author} {\bibfnamefont {P.}~\bibnamefont
  {Demol}}, \bibinfo {author} {\bibfnamefont {T.}~\bibnamefont {Duguet}},
  \bibinfo {author} {\bibfnamefont {A.}~\bibnamefont {Ekstr\"om}}, \bibinfo
  {author} {\bibfnamefont {M.}~\bibnamefont {Frosini}}, \bibinfo {author}
  {\bibfnamefont {K.}~\bibnamefont {Hebeler}}, \bibinfo {author} {\bibfnamefont
  {S.}~\bibnamefont {K\"onig}}, \bibinfo {author} {\bibfnamefont
  {D.}~\bibnamefont {Lee}}, \bibinfo {author} {\bibfnamefont {A.}~\bibnamefont
  {Schwenk}}, \bibinfo {author} {\bibfnamefont {V.}~\bibnamefont {Som\`a}},\
  and\ \bibinfo {author} {\bibfnamefont {A.}~\bibnamefont {Tichai}},\
  }\bibfield  {title} {\bibinfo {title} {{Improved many-body expansions from
  eigenvector continuation}},\ }\href
  {https://doi.org/10.1103/PhysRevC.101.041302} {\bibfield  {journal} {\bibinfo
   {journal} {Phys. Rev. C}\ }\textbf {\bibinfo {volume} {101}},\ \bibinfo
  {pages} {041302} (\bibinfo {year} {2020})},\ \Eprint
  {https://arxiv.org/abs/1911.12578} {arXiv:1911.12578} \BibitemShut {NoStop}%
\bibitem [{\citenamefont {Ekström}\ and\ \citenamefont
  {Hagen}(2019)}]{Ekstrom:2019lss}%
  \BibitemOpen
  \bibfield  {author} {\bibinfo {author} {\bibfnamefont {A.}~\bibnamefont
  {Ekström}}\ and\ \bibinfo {author} {\bibfnamefont {G.}~\bibnamefont
  {Hagen}},\ }\bibfield  {title} {\bibinfo {title} {{Global sensitivity
  analysis of bulk properties of an atomic nucleus}},\ }\href
  {https://doi.org/10.1103/PhysRevLett.123.252501} {\bibfield  {journal}
  {\bibinfo  {journal} {Phys. Rev. Lett.}\ }\textbf {\bibinfo {volume} {123}},\
  \bibinfo {pages} {252501} (\bibinfo {year} {2019})},\ \Eprint
  {https://arxiv.org/abs/1910.02922} {arXiv:1910.02922 [nucl-th]} \BibitemShut
  {NoStop}%
\bibitem [{\citenamefont {Demol}\ \emph {et~al.}(2021)\citenamefont {Demol},
  \citenamefont {Frosini}, \citenamefont {Tichai}, \citenamefont {Som\`a},\
  and\ \citenamefont {Duguet}}]{Demol:2020mzd}%
  \BibitemOpen
  \bibfield  {author} {\bibinfo {author} {\bibfnamefont {P.}~\bibnamefont
  {Demol}}, \bibinfo {author} {\bibfnamefont {M.}~\bibnamefont {Frosini}},
  \bibinfo {author} {\bibfnamefont {A.}~\bibnamefont {Tichai}}, \bibinfo
  {author} {\bibfnamefont {V.}~\bibnamefont {Som\`a}},\ and\ \bibinfo {author}
  {\bibfnamefont {T.}~\bibnamefont {Duguet}},\ }\bibfield  {title} {\bibinfo
  {title} {{Bogoliubov many-body perturbation theory under constraint}},\
  }\href {https://doi.org/10.1016/j.aop.2020.168358} {\bibfield  {journal}
  {\bibinfo  {journal} {Annals Phys.}\ }\textbf {\bibinfo {volume} {424}},\
  \bibinfo {pages} {168358} (\bibinfo {year} {2021})},\ \Eprint
  {https://arxiv.org/abs/2002.02724} {arXiv:2002.02724 [nucl-th]} \BibitemShut
  {NoStop}%
\bibitem [{\citenamefont {Yoshida}\ and\ \citenamefont
  {Shimizu}(2022)}]{Yoshida:2021jbl}%
  \BibitemOpen
  \bibfield  {author} {\bibinfo {author} {\bibfnamefont {S.}~\bibnamefont
  {Yoshida}}\ and\ \bibinfo {author} {\bibfnamefont {N.}~\bibnamefont
  {Shimizu}},\ }\bibfield  {title} {\bibinfo {title} {{Constructing approximate
  shell-model wavefunctions by eigenvector continuation}},\ }\href
  {https://doi.org/10.1093/ptep/ptac057} {\bibfield  {journal} {\bibinfo
  {journal} {PTEP}\ }\textbf {\bibinfo {volume} {2022}},\ \bibinfo {pages}
  {053D02} (\bibinfo {year} {2022})},\ \Eprint
  {https://arxiv.org/abs/2105.08256} {arXiv:2105.08256} \BibitemShut {NoStop}%
\bibitem [{\citenamefont {Anderson}\ \emph {et~al.}(2022)\citenamefont
  {Anderson}, \citenamefont {O'Donnell},\ and\ \citenamefont
  {Piekarewicz}}]{Anderson:2022jhq}%
  \BibitemOpen
  \bibfield  {author} {\bibinfo {author} {\bibfnamefont {A.~L.}\ \bibnamefont
  {Anderson}}, \bibinfo {author} {\bibfnamefont {G.~L.}\ \bibnamefont
  {O'Donnell}},\ and\ \bibinfo {author} {\bibfnamefont {J.}~\bibnamefont
  {Piekarewicz}},\ }\bibfield  {title} {\bibinfo {title} {{Applications of
  reduced-basis methods to the nuclear single-particle spectrum}},\ }\href
  {https://doi.org/10.1103/PhysRevC.106.L031302} {\bibfield  {journal}
  {\bibinfo  {journal} {Phys. Rev. C}\ }\textbf {\bibinfo {volume} {106}},\
  \bibinfo {pages} {L031302} (\bibinfo {year} {2022})},\ \Eprint
  {https://arxiv.org/abs/2206.14889} {arXiv:2206.14889 [nucl-th]} \BibitemShut
  {NoStop}%
\bibitem [{\citenamefont {Giuliani}\ \emph {et~al.}(2023)\citenamefont
  {Giuliani}, \citenamefont {Godbey}, \citenamefont {Bonilla}, \citenamefont
  {Viens},\ and\ \citenamefont {Piekarewicz}}]{Giuliani:2022yna}%
  \BibitemOpen
  \bibfield  {author} {\bibinfo {author} {\bibfnamefont {P.}~\bibnamefont
  {Giuliani}}, \bibinfo {author} {\bibfnamefont {K.}~\bibnamefont {Godbey}},
  \bibinfo {author} {\bibfnamefont {E.}~\bibnamefont {Bonilla}}, \bibinfo
  {author} {\bibfnamefont {F.}~\bibnamefont {Viens}},\ and\ \bibinfo {author}
  {\bibfnamefont {J.}~\bibnamefont {Piekarewicz}},\ }\bibfield  {title}
  {\bibinfo {title} {{Bayes goes fast: Uncertainty Quantification for a
  Covariant Energy Density Functional emulated by the Reduced Basis Method}},\
  }\bibfield  {journal} {\bibinfo  {journal} {Front. Phys.}\ }\textbf {\bibinfo
  {volume} {10}},\ \href {https://doi.org/10.3389/fphy.2022.1054524}
  {10.3389/fphy.2022.1054524} (\bibinfo {year} {2023}),\ \Eprint
  {https://arxiv.org/abs/2209.13039} {arXiv:2209.13039} \BibitemShut {NoStop}%
\bibitem [{\citenamefont {Yapa}\ \emph {et~al.}(2023)\citenamefont {Yapa},
  \citenamefont {Fossez},\ and\ \citenamefont {K\"onig}}]{Yapa:2023xyf}%
  \BibitemOpen
  \bibfield  {author} {\bibinfo {author} {\bibfnamefont {N.}~\bibnamefont
  {Yapa}}, \bibinfo {author} {\bibfnamefont {K.}~\bibnamefont {Fossez}},\ and\
  \bibinfo {author} {\bibfnamefont {S.}~\bibnamefont {K\"onig}},\ }\bibfield
  {title} {\bibinfo {title} {{Eigenvector continuation for emulating and
  extrapolating two-body resonances}},\ }\href
  {https://doi.org/10.1103/PhysRevC.107.064316} {\bibfield  {journal} {\bibinfo
   {journal} {Phys. Rev. C}\ }\textbf {\bibinfo {volume} {107}},\ \bibinfo
  {pages} {064316} (\bibinfo {year} {2023})},\ \Eprint
  {https://arxiv.org/abs/2303.06139} {arXiv:2303.06139 [nucl-th]} \BibitemShut
  {NoStop}%
\bibitem [{\citenamefont {Liu}\ \emph {et~al.}(2024)\citenamefont {Liu},
  \citenamefont {Lei},\ and\ \citenamefont {Ren}}]{Liu:2024pqp}%
  \BibitemOpen
  \bibfield  {author} {\bibinfo {author} {\bibfnamefont {J.}~\bibnamefont
  {Liu}}, \bibinfo {author} {\bibfnamefont {J.}~\bibnamefont {Lei}},\ and\
  \bibinfo {author} {\bibfnamefont {Z.}~\bibnamefont {Ren}},\ }\bibfield
  {title} {\bibinfo {title} {{A complex scaling method for efficient and
  accurate scattering emulation in nuclear reactions}},\ }\href
  {https://doi.org/10.1016/j.physletb.2024.139070} {\bibfield  {journal}
  {\bibinfo  {journal} {Phys. Lett. B}\ }\textbf {\bibinfo {volume} {858}},\
  \bibinfo {pages} {139070} (\bibinfo {year} {2024})},\ \Eprint
  {https://arxiv.org/abs/2408.07954} {arXiv:2408.07954 [nucl-th]} \BibitemShut
  {NoStop}%
\bibitem [{\citenamefont {Yapa}\ \emph {et~al.}(2025)\citenamefont {Yapa},
  \citenamefont {K{\"o}nig},\ and\ \citenamefont {Fossez}}]{Yapa:2024lya}%
  \BibitemOpen
  \bibfield  {author} {\bibinfo {author} {\bibfnamefont {N.}~\bibnamefont
  {Yapa}}, \bibinfo {author} {\bibfnamefont {S.}~\bibnamefont {K{\"o}nig}},\
  and\ \bibinfo {author} {\bibfnamefont {K.}~\bibnamefont {Fossez}},\
  }\bibfield  {title} {\bibinfo {title} {{Toward scalable bound-to-resonance
  extrapolations for few- and many-body systems}},\ }\href
  {https://doi.org/10.1103/PhysRevC.111.064318} {\bibfield  {journal} {\bibinfo
   {journal} {Phys. Rev. C}\ }\textbf {\bibinfo {volume} {111}},\ \bibinfo
  {pages} {064318} (\bibinfo {year} {2025})},\ \Eprint
  {https://arxiv.org/abs/2409.03116} {arXiv:2409.03116 [nucl-th]} \BibitemShut
  {NoStop}%
\bibitem [{\citenamefont {Furnstahl}\ \emph {et~al.}(2020)\citenamefont
  {Furnstahl}, \citenamefont {Garcia}, \citenamefont {Millican},\ and\
  \citenamefont {Zhang}}]{Furnstahl:2020abp}%
  \BibitemOpen
  \bibfield  {author} {\bibinfo {author} {\bibfnamefont {R.~J.}\ \bibnamefont
  {Furnstahl}}, \bibinfo {author} {\bibfnamefont {A.~J.}\ \bibnamefont
  {Garcia}}, \bibinfo {author} {\bibfnamefont {P.~J.}\ \bibnamefont
  {Millican}},\ and\ \bibinfo {author} {\bibfnamefont {X.}~\bibnamefont
  {Zhang}},\ }\bibfield  {title} {\bibinfo {title} {{Efficient emulators for
  scattering using eigenvector continuation}},\ }\href
  {https://doi.org/10.1016/j.physletb.2020.135719} {\bibfield  {journal}
  {\bibinfo  {journal} {Phys. Lett. B}\ }\textbf {\bibinfo {volume} {809}},\
  \bibinfo {pages} {135719} (\bibinfo {year} {2020})},\ \Eprint
  {https://arxiv.org/abs/2007.03635} {arXiv:2007.03635 [nucl-th]} \BibitemShut
  {NoStop}%
\bibitem [{\citenamefont {Drischler}\ \emph {et~al.}(2021)\citenamefont
  {Drischler}, \citenamefont {Quinonez}, \citenamefont {Giuliani},
  \citenamefont {Lovell},\ and\ \citenamefont {Nunes}}]{Drischler:2021qoy}%
  \BibitemOpen
  \bibfield  {author} {\bibinfo {author} {\bibfnamefont {C.}~\bibnamefont
  {Drischler}}, \bibinfo {author} {\bibfnamefont {M.}~\bibnamefont {Quinonez}},
  \bibinfo {author} {\bibfnamefont {P.~G.}\ \bibnamefont {Giuliani}}, \bibinfo
  {author} {\bibfnamefont {A.~E.}\ \bibnamefont {Lovell}},\ and\ \bibinfo
  {author} {\bibfnamefont {F.~M.}\ \bibnamefont {Nunes}},\ }\bibfield  {title}
  {\bibinfo {title} {{Toward emulating nuclear reactions using eigenvector
  continuation}},\ }\href {https://doi.org/10.1016/j.physletb.2021.136777}
  {\bibfield  {journal} {\bibinfo  {journal} {Phys. Lett. B}\ }\textbf
  {\bibinfo {volume} {823}},\ \bibinfo {pages} {136777} (\bibinfo {year}
  {2021})},\ \Eprint {https://arxiv.org/abs/2108.08269} {arXiv:2108.08269
  [nucl-th]} \BibitemShut {NoStop}%
\bibitem [{\citenamefont {Melendez}\ \emph {et~al.}(2021)\citenamefont
  {Melendez}, \citenamefont {Drischler}, \citenamefont {Garcia}, \citenamefont
  {Furnstahl},\ and\ \citenamefont {Zhang}}]{Melendez:2021lyq}%
  \BibitemOpen
  \bibfield  {author} {\bibinfo {author} {\bibfnamefont {J.~A.}\ \bibnamefont
  {Melendez}}, \bibinfo {author} {\bibfnamefont {C.}~\bibnamefont {Drischler}},
  \bibinfo {author} {\bibfnamefont {A.~J.}\ \bibnamefont {Garcia}}, \bibinfo
  {author} {\bibfnamefont {R.~J.}\ \bibnamefont {Furnstahl}},\ and\ \bibinfo
  {author} {\bibfnamefont {X.}~\bibnamefont {Zhang}},\ }\bibfield  {title}
  {\bibinfo {title} {{Fast \& accurate emulation of two-body scattering
  observables without wave functions}},\ }\href
  {https://doi.org/10.1016/j.physletb.2021.136608} {\bibfield  {journal}
  {\bibinfo  {journal} {Phys. Lett. B}\ }\textbf {\bibinfo {volume} {821}},\
  \bibinfo {pages} {136608} (\bibinfo {year} {2021})},\ \Eprint
  {https://arxiv.org/abs/2106.15608} {arXiv:2106.15608 [nucl-th]} \BibitemShut
  {NoStop}%
\bibitem [{\citenamefont {Zhang}\ and\ \citenamefont
  {Furnstahl}(2022)}]{Zhang:2021jmi}%
  \BibitemOpen
  \bibfield  {author} {\bibinfo {author} {\bibfnamefont {X.}~\bibnamefont
  {Zhang}}\ and\ \bibinfo {author} {\bibfnamefont {R.~J.}\ \bibnamefont
  {Furnstahl}},\ }\bibfield  {title} {\bibinfo {title} {{Fast emulation of
  quantum three-body scattering}},\ }\href
  {https://doi.org/10.1103/PhysRevC.105.064004} {\bibfield  {journal} {\bibinfo
   {journal} {Phys. Rev. C}\ }\textbf {\bibinfo {volume} {105}},\ \bibinfo
  {pages} {064004} (\bibinfo {year} {2022})},\ \Eprint
  {https://arxiv.org/abs/2110.04269} {arXiv:2110.04269 [nucl-th]} \BibitemShut
  {NoStop}%
\bibitem [{\citenamefont {Bai}\ and\ \citenamefont {Ren}(2021)}]{Bai:2021xok}%
  \BibitemOpen
  \bibfield  {author} {\bibinfo {author} {\bibfnamefont {D.}~\bibnamefont
  {Bai}}\ and\ \bibinfo {author} {\bibfnamefont {Z.}~\bibnamefont {Ren}},\
  }\bibfield  {title} {\bibinfo {title} {{Generalizing the calculable
  $R$-matrix theory and eigenvector continuation to the incoming wave boundary
  condition}},\ }\href {https://doi.org/10.1103/PhysRevC.103.014612} {\bibfield
   {journal} {\bibinfo  {journal} {Phys. Rev. C}\ }\textbf {\bibinfo {volume}
  {103}},\ \bibinfo {pages} {014612} (\bibinfo {year} {2021})},\ \Eprint
  {https://arxiv.org/abs/2101.06336} {arXiv:2101.06336 [nucl-th]} \BibitemShut
  {NoStop}%
\bibitem [{\citenamefont {Drischler}\ and\ \citenamefont
  {Zhang}(2022)}]{Drischler:2022yfb}%
  \BibitemOpen
  \bibfield  {author} {\bibinfo {author} {\bibfnamefont {C.}~\bibnamefont
  {Drischler}}\ and\ \bibinfo {author} {\bibfnamefont {X.}~\bibnamefont
  {Zhang}},\ }\bibfield  {title} {\bibinfo {title} {Few-body emulators based on
  eigenvector continuation},\ }in\ \href
  {https://doi.org/10.1007/s00601-022-01749-x} {\emph {\bibinfo {booktitle}
  {{Nuclear Forces for Precision Nuclear Physics: A Collection of
  Perspectives}}}},\ Vol.~\bibinfo {volume} {63},\ \bibinfo {editor} {edited
  by\ \bibinfo {editor} {\bibfnamefont {I.}~\bibnamefont {Tews}}, \bibinfo
  {editor} {\bibfnamefont {Z.}~\bibnamefont {Davoudi}}, \bibinfo {editor}
  {\bibfnamefont {A.}~\bibnamefont {Ekstr{\"o}m}},\ and\ \bibinfo {editor}
  {\bibfnamefont {J.~D.}\ \bibnamefont {Holt}}}\ (\bibinfo  {publisher}
  {Springer-Verlag GmbH Austria},\ \bibinfo {year} {2022})\ Chap.~\bibinfo
  {chapter} {8}, p.~\bibinfo {pages} {67},\ \Eprint
  {https://arxiv.org/abs/2202.01105} {arXiv:2202.01105} \BibitemShut {NoStop}%
\bibitem [{\citenamefont {Bai}(2022)}]{Bai:2022hjg}%
  \BibitemOpen
  \bibfield  {author} {\bibinfo {author} {\bibfnamefont {D.}~\bibnamefont
  {Bai}},\ }\bibfield  {title} {\bibinfo {title} {{New extensions of
  eigenvector continuation R-matrix theory based on analyticity in momentum and
  angular momentum}},\ }\href {https://doi.org/10.1103/PhysRevC.106.024611}
  {\bibfield  {journal} {\bibinfo  {journal} {Phys. Rev. C}\ }\textbf {\bibinfo
  {volume} {106}},\ \bibinfo {pages} {024611} (\bibinfo {year}
  {2022})}\BibitemShut {NoStop}%
\bibitem [{\citenamefont {Garcia}\ \emph {et~al.}(2023)\citenamefont {Garcia},
  \citenamefont {Drischler}, \citenamefont {Furnstahl}, \citenamefont
  {Melendez},\ and\ \citenamefont {Zhang}}]{Garcia:2023slj}%
  \BibitemOpen
  \bibfield  {author} {\bibinfo {author} {\bibfnamefont {A.~J.}\ \bibnamefont
  {Garcia}}, \bibinfo {author} {\bibfnamefont {C.}~\bibnamefont {Drischler}},
  \bibinfo {author} {\bibfnamefont {R.~J.}\ \bibnamefont {Furnstahl}}, \bibinfo
  {author} {\bibfnamefont {J.~A.}\ \bibnamefont {Melendez}},\ and\ \bibinfo
  {author} {\bibfnamefont {X.}~\bibnamefont {Zhang}},\ }\bibfield  {title}
  {\bibinfo {title} {{Wave-function-based emulation for nucleon-nucleon
  scattering in momentum space}},\ }\href
  {https://doi.org/10.1103/PhysRevC.107.054001} {\bibfield  {journal} {\bibinfo
   {journal} {Phys. Rev. C}\ }\textbf {\bibinfo {volume} {107}},\ \bibinfo
  {pages} {054001} (\bibinfo {year} {2023})},\ \Eprint
  {https://arxiv.org/abs/2301.05093} {arXiv:2301.05093 [nucl-th]} \BibitemShut
  {NoStop}%
\bibitem [{\citenamefont {Odell}\ \emph {et~al.}(2024)\citenamefont {Odell},
  \citenamefont {Giuliani}, \citenamefont {Beyer}, \citenamefont
  {Catacora-Rios}, \citenamefont {Chan}, \citenamefont {Bonilla}, \citenamefont
  {Furnstahl}, \citenamefont {Godbey},\ and\ \citenamefont
  {Nunes}}]{Odell:2023cun}%
  \BibitemOpen
  \bibfield  {author} {\bibinfo {author} {\bibfnamefont {D.}~\bibnamefont
  {Odell}}, \bibinfo {author} {\bibfnamefont {P.}~\bibnamefont {Giuliani}},
  \bibinfo {author} {\bibfnamefont {K.}~\bibnamefont {Beyer}}, \bibinfo
  {author} {\bibfnamefont {M.}~\bibnamefont {Catacora-Rios}}, \bibinfo {author}
  {\bibfnamefont {M.~Y.~H.}\ \bibnamefont {Chan}}, \bibinfo {author}
  {\bibfnamefont {E.}~\bibnamefont {Bonilla}}, \bibinfo {author} {\bibfnamefont
  {R.~J.}\ \bibnamefont {Furnstahl}}, \bibinfo {author} {\bibfnamefont
  {K.}~\bibnamefont {Godbey}},\ and\ \bibinfo {author} {\bibfnamefont {F.~M.}\
  \bibnamefont {Nunes}},\ }\bibfield  {title} {\bibinfo {title} {{ROSE: A
  reduced-order scattering emulator for optical models}},\ }\href
  {https://doi.org/10.1103/PhysRevC.109.044612} {\bibfield  {journal} {\bibinfo
   {journal} {Phys. Rev. C}\ }\textbf {\bibinfo {volume} {109}},\ \bibinfo
  {pages} {044612} (\bibinfo {year} {2024})},\ \Eprint
  {https://arxiv.org/abs/2312.12426} {arXiv:2312.12426 [physics.comp-ph]}
  \BibitemShut {NoStop}%
\bibitem [{\citenamefont {Antoulas}(2005)}]{Antoulas2005book}%
  \BibitemOpen
  \bibfield  {author} {\bibinfo {author} {\bibfnamefont {A.~C.}\ \bibnamefont
  {Antoulas}},\ }\href {https://doi.org/10.1137/1.9780898718713} {\emph
  {\bibinfo {title} {Approximation of Large-Scale Dynamical Systems}}}\
  (\bibinfo  {publisher} {Society for Industrial and Applied Mathematics},\
  \bibinfo {year} {2005})\ \Eprint
  {https://arxiv.org/abs/https://epubs.siam.org/doi/pdf/10.1137/1.9780898718713}
  {https://epubs.siam.org/doi/pdf/10.1137/1.9780898718713} \BibitemShut
  {NoStop}%
\bibitem [{\citenamefont {Van~Beeumen}\ \emph {et~al.}(2017)\citenamefont
  {Van~Beeumen}, \citenamefont {Williams-Young}, \citenamefont {Kasper},
  \citenamefont {Yang}, \citenamefont {Ng},\ and\ \citenamefont
  {Li}}]{Beeumen2017}%
  \BibitemOpen
  \bibfield  {author} {\bibinfo {author} {\bibfnamefont {R.}~\bibnamefont
  {Van~Beeumen}}, \bibinfo {author} {\bibfnamefont {D.~B.}\ \bibnamefont
  {Williams-Young}}, \bibinfo {author} {\bibfnamefont {J.~M.}\ \bibnamefont
  {Kasper}}, \bibinfo {author} {\bibfnamefont {C.}~\bibnamefont {Yang}},
  \bibinfo {author} {\bibfnamefont {E.~G.}\ \bibnamefont {Ng}},\ and\ \bibinfo
  {author} {\bibfnamefont {X.}~\bibnamefont {Li}},\ }\bibfield  {title}
  {\bibinfo {title} {Model order reduction algorithm for estimating the
  absorption spectrum},\ }\href {https://doi.org/10.1021/acs.jctc.7b00402}
  {\bibfield  {journal} {\bibinfo  {journal} {Journal of Chemical Theory and
  Computation}\ }\textbf {\bibinfo {volume} {13}},\ \bibinfo {pages} {4950}
  (\bibinfo {year} {2017})},\ \bibinfo {note} {pMID: 28862869},\ \Eprint
  {https://arxiv.org/abs/https://doi.org/10.1021/acs.jctc.7b00402}
  {https://doi.org/10.1021/acs.jctc.7b00402} \BibitemShut {NoStop}%
\bibitem [{\citenamefont {Peng}\ \emph {et~al.}(2019)\citenamefont {Peng},
  \citenamefont {Van~Beeumen}, \citenamefont {Williams-Young}, \citenamefont
  {Kowalski},\ and\ \citenamefont {Yang}}]{Peng2019}%
  \BibitemOpen
  \bibfield  {author} {\bibinfo {author} {\bibfnamefont {B.}~\bibnamefont
  {Peng}}, \bibinfo {author} {\bibfnamefont {R.}~\bibnamefont {Van~Beeumen}},
  \bibinfo {author} {\bibfnamefont {D.~B.}\ \bibnamefont {Williams-Young}},
  \bibinfo {author} {\bibfnamefont {K.}~\bibnamefont {Kowalski}},\ and\
  \bibinfo {author} {\bibfnamefont {C.}~\bibnamefont {Yang}},\ }\bibfield
  {title} {\bibinfo {title} {Approximate green’s function coupled cluster
  method employing effective dimension reduction},\ }\href
  {https://doi.org/10.1021/acs.jctc.9b00172} {\bibfield  {journal} {\bibinfo
  {journal} {Journal of Chemical Theory and Computation}\ }\textbf {\bibinfo
  {volume} {15}},\ \bibinfo {pages} {3185} (\bibinfo {year} {2019})},\ \bibinfo
  {note} {pMID: 30951302},\ \Eprint
  {https://arxiv.org/abs/https://doi.org/10.1021/acs.jctc.9b00172}
  {https://doi.org/10.1021/acs.jctc.9b00172} \BibitemShut {NoStop}%
\bibitem [{\citenamefont {Gl{\"o}ckle}(1983)}]{Glockle:1983}%
  \BibitemOpen
  \bibfield  {author} {\bibinfo {author} {\bibfnamefont {W.}~\bibnamefont
  {Gl{\"o}ckle}},\ }\href
  {https://doi.org/https://doi.org/10.1007/978-3-642-82081-6} {\emph {\bibinfo
  {title} {The Quantum Mechanical Few-Body Problem}}}\ (\bibinfo  {publisher}
  {Springer-Verlag},\ \bibinfo {address} {Berlin},\ \bibinfo {year}
  {1983})\BibitemShut {NoStop}%
\bibitem [{\citenamefont {Lazauskas}\ and\ \citenamefont
  {Carbonell}(2011)}]{Lazauskas:2011uj}%
  \BibitemOpen
  \bibfield  {author} {\bibinfo {author} {\bibfnamefont {R.}~\bibnamefont
  {Lazauskas}}\ and\ \bibinfo {author} {\bibfnamefont {J.}~\bibnamefont
  {Carbonell}},\ }\bibfield  {title} {\bibinfo {title} {{Application of
  complex-scaling method for few-body scattering}},\ }\href
  {https://doi.org/10.1103/PhysRevC.84.034002} {\bibfield  {journal} {\bibinfo
  {journal} {Phys. Rev. C}\ }\textbf {\bibinfo {volume} {84}},\ \bibinfo
  {pages} {034002} (\bibinfo {year} {2011})},\ \Eprint
  {https://arxiv.org/abs/1104.2016} {arXiv:1104.2016 [nucl-th]} \BibitemShut
  {NoStop}%
\bibitem [{\citenamefont {Lazauskas}(2012)}]{Lazauskas:2012jc}%
  \BibitemOpen
  \bibfield  {author} {\bibinfo {author} {\bibfnamefont {R.}~\bibnamefont
  {Lazauskas}},\ }\bibfield  {title} {\bibinfo {title} {{Application of the
  complex-scaling method to four-nucleon scattering above break-up
  threshold}},\ }\href {https://doi.org/10.1103/PhysRevC.86.044002} {\bibfield
  {journal} {\bibinfo  {journal} {Phys. Rev. C}\ }\textbf {\bibinfo {volume}
  {86}},\ \bibinfo {pages} {044002} (\bibinfo {year} {2012})}\BibitemShut
  {NoStop}%
\bibitem [{\citenamefont {Papadimitriou}\ and\ \citenamefont
  {Vary}(2015)}]{Papadimitriou:2015rca}%
  \BibitemOpen
  \bibfield  {author} {\bibinfo {author} {\bibfnamefont {G.}~\bibnamefont
  {Papadimitriou}}\ and\ \bibinfo {author} {\bibfnamefont {J.~P.}\ \bibnamefont
  {Vary}},\ }\bibfield  {title} {\bibinfo {title} {{Nucleon\textendash{}nucleon
  resonances at intermediate energies using a complex energy formalism}},\
  }\href {https://doi.org/10.1016/j.physletb.2015.04.058} {\bibfield  {journal}
  {\bibinfo  {journal} {Phys. Lett. B}\ }\textbf {\bibinfo {volume} {746}},\
  \bibinfo {pages} {121} (\bibinfo {year} {2015})},\ \Eprint
  {https://arxiv.org/abs/1503.05277} {arXiv:1503.05277 [nucl-th]} \BibitemShut
  {NoStop}%
\bibitem [{\citenamefont {Lazauskas}(2015)}]{Lazauskas:2015ula}%
  \BibitemOpen
  \bibfield  {author} {\bibinfo {author} {\bibfnamefont {R.}~\bibnamefont
  {Lazauskas}},\ }\bibfield  {title} {\bibinfo {title} {{Modern nuclear force
  predictions for n-H3 scattering above the three- and four-nucleon breakup
  thresholds}},\ }\href {https://doi.org/10.1103/PhysRevC.91.041001} {\bibfield
   {journal} {\bibinfo  {journal} {Phys. Rev. C}\ }\textbf {\bibinfo {volume}
  {91}},\ \bibinfo {pages} {041001} (\bibinfo {year} {2015})}\BibitemShut
  {NoStop}%
\bibitem [{\citenamefont {Lazauskas}\ and\ \citenamefont
  {Carbonell}(2020)}]{Lazauskas:2019hil}%
  \BibitemOpen
  \bibfield  {author} {\bibinfo {author} {\bibfnamefont {R.}~\bibnamefont
  {Lazauskas}}\ and\ \bibinfo {author} {\bibfnamefont {J.}~\bibnamefont
  {Carbonell}},\ }\bibfield  {title} {\bibinfo {title} {{Description of Four-
  and Five-Nucleon Systems by Solving Faddeev-Yakubovsky Equations in
  Configuration Space}},\ }\href {https://doi.org/10.3389/fphy.2019.00251}
  {\bibfield  {journal} {\bibinfo  {journal} {Front. in Phys.}\ }\textbf
  {\bibinfo {volume} {7}},\ \bibinfo {pages} {251} (\bibinfo {year} {2020})},\
  \Eprint {https://arxiv.org/abs/2002.05876} {arXiv:2002.05876 [nucl-th]}
  \BibitemShut {NoStop}%
\bibitem [{\citenamefont {Berggren}(1968)}]{Berggren:1968zz}%
  \BibitemOpen
  \bibfield  {author} {\bibinfo {author} {\bibfnamefont {T.}~\bibnamefont
  {Berggren}},\ }\bibfield  {title} {\bibinfo {title} {{On the use of resonant
  states in eigenfunction expansions of scattering and reaction amplitudes}},\
  }\href {https://doi.org/10.1016/0375-9474(68)90593-9} {\bibfield  {journal}
  {\bibinfo  {journal} {Nucl. Phys. A}\ }\textbf {\bibinfo {volume} {109}},\
  \bibinfo {pages} {265} (\bibinfo {year} {1968})}\BibitemShut {NoStop}%
\bibitem [{\citenamefont {Berggren}\ and\ \citenamefont
  {Lind}(1993)}]{Berggren1993}%
  \BibitemOpen
  \bibfield  {author} {\bibinfo {author} {\bibfnamefont {T.}~\bibnamefont
  {Berggren}}\ and\ \bibinfo {author} {\bibfnamefont {P.}~\bibnamefont
  {Lind}},\ }\bibfield  {title} {\bibinfo {title} {Resonant state expansion of
  the resolvent},\ }\href {https://doi.org/10.1103/PhysRevC.47.768} {\bibfield
  {journal} {\bibinfo  {journal} {Phys. Rev. C}\ }\textbf {\bibinfo {volume}
  {47}},\ \bibinfo {pages} {768} (\bibinfo {year} {1993})}\BibitemShut
  {NoStop}%
\bibitem [{\citenamefont {Michel}\ and\ \citenamefont
  {P\l{}oszajczak}(2021)}]{Michel:2021jkx}%
  \BibitemOpen
  \bibfield  {author} {\bibinfo {author} {\bibfnamefont {N.}~\bibnamefont
  {Michel}}\ and\ \bibinfo {author} {\bibfnamefont {M.}~\bibnamefont
  {P\l{}oszajczak}},\ }\href {https://doi.org/10.1007/978-3-030-69356-5} {\emph
  {\bibinfo {title} {{Gamow Shell Model: The Unified Theory of Nuclear
  Structure and Reactions}}}},\ Vol.\ \bibinfo {volume} {983}\ (\bibinfo
  {publisher} {Springer Cham},\ \bibinfo {year} {2021})\BibitemShut {NoStop}%
\bibitem [{\citenamefont
  {Schlessinger}(1968{\natexlab{a}})}]{Schlessinger:1968vsk}%
  \BibitemOpen
  \bibfield  {author} {\bibinfo {author} {\bibfnamefont {L.}~\bibnamefont
  {Schlessinger}},\ }\bibfield  {title} {\bibinfo {title} {{Use of Analyticity
  in the Calculation of Nonrelativistic Scattering Amplitudes}},\ }\href
  {https://doi.org/10.1103/PhysRev.167.1411} {\bibfield  {journal} {\bibinfo
  {journal} {Phys. Rev.}\ }\textbf {\bibinfo {volume} {167}},\ \bibinfo {pages}
  {1411} (\bibinfo {year} {1968}{\natexlab{a}})}\BibitemShut {NoStop}%
\bibitem [{\citenamefont
  {Schlessinger}(1968{\natexlab{b}})}]{Schlessinger:1968zz}%
  \BibitemOpen
  \bibfield  {author} {\bibinfo {author} {\bibfnamefont {L.}~\bibnamefont
  {Schlessinger}},\ }\bibfield  {title} {\bibinfo {title} {{Calculation of Some
  Three-Body Scattering Amplitudes}},\ }\href
  {https://doi.org/10.1103/PhysRev.171.1523} {\bibfield  {journal} {\bibinfo
  {journal} {Phys. Rev.}\ }\textbf {\bibinfo {volume} {171}},\ \bibinfo {pages}
  {1523} (\bibinfo {year} {1968}{\natexlab{b}})}\BibitemShut {NoStop}%
\bibitem [{\citenamefont {McDonald}\ and\ \citenamefont
  {Nuttall}(1969)}]{McDonald:1969zza}%
  \BibitemOpen
  \bibfield  {author} {\bibinfo {author} {\bibfnamefont {F.~A.}\ \bibnamefont
  {McDonald}}\ and\ \bibinfo {author} {\bibfnamefont {J.}~\bibnamefont
  {Nuttall}},\ }\bibfield  {title} {\bibinfo {title} {{Complex-Energy Method
  for Elastic e-H Scattering above the Ionization Threshold}},\ }\href
  {https://doi.org/10.1103/PhysRevLett.23.361} {\bibfield  {journal} {\bibinfo
  {journal} {Phys. Rev. Lett.}\ }\textbf {\bibinfo {volume} {23}},\ \bibinfo
  {pages} {361} (\bibinfo {year} {1969})}\BibitemShut {NoStop}%
\bibitem [{\citenamefont {Uzu}\ \emph {et~al.}(2003)\citenamefont {Uzu},
  \citenamefont {Kamada},\ and\ \citenamefont {Koike}}]{Uzu:2003ms}%
  \BibitemOpen
  \bibfield  {author} {\bibinfo {author} {\bibfnamefont {E.}~\bibnamefont
  {Uzu}}, \bibinfo {author} {\bibfnamefont {H.}~\bibnamefont {Kamada}},\ and\
  \bibinfo {author} {\bibfnamefont {Y.}~\bibnamefont {Koike}},\ }\bibfield
  {title} {\bibinfo {title} {{Complex energy method in four body
  Faddeev-Yakubovsky equations}},\ }\href
  {https://doi.org/10.1103/PhysRevC.68.061001} {\bibfield  {journal} {\bibinfo
  {journal} {Phys. Rev. C}\ }\textbf {\bibinfo {volume} {68}},\ \bibinfo
  {pages} {061001} (\bibinfo {year} {2003})},\ \Eprint
  {https://arxiv.org/abs/nucl-th/0310001} {arXiv:nucl-th/0310001} \BibitemShut
  {NoStop}%
\bibitem [{\citenamefont {Deltuva}\ and\ \citenamefont
  {Fonseca}(2012)}]{Deltuva:2012fa}%
  \BibitemOpen
  \bibfield  {author} {\bibinfo {author} {\bibfnamefont {A.}~\bibnamefont
  {Deltuva}}\ and\ \bibinfo {author} {\bibfnamefont {A.~C.}\ \bibnamefont
  {Fonseca}},\ }\bibfield  {title} {\bibinfo {title} {{Neutron-$^3H$ scattering
  above the four-nucleon breakup threshold}},\ }\href
  {https://doi.org/10.1103/PhysRevC.86.011001} {\bibfield  {journal} {\bibinfo
  {journal} {Phys. Rev. C}\ }\textbf {\bibinfo {volume} {86}},\ \bibinfo
  {pages} {011001} (\bibinfo {year} {2012})},\ \Eprint
  {https://arxiv.org/abs/1206.4574} {arXiv:1206.4574 [nucl-th]} \BibitemShut
  {NoStop}%
\bibitem [{\citenamefont {Deltuva}\ and\ \citenamefont
  {Fonseca}(2013{\natexlab{a}})}]{Deltuva:2013qf}%
  \BibitemOpen
  \bibfield  {author} {\bibinfo {author} {\bibfnamefont {A.}~\bibnamefont
  {Deltuva}}\ and\ \bibinfo {author} {\bibfnamefont {A.~C.}\ \bibnamefont
  {Fonseca}},\ }\bibfield  {title} {\bibinfo {title} {{$^3$H production via
  neutron-neutron-deuteron recombination}},\ }\href
  {https://doi.org/10.1103/PhysRevC.87.014002} {\bibfield  {journal} {\bibinfo
  {journal} {Phys. Rev. C}\ }\textbf {\bibinfo {volume} {87}},\ \bibinfo
  {pages} {014002} (\bibinfo {year} {2013}{\natexlab{a}})},\ \Eprint
  {https://arxiv.org/abs/1301.1905} {arXiv:1301.1905 [nucl-th]} \BibitemShut
  {NoStop}%
\bibitem [{\citenamefont {Deltuva}\ and\ \citenamefont
  {Fonseca}(2013{\natexlab{b}})}]{Deltuva:2013mda}%
  \BibitemOpen
  \bibfield  {author} {\bibinfo {author} {\bibfnamefont {A.}~\bibnamefont
  {Deltuva}}\ and\ \bibinfo {author} {\bibfnamefont {A.~C.}\ \bibnamefont
  {Fonseca}},\ }\bibfield  {title} {\bibinfo {title} {{Calculation of
  proton-$^3$He elastic scattering between 7 and 35 MeV}},\ }\href
  {https://doi.org/10.1103/PhysRevC.87.054002} {\bibfield  {journal} {\bibinfo
  {journal} {Phys. Rev. C}\ }\textbf {\bibinfo {volume} {87}},\ \bibinfo
  {pages} {054002} (\bibinfo {year} {2013}{\natexlab{b}})},\ \Eprint
  {https://arxiv.org/abs/1304.5410} {arXiv:1304.5410 [nucl-th]} \BibitemShut
  {NoStop}%
\bibitem [{\citenamefont {Deltuva}\ and\ \citenamefont
  {Fonseca}(2014)}]{Deltuva:2014pda}%
  \BibitemOpen
  \bibfield  {author} {\bibinfo {author} {\bibfnamefont {A.}~\bibnamefont
  {Deltuva}}\ and\ \bibinfo {author} {\bibfnamefont {A.~C.}\ \bibnamefont
  {Fonseca}},\ }\bibfield  {title} {\bibinfo {title} {{Calculation of
  neutron-$^3He$ scattering up to 30 MeV}},\ }\href
  {https://doi.org/10.1103/PhysRevC.90.044002} {\bibfield  {journal} {\bibinfo
  {journal} {Phys. Rev. C}\ }\textbf {\bibinfo {volume} {90}},\ \bibinfo
  {pages} {044002} (\bibinfo {year} {2014})},\ \Eprint
  {https://arxiv.org/abs/1409.7318} {arXiv:1409.7318 [nucl-th]} \BibitemShut
  {NoStop}%
\bibitem [{\citenamefont {Efros}\ \emph {et~al.}(1994)\citenamefont {Efros},
  \citenamefont {Leidemann},\ and\ \citenamefont {Orlandini}}]{Efros:1994iq}%
  \BibitemOpen
  \bibfield  {author} {\bibinfo {author} {\bibfnamefont {V.~D.}\ \bibnamefont
  {Efros}}, \bibinfo {author} {\bibfnamefont {W.}~\bibnamefont {Leidemann}},\
  and\ \bibinfo {author} {\bibfnamefont {G.}~\bibnamefont {Orlandini}},\
  }\bibfield  {title} {\bibinfo {title} {{Response functions from integral
  transforms with a Lorentz kernel}},\ }\href
  {https://doi.org/10.1016/0370-2693(94)91355-2} {\bibfield  {journal}
  {\bibinfo  {journal} {Phys. Lett. B}\ }\textbf {\bibinfo {volume} {338}},\
  \bibinfo {pages} {130} (\bibinfo {year} {1994})},\ \Eprint
  {https://arxiv.org/abs/nucl-th/9409004} {arXiv:nucl-th/9409004} \BibitemShut
  {NoStop}%
\bibitem [{\citenamefont {Efros}\ \emph {et~al.}(2007)\citenamefont {Efros},
  \citenamefont {Leidemann}, \citenamefont {Orlandini},\ and\ \citenamefont
  {Barnea}}]{Efros:2007nq}%
  \BibitemOpen
  \bibfield  {author} {\bibinfo {author} {\bibfnamefont {V.~D.}\ \bibnamefont
  {Efros}}, \bibinfo {author} {\bibfnamefont {W.}~\bibnamefont {Leidemann}},
  \bibinfo {author} {\bibfnamefont {G.}~\bibnamefont {Orlandini}},\ and\
  \bibinfo {author} {\bibfnamefont {N.}~\bibnamefont {Barnea}},\ }\bibfield
  {title} {\bibinfo {title} {{The Lorentz Integral Transform (LIT) method and
  its applications to perturbation induced reactions}},\ }\href
  {https://doi.org/10.1088/0954-3899/34/12/R02} {\bibfield  {journal} {\bibinfo
   {journal} {J. Phys. G}\ }\textbf {\bibinfo {volume} {34}},\ \bibinfo {pages}
  {R459} (\bibinfo {year} {2007})},\ \Eprint {https://arxiv.org/abs/0708.2803}
  {arXiv:0708.2803 [nucl-th]} \BibitemShut {NoStop}%
\bibitem [{\citenamefont {Orlandini}\ \emph {et~al.}(2014)\citenamefont
  {Orlandini}, \citenamefont {Bacca}, \citenamefont {Barnea}, \citenamefont
  {Hagen}, \citenamefont {Miorelli},\ and\ \citenamefont
  {Papenbrock}}]{Orlandini:2013eya}%
  \BibitemOpen
  \bibfield  {author} {\bibinfo {author} {\bibfnamefont {G.}~\bibnamefont
  {Orlandini}}, \bibinfo {author} {\bibfnamefont {S.}~\bibnamefont {Bacca}},
  \bibinfo {author} {\bibfnamefont {N.}~\bibnamefont {Barnea}}, \bibinfo
  {author} {\bibfnamefont {G.}~\bibnamefont {Hagen}}, \bibinfo {author}
  {\bibfnamefont {M.}~\bibnamefont {Miorelli}},\ and\ \bibinfo {author}
  {\bibfnamefont {T.}~\bibnamefont {Papenbrock}},\ }\bibfield  {title}
  {\bibinfo {title} {{Coupling the Lorentz Integral Transform (LIT) and the
  Coupled Cluster (CC) Methods: A Way Towards Continuum Spectra of
  'Not-So-Few-Body' Systems}},\ }\href
  {https://doi.org/10.1007/s00601-013-0772-4} {\bibfield  {journal} {\bibinfo
  {journal} {Few Body Syst.}\ }\textbf {\bibinfo {volume} {55}},\ \bibinfo
  {pages} {907} (\bibinfo {year} {2014})},\ \Eprint
  {https://arxiv.org/abs/1311.2141} {arXiv:1311.2141 [nucl-th]} \BibitemShut
  {NoStop}%
\bibitem [{\citenamefont {Sobczyk}\ \emph {et~al.}(2021)\citenamefont
  {Sobczyk}, \citenamefont {Acharya}, \citenamefont {Bacca},\ and\
  \citenamefont {Hagen}}]{Sobczyk:2021dwm}%
  \BibitemOpen
  \bibfield  {author} {\bibinfo {author} {\bibfnamefont {J.~E.}\ \bibnamefont
  {Sobczyk}}, \bibinfo {author} {\bibfnamefont {B.}~\bibnamefont {Acharya}},
  \bibinfo {author} {\bibfnamefont {S.}~\bibnamefont {Bacca}},\ and\ \bibinfo
  {author} {\bibfnamefont {G.}~\bibnamefont {Hagen}},\ }\bibfield  {title}
  {\bibinfo {title} {{Ab initio computation of the longitudinal response
  function in $^{40}$Ca}},\ }\href
  {https://doi.org/10.1103/PhysRevLett.127.072501} {\bibfield  {journal}
  {\bibinfo  {journal} {Phys. Rev. Lett.}\ }\textbf {\bibinfo {volume} {127}},\
  \bibinfo {pages} {072501} (\bibinfo {year} {2021})},\ \Eprint
  {https://arxiv.org/abs/2103.06786} {arXiv:2103.06786 [nucl-th]} \BibitemShut
  {NoStop}%
\bibitem [{\citenamefont {Sobczyk}\ \emph {et~al.}(2024)\citenamefont
  {Sobczyk}, \citenamefont {Acharya}, \citenamefont {Bacca},\ and\
  \citenamefont {Hagen}}]{Sobczyk:2023sxh}%
  \BibitemOpen
  \bibfield  {author} {\bibinfo {author} {\bibfnamefont {J.~E.}\ \bibnamefont
  {Sobczyk}}, \bibinfo {author} {\bibfnamefont {B.}~\bibnamefont {Acharya}},
  \bibinfo {author} {\bibfnamefont {S.}~\bibnamefont {Bacca}},\ and\ \bibinfo
  {author} {\bibfnamefont {G.}~\bibnamefont {Hagen}},\ }\bibfield  {title}
  {\bibinfo {title} {{Ca40 transverse response function from coupled-cluster
  theory}},\ }\href {https://doi.org/10.1103/PhysRevC.109.025502} {\bibfield
  {journal} {\bibinfo  {journal} {Phys. Rev. C}\ }\textbf {\bibinfo {volume}
  {109}},\ \bibinfo {pages} {025502} (\bibinfo {year} {2024})},\ \Eprint
  {https://arxiv.org/abs/2310.03109} {arXiv:2310.03109 [nucl-th]} \BibitemShut
  {NoStop}%
\bibitem [{\citenamefont {Bonaiti}\ \emph {et~al.}(2024)\citenamefont
  {Bonaiti}, \citenamefont {Bacca}, \citenamefont {Hagen},\ and\ \citenamefont
  {Jansen}}]{Bonaiti:2024fft}%
  \BibitemOpen
  \bibfield  {author} {\bibinfo {author} {\bibfnamefont {F.}~\bibnamefont
  {Bonaiti}}, \bibinfo {author} {\bibfnamefont {S.}~\bibnamefont {Bacca}},
  \bibinfo {author} {\bibfnamefont {G.}~\bibnamefont {Hagen}},\ and\ \bibinfo
  {author} {\bibfnamefont {G.~R.}\ \bibnamefont {Jansen}},\ }\bibfield  {title}
  {\bibinfo {title} {{Electromagnetic observables of open-shell nuclei from
  coupled-cluster theory}},\ }\href
  {https://doi.org/10.1103/PhysRevC.110.044306} {\bibfield  {journal} {\bibinfo
   {journal} {Phys. Rev. C}\ }\textbf {\bibinfo {volume} {110}},\ \bibinfo
  {pages} {044306} (\bibinfo {year} {2024})},\ \Eprint
  {https://arxiv.org/abs/2405.05608} {arXiv:2405.05608 [nucl-th]} \BibitemShut
  {NoStop}%
\bibitem [{\citenamefont {{{BUQEYE collaboration}}}(2022)}]{BUQEYEsoftware}%
  \BibitemOpen
  \bibfield  {author} {\bibinfo {author} {\bibnamefont {{{BUQEYE
  collaboration}}}},\ }\href {\mbox{https://buqeye.github.io/software/}} {}
  (\bibinfo {year} {2022}),\ \bibinfo {note}
  {\url{https://buqeye.github.io/software/}}\BibitemShut {NoStop}%
\bibitem [{\citenamefont {Fetter}\ and\ \citenamefont
  {Walecka}(1972)}]{FETTER71}%
  \BibitemOpen
  \bibfield  {author} {\bibinfo {author} {\bibfnamefont {A.~L.}\ \bibnamefont
  {Fetter}}\ and\ \bibinfo {author} {\bibfnamefont {J.~D.}\ \bibnamefont
  {Walecka}},\ }\href@noop {} {\emph {\bibinfo {title} {Quantum Many-Particle
  Systems}}}\ (\bibinfo  {publisher} {McGraw--Hill},\ \bibinfo {address} {New
  York},\ \bibinfo {year} {1972})\BibitemShut {NoStop}%
\bibitem [{\citenamefont {Pastore}\ \emph {et~al.}(2008)\citenamefont
  {Pastore}, \citenamefont {Schiavilla},\ and\ \citenamefont
  {Goity}}]{Pastore:2008ui}%
  \BibitemOpen
  \bibfield  {author} {\bibinfo {author} {\bibfnamefont {S.}~\bibnamefont
  {Pastore}}, \bibinfo {author} {\bibfnamefont {R.}~\bibnamefont
  {Schiavilla}},\ and\ \bibinfo {author} {\bibfnamefont {J.~L.}\ \bibnamefont
  {Goity}},\ }\bibfield  {title} {\bibinfo {title} {{Electromagnetic two-body
  currents of one- and two-pion range}},\ }\href
  {https://doi.org/10.1103/PhysRevC.78.064002} {\bibfield  {journal} {\bibinfo
  {journal} {Phys. Rev. C}\ }\textbf {\bibinfo {volume} {78}},\ \bibinfo
  {pages} {064002} (\bibinfo {year} {2008})},\ \Eprint
  {https://arxiv.org/abs/0810.1941} {arXiv:0810.1941 [nucl-th]} \BibitemShut
  {NoStop}%
\bibitem [{\citenamefont {Walecka}(1995)}]{Walecka1995}%
  \BibitemOpen
  \bibfield  {author} {\bibinfo {author} {\bibfnamefont {J.}~\bibnamefont
  {Walecka}},\ }\href@noop {} {\emph {\bibinfo {title} {Theoretical nuclear and
  subnuclear Physics}}}\ (\bibinfo  {publisher} {Oxford University Press},\
  \bibinfo {address} {New York},\ \bibinfo {year} {1995})\BibitemShut {NoStop}%
\bibitem [{\citenamefont {Goldberger}\ and\ \citenamefont
  {Watson}(1964)}]{Goldberger1964}%
  \BibitemOpen
  \bibfield  {author} {\bibinfo {author} {\bibfnamefont {M.}~\bibnamefont
  {Goldberger}}\ and\ \bibinfo {author} {\bibfnamefont {K.}~\bibnamefont
  {Watson}},\ }\href@noop {} {\emph {\bibinfo {title} {Collision Theory}}}\
  (\bibinfo  {publisher} {Wiley},\ \bibinfo {address} {New York},\ \bibinfo
  {year} {1964})\BibitemShut {NoStop}%
\bibitem [{\citenamefont {Newton}(1982)}]{newton1982scattering}%
  \BibitemOpen
  \bibfield  {author} {\bibinfo {author} {\bibfnamefont {R.~G.}\ \bibnamefont
  {Newton}},\ }\href@noop {} {\emph {\bibinfo {title} {Scattering theory of
  waves and particles}}}\ (\bibinfo  {publisher} {Springer Berlin},\ \bibinfo
  {address} {Heidelberg},\ \bibinfo {year} {1982})\BibitemShut {NoStop}%
\bibitem [{\citenamefont {Rotureau}\ \emph {et~al.}(2017)\citenamefont
  {Rotureau}, \citenamefont {Danielewicz}, \citenamefont {Hagen}, \citenamefont
  {Nunes},\ and\ \citenamefont {Papenbrock}}]{Rotureau:2016jpf}%
  \BibitemOpen
  \bibfield  {author} {\bibinfo {author} {\bibfnamefont {J.}~\bibnamefont
  {Rotureau}}, \bibinfo {author} {\bibfnamefont {P.}~\bibnamefont
  {Danielewicz}}, \bibinfo {author} {\bibfnamefont {G.}~\bibnamefont {Hagen}},
  \bibinfo {author} {\bibfnamefont {F.}~\bibnamefont {Nunes}},\ and\ \bibinfo
  {author} {\bibfnamefont {T.}~\bibnamefont {Papenbrock}},\ }\bibfield  {title}
  {\bibinfo {title} {{Optical potential from first principles}},\ }\href
  {https://doi.org/10.1103/PhysRevC.95.024315} {\bibfield  {journal} {\bibinfo
  {journal} {Phys. Rev. C}\ }\textbf {\bibinfo {volume} {95}},\ \bibinfo
  {pages} {024315} (\bibinfo {year} {2017})},\ \Eprint
  {https://arxiv.org/abs/1611.04554} {arXiv:1611.04554 [nucl-th]} \BibitemShut
  {NoStop}%
\bibitem [{\citenamefont {Burrows}\ \emph {et~al.}(2024)\citenamefont
  {Burrows}, \citenamefont {Launey}, \citenamefont {Mercenne}, \citenamefont
  {Baker}, \citenamefont {Sargsyan}, \citenamefont {Dytrych},\ and\
  \citenamefont {Langr}}]{Burrows:2023ygq}%
  \BibitemOpen
  \bibfield  {author} {\bibinfo {author} {\bibfnamefont {M.}~\bibnamefont
  {Burrows}}, \bibinfo {author} {\bibfnamefont {K.~D.}\ \bibnamefont {Launey}},
  \bibinfo {author} {\bibfnamefont {A.}~\bibnamefont {Mercenne}}, \bibinfo
  {author} {\bibfnamefont {R.~B.}\ \bibnamefont {Baker}}, \bibinfo {author}
  {\bibfnamefont {G.~H.}\ \bibnamefont {Sargsyan}}, \bibinfo {author}
  {\bibfnamefont {T.}~\bibnamefont {Dytrych}},\ and\ \bibinfo {author}
  {\bibfnamefont {D.}~\bibnamefont {Langr}},\ }\bibfield  {title} {\bibinfo
  {title} {{Ab initio translationally invariant nucleon-nucleus optical
  potentials}},\ }\href {https://doi.org/10.1103/PhysRevC.109.014616}
  {\bibfield  {journal} {\bibinfo  {journal} {Phys. Rev. C}\ }\textbf {\bibinfo
  {volume} {109}},\ \bibinfo {pages} {014616} (\bibinfo {year} {2024})},\
  \Eprint {https://arxiv.org/abs/2307.00202} {arXiv:2307.00202 [nucl-th]}
  \BibitemShut {NoStop}%
\bibitem [{\citenamefont {Baye}(2015)}]{Baye:2015xoi}%
  \BibitemOpen
  \bibfield  {author} {\bibinfo {author} {\bibfnamefont {D.}~\bibnamefont
  {Baye}},\ }\bibfield  {title} {\bibinfo {title} {{The Lagrange-mesh
  method}},\ }\href {https://doi.org/10.1016/j.physrep.2014.11.006} {\bibfield
  {journal} {\bibinfo  {journal} {Phys. Rept.}\ }\textbf {\bibinfo {volume}
  {565}},\ \bibinfo {pages} {1} (\bibinfo {year} {2015})}\BibitemShut {NoStop}%
\bibitem [{\citenamefont {Luscher}(1991)}]{Luscher:1990ux}%
  \BibitemOpen
  \bibfield  {author} {\bibinfo {author} {\bibfnamefont {M.}~\bibnamefont
  {Luscher}},\ }\bibfield  {title} {\bibinfo {title} {{Two particle states on a
  torus and their relation to the scattering matrix}},\ }\href
  {https://doi.org/10.1016/0550-3213(91)90366-6} {\bibfield  {journal}
  {\bibinfo  {journal} {Nucl. Phys.}\ }\textbf {\bibinfo {volume} {B354}},\
  \bibinfo {pages} {531} (\bibinfo {year} {1991})}\BibitemShut {NoStop}%
\bibitem [{\citenamefont {Busch}\ \emph {et~al.}(1998)\citenamefont {Busch},
  \citenamefont {Englert}, \citenamefont {Rzazewski},\ and\ \citenamefont
  {Wilkens}}]{Busch1998}%
  \BibitemOpen
  \bibfield  {author} {\bibinfo {author} {\bibfnamefont {T.}~\bibnamefont
  {Busch}}, \bibinfo {author} {\bibfnamefont {B.-G.}\ \bibnamefont {Englert}},
  \bibinfo {author} {\bibfnamefont {K.}~\bibnamefont {Rzazewski}},\ and\
  \bibinfo {author} {\bibfnamefont {M.}~\bibnamefont {Wilkens}},\ }\bibfield
  {title} {{\selectlanguage {English}\bibinfo {title} {Two cold atoms in a
  harmonic trap}},\ }\href {https://doi.org/10.1023/A:1018705520999} {\bibfield
   {journal} {\bibinfo  {journal} {Foundations of Physics}\ }\textbf {\bibinfo
  {volume} {28}},\ \bibinfo {pages} {549} (\bibinfo {year} {1998})}\BibitemShut
  {NoStop}%
\bibitem [{\citenamefont {Stetcu}\ \emph {et~al.}(2010)\citenamefont {Stetcu},
  \citenamefont {Rotureau}, \citenamefont {Barrett},\ and\ \citenamefont {van
  Kolck}}]{Stetcu:2010xq}%
  \BibitemOpen
  \bibfield  {author} {\bibinfo {author} {\bibfnamefont {I.}~\bibnamefont
  {Stetcu}}, \bibinfo {author} {\bibfnamefont {J.}~\bibnamefont {Rotureau}},
  \bibinfo {author} {\bibfnamefont {B.~R.}\ \bibnamefont {Barrett}},\ and\
  \bibinfo {author} {\bibfnamefont {U.}~\bibnamefont {van Kolck}},\ }\bibfield
  {title} {\bibinfo {title} {{An Effective field theory approach to two trapped
  particles}},\ }\href {https://doi.org/10.1016/j.aop.2010.02.008} {\bibfield
  {journal} {\bibinfo  {journal} {Annals Phys.}\ }\textbf {\bibinfo {volume}
  {325}},\ \bibinfo {pages} {1644} (\bibinfo {year} {2010})},\ \Eprint
  {https://arxiv.org/abs/1001.5071} {arXiv:1001.5071 [cond-mat.quant-gas]}
  \BibitemShut {NoStop}%
\bibitem [{\citenamefont {Rotureau}\ \emph {et~al.}(2012)\citenamefont
  {Rotureau}, \citenamefont {Stetcu}, \citenamefont {Barrett},\ and\
  \citenamefont {van Kolck}}]{Rotureau:2011vf}%
  \BibitemOpen
  \bibfield  {author} {\bibinfo {author} {\bibfnamefont {J.}~\bibnamefont
  {Rotureau}}, \bibinfo {author} {\bibfnamefont {I.}~\bibnamefont {Stetcu}},
  \bibinfo {author} {\bibfnamefont {B.~R.}\ \bibnamefont {Barrett}},\ and\
  \bibinfo {author} {\bibfnamefont {U.}~\bibnamefont {van Kolck}},\ }\bibfield
  {title} {\bibinfo {title} {{Two and Three Nucleons in a Trap and the
  Continuum Limit}},\ }\href {https://doi.org/10.1103/PhysRevC.85.034003}
  {\bibfield  {journal} {\bibinfo  {journal} {Phys. Rev.}\ }\textbf {\bibinfo
  {volume} {C85}},\ \bibinfo {pages} {034003} (\bibinfo {year} {2012})},\
  \Eprint {https://arxiv.org/abs/1112.0267} {arXiv:1112.0267 [nucl-th]}
  \BibitemShut {NoStop}%
\bibitem [{\citenamefont {Zhang}(2020)}]{Zhang:2019cai}%
  \BibitemOpen
  \bibfield  {author} {\bibinfo {author} {\bibfnamefont {X.}~\bibnamefont
  {Zhang}},\ }\bibfield  {title} {\bibinfo {title} {{Extracting free-space
  observables from trapped interacting clusters}},\ }\href
  {https://doi.org/10.1103/PhysRevC.101.051602} {\bibfield  {journal} {\bibinfo
   {journal} {Phys. Rev. C}\ }\textbf {\bibinfo {volume} {101}},\ \bibinfo
  {pages} {051602} (\bibinfo {year} {2020})},\ \Eprint
  {https://arxiv.org/abs/1905.05275} {arXiv:1905.05275} \BibitemShut {NoStop}%
\bibitem [{\citenamefont {Zhang}\ \emph {et~al.}(2020)\citenamefont {Zhang},
  \citenamefont {Stroberg}, \citenamefont {Navr\'atil}, \citenamefont {Gwak},
  \citenamefont {Melendez}, \citenamefont {Furnstahl},\ and\ \citenamefont
  {Holt}}]{Zhang:2020rhz}%
  \BibitemOpen
  \bibfield  {author} {\bibinfo {author} {\bibfnamefont {X.}~\bibnamefont
  {Zhang}}, \bibinfo {author} {\bibfnamefont {S.~R.}\ \bibnamefont {Stroberg}},
  \bibinfo {author} {\bibfnamefont {P.}~\bibnamefont {Navr\'atil}}, \bibinfo
  {author} {\bibfnamefont {C.}~\bibnamefont {Gwak}}, \bibinfo {author}
  {\bibfnamefont {J.~A.}\ \bibnamefont {Melendez}}, \bibinfo {author}
  {\bibfnamefont {R.~J.}\ \bibnamefont {Furnstahl}},\ and\ \bibinfo {author}
  {\bibfnamefont {J.~D.}\ \bibnamefont {Holt}},\ }\bibfield  {title} {\bibinfo
  {title} {{Ab Initio Calculations of Low-Energy Nuclear Scattering Using
  Confining Potential Traps}},\ }\href
  {https://doi.org/10.1103/PhysRevLett.125.112503} {\bibfield  {journal}
  {\bibinfo  {journal} {Phys. Rev. Lett.}\ }\textbf {\bibinfo {volume} {125}},\
  \bibinfo {pages} {112503} (\bibinfo {year} {2020})},\ \Eprint
  {https://arxiv.org/abs/2004.13575} {arXiv:2004.13575 [nucl-th]} \BibitemShut
  {NoStop}%
\bibitem [{\citenamefont {Guo}\ and\ \citenamefont {Long}(2022)}]{Guo:2021uig}%
  \BibitemOpen
  \bibfield  {author} {\bibinfo {author} {\bibfnamefont {P.}~\bibnamefont
  {Guo}}\ and\ \bibinfo {author} {\bibfnamefont {B.}~\bibnamefont {Long}},\
  }\bibfield  {title} {\bibinfo {title} {{Nuclear reactions in artificial
  traps}},\ }\href {https://doi.org/10.1088/1361-6471/ac59d5} {\bibfield
  {journal} {\bibinfo  {journal} {J. Phys. G}\ }\textbf {\bibinfo {volume}
  {49}},\ \bibinfo {pages} {055104} (\bibinfo {year} {2022})},\ \Eprint
  {https://arxiv.org/abs/2101.03901} {arXiv:2101.03901 [nucl-th]} \BibitemShut
  {NoStop}%
\bibitem [{\citenamefont {Zhang}\ \emph
  {et~al.}(2024{\natexlab{a}})\citenamefont {Zhang}, \citenamefont {Bai},
  \citenamefont {Wang},\ and\ \citenamefont {Ren}}]{Zhang:2024mot}%
  \BibitemOpen
  \bibfield  {author} {\bibinfo {author} {\bibfnamefont {H.}~\bibnamefont
  {Zhang}}, \bibinfo {author} {\bibfnamefont {D.}~\bibnamefont {Bai}}, \bibinfo
  {author} {\bibfnamefont {Z.}~\bibnamefont {Wang}},\ and\ \bibinfo {author}
  {\bibfnamefont {Z.}~\bibnamefont {Ren}},\ }\bibfield  {title} {\bibinfo
  {title} {{Charged particle scattering in harmonic traps}},\ }\href
  {https://doi.org/10.1016/j.physletb.2024.138490} {\bibfield  {journal}
  {\bibinfo  {journal} {Phys. Lett. B}\ }\textbf {\bibinfo {volume} {850}},\
  \bibinfo {pages} {138490} (\bibinfo {year} {2024}{\natexlab{a}})}\BibitemShut
  {NoStop}%
\bibitem [{\citenamefont {Zhang}\ \emph
  {et~al.}(2024{\natexlab{b}})\citenamefont {Zhang}, \citenamefont {Bai},
  \citenamefont {Wang},\ and\ \citenamefont {Ren}}]{Zhang:2024vmz}%
  \BibitemOpen
  \bibfield  {author} {\bibinfo {author} {\bibfnamefont {H.}~\bibnamefont
  {Zhang}}, \bibinfo {author} {\bibfnamefont {D.}~\bibnamefont {Bai}}, \bibinfo
  {author} {\bibfnamefont {Z.}~\bibnamefont {Wang}},\ and\ \bibinfo {author}
  {\bibfnamefont {Z.}~\bibnamefont {Ren}},\ }\bibfield  {title} {\bibinfo
  {title} {{Microscopic cluster model in harmonic oscillator traps}},\ }\href
  {https://doi.org/10.1103/PhysRevC.109.034307} {\bibfield  {journal} {\bibinfo
   {journal} {Phys. Rev. C}\ }\textbf {\bibinfo {volume} {109}},\ \bibinfo
  {pages} {034307} (\bibinfo {year} {2024}{\natexlab{b}})}\BibitemShut
  {NoStop}%
\bibitem [{\citenamefont {Zhang}\ \emph
  {et~al.}(2024{\natexlab{c}})\citenamefont {Zhang}, \citenamefont {Bai},\ and\
  \citenamefont {Ren}}]{Zhang:2024vch}%
  \BibitemOpen
  \bibfield  {author} {\bibinfo {author} {\bibfnamefont {H.}~\bibnamefont
  {Zhang}}, \bibinfo {author} {\bibfnamefont {D.}~\bibnamefont {Bai}},\ and\
  \bibinfo {author} {\bibfnamefont {Z.}~\bibnamefont {Ren}},\ }\bibfield
  {title} {\bibinfo {title} {{Coupled-channels reactions for charged particles
  in harmonic traps}},\ }\href {https://doi.org/10.1103/PhysRevC.110.034308}
  {\bibfield  {journal} {\bibinfo  {journal} {Phys. Rev. C}\ }\textbf {\bibinfo
  {volume} {110}},\ \bibinfo {pages} {034308} (\bibinfo {year}
  {2024}{\natexlab{c}})}\BibitemShut {NoStop}%
\bibitem [{\citenamefont {Bagnarol}\ \emph {et~al.}(2025)\citenamefont
  {Bagnarol}, \citenamefont {Barnea}, \citenamefont {Rojik},\ and\
  \citenamefont {Schafer}}]{Bagnarol:2024rhq}%
  \BibitemOpen
  \bibfield  {author} {\bibinfo {author} {\bibfnamefont {M.}~\bibnamefont
  {Bagnarol}}, \bibinfo {author} {\bibfnamefont {N.}~\bibnamefont {Barnea}},
  \bibinfo {author} {\bibfnamefont {M.}~\bibnamefont {Rojik}},\ and\ \bibinfo
  {author} {\bibfnamefont {M.}~\bibnamefont {Schafer}},\ }\bibfield  {title}
  {\bibinfo {title} {{Accurate calculation of low energy scattering phase
  shifts of charged particles in a harmonic oscillator trap}},\ }\href
  {https://doi.org/10.1016/j.physletb.2024.139230} {\bibfield  {journal}
  {\bibinfo  {journal} {Phys. Lett. B}\ }\textbf {\bibinfo {volume} {861}},\
  \bibinfo {pages} {139230} (\bibinfo {year} {2025})},\ \Eprint
  {https://arxiv.org/abs/2410.02602} {arXiv:2410.02602 [nucl-th]} \BibitemShut
  {NoStop}%
\bibitem [{\citenamefont {Kievsky}\ \emph {et~al.}(2012)\citenamefont
  {Kievsky}, \citenamefont {Viviani},\ and\ \citenamefont
  {Marcucci}}]{Kievsky:2011gp}%
  \BibitemOpen
  \bibfield  {author} {\bibinfo {author} {\bibfnamefont {A.}~\bibnamefont
  {Kievsky}}, \bibinfo {author} {\bibfnamefont {M.}~\bibnamefont {Viviani}},\
  and\ \bibinfo {author} {\bibfnamefont {L.~E.}\ \bibnamefont {Marcucci}},\
  }\bibfield  {title} {\bibinfo {title} {{Theoretical description of three- and
  four-nucleon scattering states using bound-state-like wave functions}},\
  }\href {https://doi.org/10.1103/PhysRevC.85.014001} {\bibfield  {journal}
  {\bibinfo  {journal} {Phys. Rev. C}\ }\textbf {\bibinfo {volume} {85}},\
  \bibinfo {pages} {014001} (\bibinfo {year} {2012})},\ \Eprint
  {https://arxiv.org/abs/1109.3976} {arXiv:1109.3976 [nucl-th]} \BibitemShut
  {NoStop}%
\bibitem [{\citenamefont {Furnstahl}\ \emph {et~al.}(2014)\citenamefont
  {Furnstahl}, \citenamefont {Papenbrock},\ and\ \citenamefont
  {More}}]{Furnstahl:2013vda}%
  \BibitemOpen
  \bibfield  {author} {\bibinfo {author} {\bibfnamefont {R.~J.}\ \bibnamefont
  {Furnstahl}}, \bibinfo {author} {\bibfnamefont {T.}~\bibnamefont
  {Papenbrock}},\ and\ \bibinfo {author} {\bibfnamefont {S.~N.}\ \bibnamefont
  {More}},\ }\bibfield  {title} {\bibinfo {title} {{Systematic expansion for
  infrared oscillator basis extrapolations}},\ }\href
  {https://doi.org/10.1103/PhysRevC.89.044301} {\bibfield  {journal} {\bibinfo
  {journal} {Phys. Rev. C}\ }\textbf {\bibinfo {volume} {89}},\ \bibinfo
  {pages} {044301} (\bibinfo {year} {2014})},\ \Eprint
  {https://arxiv.org/abs/1312.6876} {arXiv:1312.6876 [nucl-th]} \BibitemShut
  {NoStop}%
\bibitem [{\citenamefont {Shirokov}\ \emph {et~al.}(2018)\citenamefont
  {Shirokov}, \citenamefont {Mazur}, \citenamefont {Mazur}, \citenamefont
  {Mazur}, \citenamefont {Shin}, \citenamefont {Kim}, \citenamefont
  {Blokhintsev},\ and\ \citenamefont {Vary}}]{Shirokov:2018nlj}%
  \BibitemOpen
  \bibfield  {author} {\bibinfo {author} {\bibfnamefont {A.~M.}\ \bibnamefont
  {Shirokov}}, \bibinfo {author} {\bibfnamefont {A.~I.}\ \bibnamefont {Mazur}},
  \bibinfo {author} {\bibfnamefont {I.~A.}\ \bibnamefont {Mazur}}, \bibinfo
  {author} {\bibfnamefont {E.~A.}\ \bibnamefont {Mazur}}, \bibinfo {author}
  {\bibfnamefont {I.~J.}\ \bibnamefont {Shin}}, \bibinfo {author}
  {\bibfnamefont {Y.}~\bibnamefont {Kim}}, \bibinfo {author} {\bibfnamefont
  {L.~D.}\ \bibnamefont {Blokhintsev}},\ and\ \bibinfo {author} {\bibfnamefont
  {J.~P.}\ \bibnamefont {Vary}},\ }\bibfield  {title} {\bibinfo {title}
  {{Nucleon-$\alpha$ Scattering and Resonances in $^5$He and $^5$Li with JISP16
  and Daejeon16 $NN$ interactions}},\ }\href
  {https://doi.org/10.1103/PhysRevC.98.044624} {\bibfield  {journal} {\bibinfo
  {journal} {Phys. Rev.C}\ }\textbf {\bibinfo {volume} {98}},\ \bibinfo {pages}
  {044624} (\bibinfo {year} {2018})},\ \Eprint
  {https://arxiv.org/abs/1808.03394} {arXiv:1808.03394 [nucl-th]} \BibitemShut
  {NoStop}%
\bibitem [{\citenamefont {Yu}\ \emph {et~al.}(2024)\citenamefont {Yu},
  \citenamefont {Yapa},\ and\ \citenamefont {K\"onig}}]{Yu:2023ucq}%
  \BibitemOpen
  \bibfield  {author} {\bibinfo {author} {\bibfnamefont {H.}~\bibnamefont
  {Yu}}, \bibinfo {author} {\bibfnamefont {N.}~\bibnamefont {Yapa}},\ and\
  \bibinfo {author} {\bibfnamefont {S.}~\bibnamefont {K\"onig}},\ }\bibfield
  {title} {\bibinfo {title} {{Complex scaling in finite volume}},\ }\href
  {https://doi.org/10.1103/PhysRevC.109.014316} {\bibfield  {journal} {\bibinfo
   {journal} {Phys. Rev. C}\ }\textbf {\bibinfo {volume} {109}},\ \bibinfo
  {pages} {014316} (\bibinfo {year} {2024})},\ \Eprint
  {https://arxiv.org/abs/2309.03196} {arXiv:2309.03196 [nucl-th]} \BibitemShut
  {NoStop}%
\bibitem [{\citenamefont {Trefethen}\ \emph {et~al.}(2021)\citenamefont
  {Trefethen}, \citenamefont {Nakatsukasa},\ and\ \citenamefont
  {Weideman}}]{Trefethen2021NodeClustering}%
  \BibitemOpen
  \bibfield  {author} {\bibinfo {author} {\bibfnamefont {L.~N.}\ \bibnamefont
  {Trefethen}}, \bibinfo {author} {\bibfnamefont {Y.}~\bibnamefont
  {Nakatsukasa}},\ and\ \bibinfo {author} {\bibfnamefont {J.~A.~C.}\
  \bibnamefont {Weideman}},\ }\bibfield  {title} {\bibinfo {title} {Exponential
  node clustering at singularities for rational approximation, quadrature, and
  pdes},\ }\href {https://doi.org/10.1007/s00211-020-01168-2} {\bibfield
  {journal} {\bibinfo  {journal} {Numerische Mathematik}\ }\textbf {\bibinfo
  {volume} {147}},\ \bibinfo {pages} {227} (\bibinfo {year}
  {2021})}\BibitemShut {NoStop}%
\bibitem [{\citenamefont {Nakatsukasa}\ \emph {et~al.}(2016)\citenamefont
  {Nakatsukasa}, \citenamefont {Matsuyanagi}, \citenamefont {Matsuo},\ and\
  \citenamefont {Yabana}}]{Nakatsukasa:2016nyc}%
  \BibitemOpen
  \bibfield  {author} {\bibinfo {author} {\bibfnamefont {T.}~\bibnamefont
  {Nakatsukasa}}, \bibinfo {author} {\bibfnamefont {K.}~\bibnamefont
  {Matsuyanagi}}, \bibinfo {author} {\bibfnamefont {M.}~\bibnamefont
  {Matsuo}},\ and\ \bibinfo {author} {\bibfnamefont {K.}~\bibnamefont
  {Yabana}},\ }\bibfield  {title} {\bibinfo {title} {{Time-dependent
  density-functional description of nuclear dynamics}},\ }\href
  {https://doi.org/10.1103/RevModPhys.88.045004} {\bibfield  {journal}
  {\bibinfo  {journal} {Rev. Mod. Phys.}\ }\textbf {\bibinfo {volume} {88}},\
  \bibinfo {pages} {045004} (\bibinfo {year} {2016})},\ \Eprint
  {https://arxiv.org/abs/1606.04717} {arXiv:1606.04717 [nucl-th]} \BibitemShut
  {NoStop}%
\bibitem [{\citenamefont {Newman}(1964)}]{Newman1964}%
  \BibitemOpen
  \bibfield  {author} {\bibinfo {author} {\bibfnamefont {D.~J.}\ \bibnamefont
  {Newman}},\ }\bibfield  {title} {\bibinfo {title} {{Rational approximation to
  $| x| $.}},\ }\href {https://doi.org/10.1307/mmj/1028999029} {\bibfield
  {journal} {\bibinfo  {journal} {Michigan Mathematical Journal}\ }\textbf
  {\bibinfo {volume} {11}},\ \bibinfo {pages} {11 } (\bibinfo {year}
  {1964})}\BibitemShut {NoStop}%
\bibitem [{\citenamefont {Nakatsukasa}\ \emph {et~al.}(2018)\citenamefont
  {Nakatsukasa}, \citenamefont {Sète},\ and\ \citenamefont
  {Trefethen}}]{Nakatsukasa_2018}%
  \BibitemOpen
  \bibfield  {author} {\bibinfo {author} {\bibfnamefont {Y.}~\bibnamefont
  {Nakatsukasa}}, \bibinfo {author} {\bibfnamefont {O.}~\bibnamefont {Sète}},\
  and\ \bibinfo {author} {\bibfnamefont {L.~N.}\ \bibnamefont {Trefethen}},\
  }\bibfield  {title} {\bibinfo {title} {The aaa algorithm for rational
  approximation},\ }\href {https://doi.org/10.1137/16m1106122} {\bibfield
  {journal} {\bibinfo  {journal} {SIAM Journal on Scientific Computing}\
  }\textbf {\bibinfo {volume} {40}},\ \bibinfo {pages} {A1494–A1522}
  (\bibinfo {year} {2018})}\BibitemShut {NoStop}%
\bibitem [{\citenamefont {Nakatsukasa}\ \emph {et~al.}(2023)\citenamefont
  {Nakatsukasa}, \citenamefont {Sete},\ and\ \citenamefont
  {Trefethen}}]{nakatsukasa2023years}%
  \BibitemOpen
  \bibfield  {author} {\bibinfo {author} {\bibfnamefont {Y.}~\bibnamefont
  {Nakatsukasa}}, \bibinfo {author} {\bibfnamefont {O.}~\bibnamefont {Sete}},\
  and\ \bibinfo {author} {\bibfnamefont {L.~N.}\ \bibnamefont {Trefethen}},\
  }\href@noop {} {\bibinfo {title} {The first five years of the aaa algorithm}}
  (\bibinfo {year} {2023}),\ \Eprint {https://arxiv.org/abs/2312.03565}
  {arXiv:2312.03565 [math.NA]} \BibitemShut {NoStop}%
\bibitem [{\citenamefont {Hofreither}(2021)}]{baryrat_2021}%
  \BibitemOpen
  \bibfield  {author} {\bibinfo {author} {\bibfnamefont {C.}~\bibnamefont
  {Hofreither}},\ }\bibfield  {title} {\bibinfo {title} {An algorithm for best
  rational approximation based on barycentric rational interpolation},\ }\href
  {https://doi.org/10.1007/s11075-020-01042-0} {\bibfield  {journal} {\bibinfo
  {journal} {Numer. Algorithms}\ }\textbf {\bibinfo {volume} {88}},\ \bibinfo
  {pages} {365–388} (\bibinfo {year} {2021})}\BibitemShut {NoStop}%
\bibitem [{\citenamefont {Xu}\ \emph {et~al.}(2022)\citenamefont {Xu},
  \citenamefont {Yan}, \citenamefont {Shi}, \citenamefont {Ankerhold},\ and\
  \citenamefont {Stockburger}}]{Xu:2022ybc}%
  \BibitemOpen
  \bibfield  {author} {\bibinfo {author} {\bibfnamefont {M.}~\bibnamefont
  {Xu}}, \bibinfo {author} {\bibfnamefont {Y.}~\bibnamefont {Yan}}, \bibinfo
  {author} {\bibfnamefont {Q.}~\bibnamefont {Shi}}, \bibinfo {author}
  {\bibfnamefont {J.}~\bibnamefont {Ankerhold}},\ and\ \bibinfo {author}
  {\bibfnamefont {J.~T.}\ \bibnamefont {Stockburger}},\ }\bibfield  {title}
  {\bibinfo {title} {{Taming Quantum Noise for Efficient Low Temperature
  Simulations of Open Quantum Systems}},\ }\href
  {https://doi.org/10.1103/PhysRevLett.129.230601} {\bibfield  {journal}
  {\bibinfo  {journal} {Phys. Rev. Lett.}\ }\textbf {\bibinfo {volume} {129}},\
  \bibinfo {pages} {230601} (\bibinfo {year} {2022})},\ \Eprint
  {https://arxiv.org/abs/2202.04059} {arXiv:2202.04059 [quant-ph]} \BibitemShut
  {NoStop}%
\bibitem [{\citenamefont {Huang}\ \emph {et~al.}(2023)\citenamefont {Huang},
  \citenamefont {Gull},\ and\ \citenamefont {Lin}}]{Huang_2023}%
  \BibitemOpen
  \bibfield  {author} {\bibinfo {author} {\bibfnamefont {Z.}~\bibnamefont
  {Huang}}, \bibinfo {author} {\bibfnamefont {E.}~\bibnamefont {Gull}},\ and\
  \bibinfo {author} {\bibfnamefont {L.}~\bibnamefont {Lin}},\ }\bibfield
  {title} {\bibinfo {title} {Robust analytic continuation of green’s
  functions via projection, pole estimation, and semidefinite relaxation},\
  }\bibfield  {journal} {\bibinfo  {journal} {Physical Review B}\ }\textbf
  {\bibinfo {volume} {107}},\ \href
  {https://doi.org/10.1103/physrevb.107.075151} {10.1103/physrevb.107.075151}
  (\bibinfo {year} {2023})\BibitemShut {NoStop}%
\bibitem [{\citenamefont {Goswami}\ \emph {et~al.}(2024)\citenamefont
  {Goswami}, \citenamefont {Barros},\ and\ \citenamefont
  {Carbone}}]{Goswami_2024}%
  \BibitemOpen
  \bibfield  {author} {\bibinfo {author} {\bibfnamefont {S.}~\bibnamefont
  {Goswami}}, \bibinfo {author} {\bibfnamefont {K.}~\bibnamefont {Barros}},\
  and\ \bibinfo {author} {\bibfnamefont {M.~R.}\ \bibnamefont {Carbone}},\
  }\bibfield  {title} {\bibinfo {title} {Physically interpretable
  approximations of many-body spectral functions},\ }\bibfield  {journal}
  {\bibinfo  {journal} {Physical Review E}\ }\textbf {\bibinfo {volume}
  {109}},\ \href {https://doi.org/10.1103/physreve.109.015302}
  {10.1103/physreve.109.015302} (\bibinfo {year} {2024})\BibitemShut {NoStop}%
\bibitem [{\citenamefont {Ying}(2022)}]{Ying_2022}%
  \BibitemOpen
  \bibfield  {author} {\bibinfo {author} {\bibfnamefont {L.}~\bibnamefont
  {Ying}},\ }\bibfield  {title} {\bibinfo {title} {Analytic continuation from
  limited noisy matsubara data},\ }\href
  {https://doi.org/10.1016/j.jcp.2022.111549} {\bibfield  {journal} {\bibinfo
  {journal} {Journal of Computational Physics}\ }\textbf {\bibinfo {volume}
  {469}},\ \bibinfo {pages} {111549} (\bibinfo {year} {2022})}\BibitemShut
  {NoStop}%
\bibitem [{\citenamefont {Pomraning}(1965)}]{Pomraning1965}%
  \BibitemOpen
  \bibfield  {author} {\bibinfo {author} {\bibfnamefont {G.~C.}\ \bibnamefont
  {Pomraning}},\ }\bibfield  {title} {\bibinfo {title} {A variational principle
  for linear systems},\ }\href {http://www.jstor.org/stable/2946446} {\bibfield
   {journal} {\bibinfo  {journal} {Journal of the Society for Industrial and
  Applied Mathematics}\ }\textbf {\bibinfo {volume} {13}},\ \bibinfo {pages}
  {511} (\bibinfo {year} {1965})}\BibitemShut {NoStop}%
\bibitem [{\citenamefont {Sarkar}\ and\ \citenamefont
  {Lee}(2021{\natexlab{b}})}]{sarkar2021selflearning}%
  \BibitemOpen
  \bibfield  {author} {\bibinfo {author} {\bibfnamefont {A.}~\bibnamefont
  {Sarkar}}\ and\ \bibinfo {author} {\bibfnamefont {D.}~\bibnamefont {Lee}},\
  }\href@noop {} {\bibinfo {title} {Self-learning emulators and eigenvector
  continuation}} (\bibinfo {year} {2021}{\natexlab{b}}),\ \Eprint
  {https://arxiv.org/abs/2107.13449} {arXiv:2107.13449 [nucl-th]} \BibitemShut
  {NoStop}%
\bibitem [{\citenamefont {Maldonado}\ \emph {et~al.}(2025)\citenamefont
  {Maldonado}, \citenamefont {Drischler}, \citenamefont {Furnstahl},\ and\
  \citenamefont {Mlinari{\'c}}}]{Maldonado:2025ftg}%
  \BibitemOpen
  \bibfield  {author} {\bibinfo {author} {\bibfnamefont {J.~M.}\ \bibnamefont
  {Maldonado}}, \bibinfo {author} {\bibfnamefont {C.}~\bibnamefont
  {Drischler}}, \bibinfo {author} {\bibfnamefont {R.~J.}\ \bibnamefont
  {Furnstahl}},\ and\ \bibinfo {author} {\bibfnamefont {P.}~\bibnamefont
  {Mlinari{\'c}}},\ }\bibfield  {title} {\bibinfo {title} {{Greedy emulators
  for nuclear two-body scattering}},\ }\href
  {https://doi.org/10.1103/k77q-f82l} {\bibfield  {journal} {\bibinfo
  {journal} {Phys. Rev. C}\ }\textbf {\bibinfo {volume} {112}},\ \bibinfo
  {pages} {024002} (\bibinfo {year} {2025})},\ \Eprint
  {https://arxiv.org/abs/2504.06092} {arXiv:2504.06092} \BibitemShut {NoStop}%
\bibitem [{\citenamefont {Tang}(1993)}]{doi:10.1080/01621459.1993.10476423}%
  \BibitemOpen
  \bibfield  {author} {\bibinfo {author} {\bibfnamefont {B.}~\bibnamefont
  {Tang}},\ }\bibfield  {title} {\bibinfo {title} {Orthogonal array-based latin
  hypercubes},\ }\href {https://doi.org/10.1080/01621459.1993.10476423}
  {\bibfield  {journal} {\bibinfo  {journal} {Journal of the American
  Statistical Association}\ }\textbf {\bibinfo {volume} {88}},\ \bibinfo
  {pages} {1392} (\bibinfo {year} {1993})}\BibitemShut {NoStop}%
\bibitem [{\citenamefont {Zhang}(2025)}]{Zhang:2024gac_SM}%
  \BibitemOpen
  \bibfield  {author} {\bibinfo {author} {\bibfnamefont {X.}~\bibnamefont
  {Zhang}},\ }\href {http://link.aps.org/supplemental/} {\bibinfo {title}
  {Supplemental material for non-hermitian quantum mechanics approach for
  extracting and emulating continuum physics based on bound-state-like
  calculations: Detailed description}} (\bibinfo {year} {2025}),\ \bibinfo
  {note} {including reference~\cite{THOMPSON197753}}\BibitemShut {NoStop}%
\bibitem [{\citenamefont {Thompson}\ \emph {et~al.}(1977)\citenamefont
  {Thompson}, \citenamefont {Lemere},\ and\ \citenamefont
  {Tang}}]{THOMPSON197753}%
  \BibitemOpen
  \bibfield  {author} {\bibinfo {author} {\bibfnamefont {D.}~\bibnamefont
  {Thompson}}, \bibinfo {author} {\bibfnamefont {M.}~\bibnamefont {Lemere}},\
  and\ \bibinfo {author} {\bibfnamefont {Y.}~\bibnamefont {Tang}},\ }\bibfield
  {title} {\bibinfo {title} {Systematic investigation of scattering problems
  with the resonating-group method},\ }\href
  {https://doi.org/https://doi.org/10.1016/0375-9474(77)90007-0} {\bibfield
  {journal} {\bibinfo  {journal} {Nucl. Phys. A}\ }\textbf {\bibinfo {volume}
  {286}},\ \bibinfo {pages} {53 } (\bibinfo {year} {1977})}\BibitemShut
  {NoStop}%
\bibitem [{\citenamefont {Zhang}\ \emph {et~al.}(2018)\citenamefont {Zhang},
  \citenamefont {Nollett},\ and\ \citenamefont {Phillips}}]{Zhang:2017yqc}%
  \BibitemOpen
  \bibfield  {author} {\bibinfo {author} {\bibfnamefont {X.}~\bibnamefont
  {Zhang}}, \bibinfo {author} {\bibfnamefont {K.~M.}\ \bibnamefont {Nollett}},\
  and\ \bibinfo {author} {\bibfnamefont {D.~R.}\ \bibnamefont {Phillips}},\
  }\bibfield  {title} {\bibinfo {title} {{Models, measurements, and effective
  field theory: Proton capture on $^7Be$ at next-to-leading order}},\ }\href
  {https://doi.org/10.1103/PhysRevC.98.034616} {\bibfield  {journal} {\bibinfo
  {journal} {Phys. Rev. C}\ }\textbf {\bibinfo {volume} {98}},\ \bibinfo
  {pages} {034616} (\bibinfo {year} {2018})},\ \Eprint
  {https://arxiv.org/abs/1708.04017} {arXiv:1708.04017 [nucl-th]} \BibitemShut
  {NoStop}%
\bibitem [{\citenamefont {Phillips}(1966)}]{Phillips:1966zza}%
  \BibitemOpen
  \bibfield  {author} {\bibinfo {author} {\bibfnamefont {A.}~\bibnamefont
  {Phillips}},\ }\bibfield  {title} {\bibinfo {title} {{Application of the
  Faddeev Equations to the Three-Nucleon Problem}},\ }\href
  {https://doi.org/10.1103/PhysRev.142.984} {\bibfield  {journal} {\bibinfo
  {journal} {Phys. Rev.}\ }\textbf {\bibinfo {volume} {142}},\ \bibinfo {pages}
  {984} (\bibinfo {year} {1966})}\BibitemShut {NoStop}%
\bibitem [{\citenamefont {Hammer}\ and\ \citenamefont
  {Phillips}(2011)}]{Hammer:2011ye}%
  \BibitemOpen
  \bibfield  {author} {\bibinfo {author} {\bibfnamefont {H.~W.}\ \bibnamefont
  {Hammer}}\ and\ \bibinfo {author} {\bibfnamefont {D.~R.}\ \bibnamefont
  {Phillips}},\ }\bibfield  {title} {\bibinfo {title} {{Electric properties of
  the Beryllium-11 system in Halo EFT}},\ }\href
  {https://doi.org/10.1016/j.nuclphysa.2011.06.028} {\bibfield  {journal}
  {\bibinfo  {journal} {Nucl. Phys.}\ }\textbf {\bibinfo {volume} {A865}},\
  \bibinfo {pages} {17} (\bibinfo {year} {2011})},\ \Eprint
  {https://arxiv.org/abs/1103.1087} {arXiv:1103.1087 [nucl-th]} \BibitemShut
  {NoStop}%
\bibitem [{\citenamefont {Hicks}\ and\ \citenamefont
  {Lee}(2023)}]{Hicks:2022ovs}%
  \BibitemOpen
  \bibfield  {author} {\bibinfo {author} {\bibfnamefont {C.}~\bibnamefont
  {Hicks}}\ and\ \bibinfo {author} {\bibfnamefont {D.}~\bibnamefont {Lee}},\
  }\bibfield  {title} {\bibinfo {title} {{Trimmed sampling algorithm for the
  noisy generalized eigenvalue problem}},\ }\href
  {https://doi.org/10.1103/PhysRevResearch.5.L022001} {\bibfield  {journal}
  {\bibinfo  {journal} {Phys. Rev. Res.}\ }\textbf {\bibinfo {volume} {5}},\
  \bibinfo {pages} {L022001} (\bibinfo {year} {2023})},\ \Eprint
  {https://arxiv.org/abs/2209.02083} {arXiv:2209.02083 [nucl-th]} \BibitemShut
  {NoStop}%
\bibitem [{\citenamefont {Cook}\ \emph {et~al.}(2025)\citenamefont {Cook},
  \citenamefont {Jammooa}, \citenamefont {Hjorth-Jensen}, \citenamefont {Lee},\
  and\ \citenamefont {Lee}}]{Cook:2024toj}%
  \BibitemOpen
  \bibfield  {author} {\bibinfo {author} {\bibfnamefont {P.}~\bibnamefont
  {Cook}}, \bibinfo {author} {\bibfnamefont {D.}~\bibnamefont {Jammooa}},
  \bibinfo {author} {\bibfnamefont {M.}~\bibnamefont {Hjorth-Jensen}}, \bibinfo
  {author} {\bibfnamefont {D.~D.}\ \bibnamefont {Lee}},\ and\ \bibinfo {author}
  {\bibfnamefont {D.}~\bibnamefont {Lee}},\ }\bibfield  {title} {\bibinfo
  {title} {{Parametric matrix models}},\ }\href
  {https://doi.org/10.1038/s41467-025-61362-4} {\bibfield  {journal} {\bibinfo
  {journal} {Nature Commun.}\ }\textbf {\bibinfo {volume} {16}},\ \bibinfo
  {pages} {5929} (\bibinfo {year} {2025})},\ \Eprint
  {https://arxiv.org/abs/2401.11694} {arXiv:2401.11694 [cs.LG]} \BibitemShut
  {NoStop}%
\bibitem [{{\relax DLMF}()}]{NIST:DLMF}%
  \BibitemOpen
  {\relax DLMF},\ \href {http://dlmf.nist.gov/} {\bibinfo {title} {{\it NIST
  Digital Library of Mathematical Functions}}},\ \bibinfo {howpublished}
  {http://dlmf.nist.gov/, Release 1.1.0 of 2020-12-15},\ \bibinfo {note}
  {f.~W.~J. Olver, A.~B. {Olde Daalhuis}, D.~W. Lozier, B.~I. Schneider, R.~F.
  Boisvert, C.~W. Clark, B.~R. Miller, B.~V. Saunders, H.~S. Cohl, and M.~A.
  McClain, eds.}\BibitemShut {Stop}%
\bibitem [{\citenamefont {Matsuyama}\ \emph {et~al.}(2007)\citenamefont
  {Matsuyama}, \citenamefont {Sato},\ and\ \citenamefont
  {Lee}}]{Matsuyama:2006rp}%
  \BibitemOpen
  \bibfield  {author} {\bibinfo {author} {\bibfnamefont {A.}~\bibnamefont
  {Matsuyama}}, \bibinfo {author} {\bibfnamefont {T.}~\bibnamefont {Sato}},\
  and\ \bibinfo {author} {\bibfnamefont {T.~S.~H.}\ \bibnamefont {Lee}},\
  }\bibfield  {title} {\bibinfo {title} {{Dynamical coupled-channel model of
  meson production reactions in the nucleon resonance region}},\ }\href
  {https://doi.org/10.1016/j.physrep.2006.12.003} {\bibfield  {journal}
  {\bibinfo  {journal} {Phys. Rept.}\ }\textbf {\bibinfo {volume} {439}},\
  \bibinfo {pages} {193} (\bibinfo {year} {2007})},\ \Eprint
  {https://arxiv.org/abs/nucl-th/0608051} {arXiv:nucl-th/0608051} \BibitemShut
  {NoStop}%
\bibitem [{\citenamefont {Larson}\ and\ \citenamefont
  {Hetherington}(1974)}]{Larson:1974zza}%
  \BibitemOpen
  \bibfield  {author} {\bibinfo {author} {\bibfnamefont {N.~M.}\ \bibnamefont
  {Larson}}\ and\ \bibinfo {author} {\bibfnamefont {J.~H.}\ \bibnamefont
  {Hetherington}},\ }\bibfield  {title} {\bibinfo {title} {{Solution of the
  Faddeev integral equation without contour rotation}},\ }\href
  {https://doi.org/10.1103/PhysRevC.9.699} {\bibfield  {journal} {\bibinfo
  {journal} {Phys. Rev. C}\ }\textbf {\bibinfo {volume} {9}},\ \bibinfo {pages}
  {699} (\bibinfo {year} {1974})}\BibitemShut {NoStop}%
\bibitem [{\citenamefont {Landau}(1996)}]{Landau:1996}%
  \BibitemOpen
  \bibfield  {author} {\bibinfo {author} {\bibfnamefont {R.~H.}\ \bibnamefont
  {Landau}},\ }\href@noop {} {\emph {\bibinfo {title} {Quantum Mechanics
  II}}},\ \bibinfo {edition} {2nd}\ ed.\ (\bibinfo  {publisher} {John Wiley \&
  Sons, Inc.},\ \bibinfo {address} {New York},\ \bibinfo {year}
  {1996})\BibitemShut {NoStop}%
\end{thebibliography}
%

\end{document}